\documentclass[%
 aip,
 amsmath,amssymb,
 reprint,%
nofootinbib]{revtex4-2}

\usepackage{graphicx}
\usepackage{dcolumn}
\usepackage{bm}
\usepackage{adjustbox}
\usepackage[utf8]{inputenc}
\usepackage[T1]{fontenc}
\usepackage{mathptmx}

\begin{document}

\preprint{AIP/123-QED}

\title{Quantum Monte Carlo and density functional theory study of strain and magnetism in 2D 1T-VSe$_2$ with charge density wave states}

\author{Daniel Wines}
\email{daniel.wines@nist.gov}
\affiliation{Material Measurement Laboratory, National Institute of Standards and Technology (NIST),
Gaithersburg, Maryland 20899, USA}

\author{Akram Ibrahim}
\affiliation{%
Department of Physics, University of Maryland Baltimore County, Baltimore, Maryland 21250, USA
}%

\author{Nishwanth Gudibandla}
\affiliation{Material Measurement Laboratory, National Institute of Standards and Technology (NIST),
Gaithersburg, MD 20899, USA}
\affiliation{%
Department of Physics, University of Maryland Baltimore County, Baltimore, Maryland 21250, USA
}%

\author{Tehseen Adel}
\affiliation{Physical Measurement Laboratory, National Institute of Standards and Technology (NIST),
Gaithersburg, Maryland 20899, USA}
\affiliation{ Department of Physical Sciences, University of Findlay, Findlay, Ohio 45840, USA}

\author{Frank M. Abel}
\affiliation{Material Measurement Laboratory, National Institute of Standards and Technology (NIST),
Gaithersburg, Maryland 20899, USA}
\affiliation{United States Naval Academy, Annapolis, Maryland 21402, USA}

\author{Sharadh Jois}
\affiliation{Laboratory for Physical Sciences,
College Park, Maryland 20740, USA}

\author{Kayahan Saritas}%

\affiliation{ 
Material Science and Technology Division, Oak Ridge National Laboratory, Oak Ridge, Tennessee 37831, USA
}%

\author{Jaron T. Krogel}%

\affiliation{ 
Material Science and Technology Division, Oak Ridge National Laboratory, Oak Ridge, Tennessee 37831, USA
}%

\author{Li Yin}%
\affiliation{ 
Department of Physics and Engineering Physics, Tulane University, New Orleans, Louisiana 70118, USA
}%

\author{Tom Berlijn}%

\affiliation{ 
Center for Nanophase Materials Sciences, Oak Ridge National Laboratory, Oak Ridge, Tennessee 37831, USA
}%

\author{Aubrey T. Hanbicki}
\affiliation{Laboratory for Physical Sciences,
College Park, Maryland 20740, USA}

\author{Gregory M. Stephen}
\affiliation{Laboratory for Physical Sciences,
College Park, Maryland 20740, USA}

\author{Adam L. Friedman}
\affiliation{Laboratory for Physical Sciences,
College Park, Maryland 20740, USA}

\author{Sergiy Krylyuk}
\affiliation{Material Measurement Laboratory, National Institute of Standards and Technology (NIST),
Gaithersburg, Maryland 20899, USA}

\author{Albert V. Davydov}
\affiliation{Material Measurement Laboratory, National Institute of Standards and Technology (NIST),
Gaithersburg, Maryland 20899, USA}

\author{Brian Donovan}
\affiliation{United States Naval Academy, Annapolis, Maryland 21402, USA}

\author{Michelle E. Jamer}
\affiliation{United States Naval Academy, Annapolis, Maryland 21402, USA}

\author{Angela R. Hight Walker}
\affiliation{Physical Measurement Laboratory, National Institute of Standards and Technology (NIST),
Gaithersburg, Maryland 20899, USA}

\author{Kamal Choudhary}
\affiliation{Material Measurement Laboratory, National Institute of Standards and Technology (NIST),
Gaithersburg, Maryland 20899, USA}

\author{Francesca Tavazza}
\affiliation{Material Measurement Laboratory, National Institute of Standards and Technology (NIST),
Gaithersburg, Maryland 20899, USA}

\author{Can Ataca*}
 \email{ataca@umbc.edu}
\affiliation{%
Department of Physics, University of Maryland Baltimore County, Baltimore, Maryland 21250, USA
}%

\date{\today}

\begin{abstract}
\textbf{Abstract}

Two-dimensional (2D) 1T-VSe$_2$ has prompted significant interest due to the discrepancies regarding alleged ferromagnetism (FM) at room temperature, charge density wave (CDW) states and the interplay between the two. We employed a combined Diffusion Monte Carlo (DMC) and density functional theory (DFT) approach to accurately investigate the magnetic properties, CDW states, and their response to strain in monolayer 1T-VSe$_2$. 
Our calculations show the delicate competition between various phases, revealing critical insights into the relationship between their energetic and structural properties. We performed classical Monte Carlo simulations informed by our DMC and DFT results, and found the magnetic transition temperature ($T_c$) of the undistorted (non-CDW) FM phase to be 228 K and the distorted (CDW) phase to be 68 K. Additionally, we studied the response of biaxial strain on the energetic stability and magnetic properties of various phases of 2D 1T-VSe$_2$ and found that small amounts of strain can increase the $T_c$, suggesting a promising route for engineering and enhancing magnetic behavior. Finally, we synthesized 1T-VSe$_2$ and performed Raman spectroscopy measurements, which were in close agreement with our calculated results, validating our computational approach. Our work emphasizes the role of highly accurate DMC methods in advancing the understanding of monolayer 1T-VSe$_2$ and provides a robust framework for future studies of 2D magnetic materials.

\end{abstract}

\maketitle

\section{Introduction}

Strongly correlated two-dimensional (2D) magnets are a puzzling class of materials from a fundamental physics perspective. With the discovery of ferromagnetism in 2D systems such as CrI$_3$ \cite{cri3} and Cr$_2$Ge$_2$Te$_6$ \cite{cr2ge2te6}, an effort to identify and understand the underlying mechanisms of 2D magnets with a finite transition temperature ($T_c$) has become a highly active area of materials science. One of the most interesting and controversial 2D ferromagnetic materials is VSe$_2$, which has a metallic 1T phase (octahedral (1T)-centered honeycombs) and a semiconducting 2H phase (trigonal prismatic (2H)-hexagonal honeycomb)\cite{ataca-mx2}. Despite discrepancies of the structural properties coupled to the energetic stability (whether or not 1T vs. 2H is more favorable) \cite{struc-phase,C9CP03726H,C6CP06732H,vse2-exp} that we were successfully able to resolve in our previous work using highly-accurate electronic structure methods (see Ref. \onlinecite{vse2-wines}), there are several remaining questions regarding 2D 1T-VSe$_2$.     

In a 2018 study, single layer 1T-VSe$_2$ was synthesized on van der Waals (vdW) substrates (graphite and MoS$_2$), where strong ferromagnetic ordering was measured above room temperature and a charge density wave (CDW) was detected with a transition temperature of 121 K \cite{vse2-1stexp}. This was the first study to demonstrate that the CDW transition of single layer 1T-VSe$_2$ is coupled with its magnetic properties \cite{vse2-1stexp}. Despite room temperature ferromagnetism being reproduced in some cases, such as for a chemically exfoliated monolayer \cite{https://doi.org/10.1002/adma.201903779}, there have been several conflicting reports of the magnetic properties and the interplay between the CDW state and magnetism \cite{vse2-1stexp,https://doi.org/10.1002/adma.201903779,PhysRevMaterials.6.014006,vse2-moment-exp,cdw,cdw2,cdw3,cdw4,PhysRevLett.121.196402,PhysRevB.101.014514,PhysRevMaterials.6.014006,original-vse2,He_2021,PhysRevB.96.235147,yilmaz2024evolutionfermisurface1tvse2}. For example, some studies have found the nonmagnetic CDW phase to be experimentally favorable, with an inherent absence of ferromagnetism \cite{cdw,cdw2,cdw3,cdw4}.  

Several theoretical studies have attempted to explain why these discrepancies in magnetic properties might occur, citing strain, vacancies, substrate choice, doping and chemical functionalization as possible reasons \cite{hydro,interface-vse2,strain-induced,PhysRevB.106.085117,WEI2023170683,D2CP01537D,D2MH00888B,D0NR04663A,wines2024improvedpropertypredictiontwodimensional}. These studies also highlight the strong sensitivity of the magnetic properties of 2D 1T-VSe$_2$ to extrinsic factors, which can provide a viable route to tune such properties. In our previous density functional theory (DFT) work, we explored the competing magnetic and nonmagnetic states in single layer 1T-VSe$_2$ with and without charge density wave \cite{PhysRevB.106.085117}. We found that there is strong competition between nonmagnetic and magnetic states in the CDW structures (with respect to the undistorted structure), with relative energies being on the the scale of one meV per formula unit (f.u.) \cite{PhysRevB.106.085117}. This implies that it is possible for antiferromagnetic ordering to compete with ferromagnetic ordering and the CDW state \cite{PhysRevB.106.085117}.

Although our previous semi-local DFT calculations \cite{PhysRevB.106.085117} provide a good qualitative assessment of how different magnetic orderings can compete in the undistorted and CDW structures, a more quantitative answer is required to accurately understand the magnetic and CDW transitions in monolayer 1T-VSe$_2$ and estimate quantities such as transition temperatures. For this reason, high-fidelity many-body techniques such as fixed-node Diffusion Monte Carlo (DMC) \cite{RevModPhys.73.33} can be utilized to accurately describe the electronic and magnetic properties of 2D 1T-VSe$_2$. DMC is a correlated electronic structure method that has a reduced sensitivity to approximations such as the exchange-correlation functional and the Hubbard $U$ \cite{PhysRevB.57.1505} correction. In addition, DMC has successfully been applied to several 2D and quasi-2D systems \cite{vse2-wines,PhysRevX.4.031003,doi:10.1063/5.0023223,wines2021pathway,mno2-qmc,https://doi.org/10.48550/arxiv.2209.10379,PhysRevMaterials.5.024002,staros,PhysRevB.106.075127,PhysRevResearch.5.033223,10.1063/5.0116092,PhysRevResearch.6.013007,PhysRevB.98.085429,D1CP02473F,10.1063/5.0030952,bluephos,chern}. In this work, we investigate the magnetic properties of monolayer 1T-VSe$_2$ in various magnetic states and geometries, with and without CDW distortions, through the lens of DMC and provide a thorough benchmark using several DFT functionals. In addition to providing accurate quantitative estimates of energy differences between different phases, we coupled our DMC results with classical Monte Carlo simulations to estimate magnetic transition temperatures. Finally, we used our DMC insights to perform DFT calculations of the Raman modes and compared to our own experimental results on a synthesized 1T-VSe$_2$ flake.

\begin{figure*}
    \centering
    \includegraphics[trim={0. 0cm 0 0cm},clip,width=0.8\textwidth]{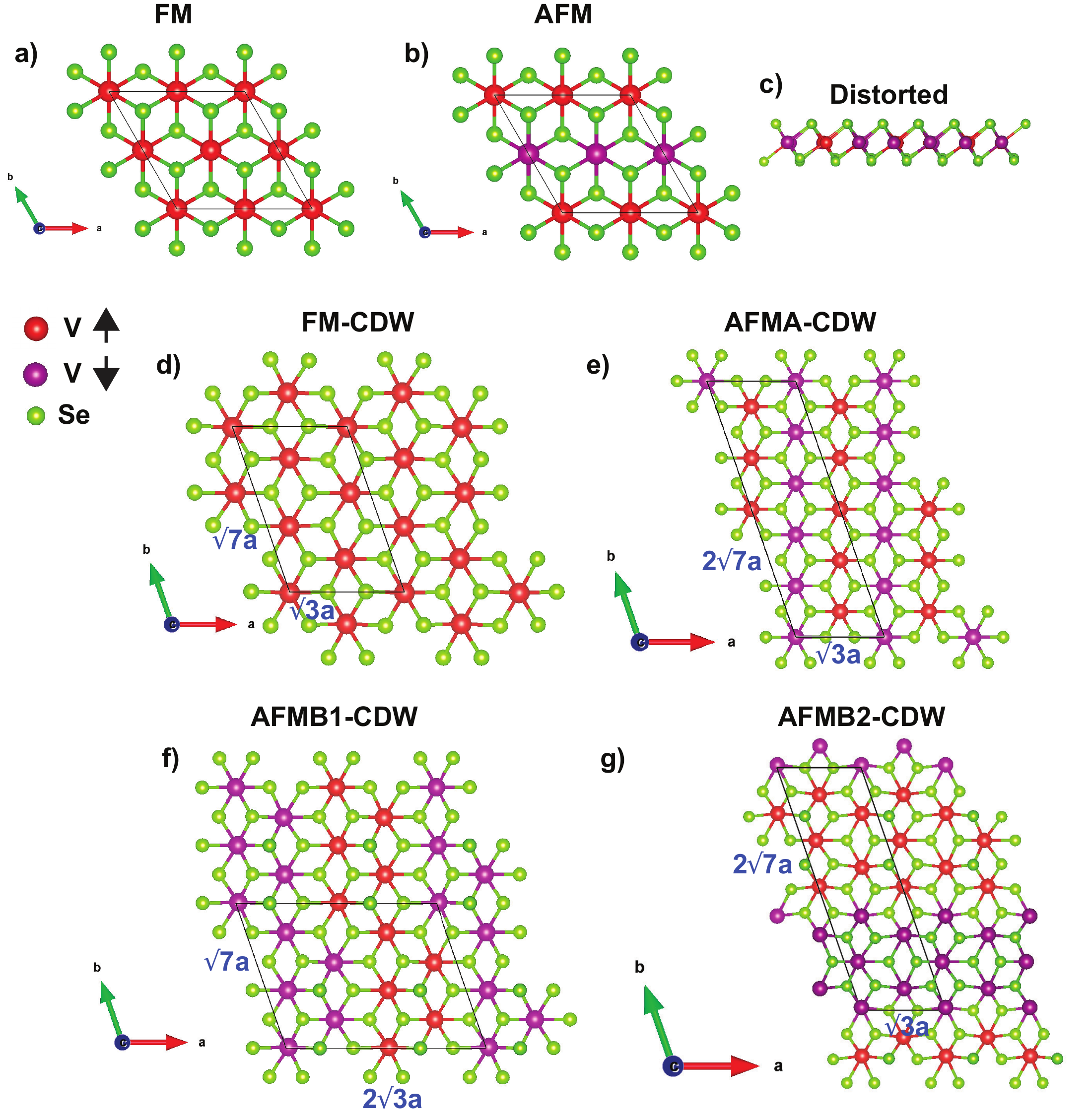}
    \caption{An overview of the various magnetic states that can exist in the undistorted and distorted (CDW) phases of monolayer VSe$_2$: a) undistorted FM, b) undistorted AFM, d) distorted FM-CDW, e) distorted AFMA-CDW, f) distorted AFMB1-CDW, and g) distorted AFMB2-CDW. c) displays an example of the distorted structure from the side view. Supercell dimensions are indicated for the distorted CDW supercells. Green indicates Se, red indicates spin-up V and purple indicates spin-down V.}
    \label{structure}
\end{figure*}

\section{Results and Discussion}

In this work, we specifically focused on monolayer 1T-VSe$_2$ in its freestanding form. Motivated by the experimental discrepancies of controversial room temperature ferromagnetism, competing magnetism and CDW states in the 1T phase, we decided to perform in-depth benchmarking at the DFT level using a variety of approximations. A summary of the different structures and magnetic configurations we studied are given in Fig. \ref{structure}. We focused on the normal (undistorted) structures of 2D 1T-VSe$_2$ in its FM, AFM (spins are anti-aligned in a stripy pattern) and nonmagnetic (NM) orientations. In addition to the normal unperturbed crystal structure of 1T-VSe$_2$, we studied the distorted $\sqrt{3}$x$\sqrt{7}$x1 supercell which is a signature of the material being in its CDW state. This distorted $\sqrt{3}$x$\sqrt{7}$x1 CDW structure has previously been studied extensively with DFT and experimentally verified. Building on the work of Ref. \onlinecite{PhysRevB.106.085117}, we performed a thorough DFT benchmark (using several different approaches) of the various magnetic states of the distorted CDW structure. These configurations included FM, NM, and various AFM configurations. Consistent with Ref. \onlinecite{PhysRevB.106.085117}, we studied the $\sqrt{3}$x$\sqrt{7}$x1 distorted AFM-A, AFM-B1 and AFM-B2 configurations (see Fig. \ref{structure}). It is important to note that the distorted AFM-A configuration is equivalent to the AFM (stripy) configuration for the undistorted structure.


In our previous work (Ref. \onlinecite{vse2-wines}), we determined the optimal geometry (including lattice parameter and bond distance) and relative phase stability of 2D 1T- and 2H-VSe$_2$ in the FM orientation with DMC and benchmarked with several DFT functionals with and without the Hubbard $U$ correction. Fig. S1a) depicts the DFT benchmarking calculations for the NM, FM, and AFM orientations of the undistorted structure. We performed DFT calculations with LDA, PBE, SCAN, r$^2$SCAN with $U$ values ranging from (0 to 3) eV, where we fully relaxed each structure in its respective magnetic orientation. As seen in Fig. S1a), which plots the energy differences with respect to the NM state for each functional, the discrepancies with each DFT method are enormous . Although qualitatively the results are somewhat similar (i.e., the FM state being lowest in energy across the board for most DFT functionals), quantitatively, the results vary drastically. In fact, the energy per formula unit (f.u.) between the NM orientation and the FM/AFM orientations can vary up to $\approx$ 0.4 eV. A quantitative energy difference between magnetic states is absolutely necessary to obtain magnetic exchange parameters ($J$) and therefore transition temperature ($T_c$) with reasonable accuracy.

\begin{table}[]
\caption{A summary of DMC and meta-GGA computed energy differences for different magnetic states of the undistorted and CDW-distorted phases of 2D 1T-VSe$_2$. DMC uncertainties (standard error of the mean) are given in shorthand form. This tabulated data can be visulized in Fig. S1. }
\begin{tabular}{l|l|l|l}
Energy Difference & DMC        
& SCAN    & r$^2$SCAN  \\
(eV/f.u.) & & & \\
\hline
E$_\textrm{FM}$ - E$_\textrm{AFM}$                      & -0.07(1)                     & -0.054 & -0.044 \\
E$_\textrm{FM}$ - E$_\textrm{NM}$                       & -0.20(1) & -0.187  & -0.166  \\
E$_\textrm{AFM}$ - E$_\textrm{NM}$                       & -0.13(1)                     & -0.133  & -0.122  \\
E$_\textrm{CDW,FM}$ - E$_\textrm{NM}$                    & -0.23(1)                     & -0.184  & -0.168  \\
E$_\textrm{CDW,NM}$ - E$_\textrm{NM}$                   & -                            & 0.010   & -0.005  \\
E$_\textrm{CDW,AFMA}$ - E$_\textrm{NM}$                  & -                            & -0.122  & -0.120  \\
E$_\textrm{CDW,AFMB1}$ - E$_\textrm{NM}$                  & -                            & -0.161  & -0.148  \\
E$_\textrm{CDW,AFMB2}$ - E$_\textrm{NM}$                  & -                            & -0.173  & -0.157 
\end{tabular}
\label{dmc-table}
\end{table}

In order to overcome these shortcomings of local and semilocal DFT, we performed DMC calculations to obtain accurate total energies of the NM, FM and AFM orientations of the normal undistorted structure of monolayer 1T-VSe$_2$. As a starting geometry for our DMC calculations, we used the structure obtained from our previous work in Ref. \onlinecite{vse2-wines}. We acknowledge that the local magnetic ordering can impact the structural properties, but for our DMC calculations we used the same structure for NM, FM and AFM calculations. In order to study the influence of using the same geometry for NM, AFM and FM on the energetics, we performed DFT calculations with the fixed FM geometry for all magnetic orientations and found a maximum energy difference of 10 meV between the fixed and relaxed structures, which is within the error bar of our DMC calculations. To reduce finite-size errors, we performed DMC calculations for two reasonably sized supercells for NM, FM and AFM and extrapolated to the thermodynamic limit. We used the following supercell sizes: 36 and 72 atoms for NM and AFM, 27 and 48 atoms for FM. The DMC energy differences and uncertainties are depicted in Fig. S1 and Table \ref{dmc-table} (in shorthand notation). From the figure and table, we see that SCAN and r$^2$SCAN (with no Hubbard $U$ correction) are in closest agreement with our DMC benchmark. For the FM configuration, we also find that SCAN and r$^2$SCAN successfully reproduce the DMC geometry that was previously obtained in Ref. \onlinecite{vse2-wines}. 

Although other semilocal DFT functionals such as PBE+$U$ ($U$ = 1, 2 eV) can do a sufficient job at reproducing our DMC geometry, they fail in terms of correctly capturing the quantitative energy differences of various magnetic states. Meta-GGA functionals (SCAN, r$^2$SCAN) can be quite successful for 2D magnetic materials such as VSe$_2$. SCAN and r$^2$SCAN include the kinetic energy density (in addition to the electron density and its gradient) in the exchange-correlation functional, which allows them to capture complicated electronic interactions. These functionals can reduce the self-interaction errors of typical GGAs and improve the description of correlation effects. SCAN satisfies all 17 exact known constraints that a meta-GGA can satisfy. r$^2$SCAN is a regularized version of SCAN designed to improve numerical stability without hindering accuracy (r$^2$SCAN is meant to obtain similar results to SCAN) \cite{r2scan}. This does so by breaking some of the exact constraints of SCAN.

Although these functionals can simultaneously capture multiple properties such as energy differences between magnetic phases and optimal geometry, SCAN and r$^2$SCAN have been widely reported to overestimate on-site magnetic moments \cite{met13040728,r2scan-mag,PhysRevB.98.094413,PhysRevB.100.045126}. In the case of 1T-VSe$_2$, we found the DMC on-site magnetic moment of the V atom to be 1.06(2) $\mu_{B}$ in Ref. \onlinecite{vse2-wines}. We found the on-site moment to be 1.40 $\mu_{B}$ with SCAN and 1.43 $\mu_{B}$ with r$^2$SCAN. 
Nonetheless, due to the success of multi-property predictions with SCAN and r$^2$SCAN with respect to our DMC benchmark calculations, we decided to use SCAN to obtain more complex quantities (that are difficult to obtain with DMC) such as magnetic anisotropy energies. For computationally demanding simulations such as for the Raman active modes where it is not computationally tractable to use SCAN or r$^2$SCAN, we used PBE+$U$ ($U$ = 1 eV) due to its relative qualitative agreement with DMC results (in terms of structural properties).


Figure S1c) depicts the results for the CDW distorted structures with various magnetic orientations. The initial supercells for these distorted structures were obtained from Ref. \onlinecite{PhysRevB.106.085117}, where the V and Se atoms were initially displaced by 0.12 and 0.18 \AA \space respectively (similarly to Ref. \onlinecite{cdw3}) prior to atomic relaxation. For our DFT benchmarking calculations, we adapted a slightly different approach for the distorted CDW structures (in contrast to the undistorted cells). For these supercells, we modified the lattice constant to be a direct multiple of the unit cell geometry we obtained in our previous work (Ref. \onlinecite{vse2-wines}) and fixed the cell dimensions. We then allowed the atomic positions (with initially displaced atoms) to relax with each respective DFT functional. Similarly to the undistorted structures in Fig. S1a), there is a massive discrepancy in relative energy between each magnetic orientation (with respect to the NM undistorted cell) of the distorted CDW structures (up to 0.5 eV). We observe that the FM orientation is the lowest energy magnetic state for all functionals depicted in Fig. S1b). In contrast to the undistorted structure in Fig. S1a), the energy differences between the FM and various AFM orientations is much smaller. Our results are in good qualitative agreement with the DFT calculations performed in Ref. \onlinecite{PhysRevB.106.085117}, but are shifted slightly in energy which can be attributed to fixing the lattice constant to the DMC obtained value. Similar to the undistorted case, we also find excellent qualitative agreement between SCAN/r$^2$SCAN and PBE+$U$ ($U$ = 1 eV) for the energy differences between FM and AFM configurations of distorted structures. To provide more insight on these DFT calculations, we performed DMC simulations for just the FM CDW distorted structure at supercells of 30 and 60 atoms and extrapolated to the thermodynamic limit. For the starting geometry of this structure, we optimized the initially displaced atomic coordinates with r$^2$SCAN and used a direct multiple of the unit cell geometry we obtained in our previous work (Ref. \onlinecite{vse2-wines}). This DMC energy difference and uncertainty is shown (with respect to the undistorted NM structure) in Fig. S1c) and Table \ref{dmc-table}. We find that the DMC energy difference lies between PBE+$U$ ($U$ = 1 eV) and PBE+$U$ ($U$ = 2 eV). Additionally, we find that the SCAN/r$^2$SCAN and LDA+$U$ ($U$ = 3 eV) energy differences are close to the DMC value. More detailed benchmarking data can be found in Table S1, S2, S3, S4 and S5.

  \begin{figure*}
    \centering
    \includegraphics[trim={0. 0cm 0 0cm},clip,width=0.9\textwidth]{fig3.eps}
    \caption{Classical Monte Carlo simulations for the (a) undistorted structure and the (b) CDW-distorted structure, utilizing the spin Hamiltonian from Eq. \ref{eq:hamiltonian}. Each panel shows both the expected normalized squared magnetization, $\langle M^2 \rangle / N^2$, with corresponding uncertainty represented by standard deviation, and the magnetic specific heat, $(\langle E^2 \rangle - \langle E \rangle^2) / (N (k_B T)^2)$, normalized by the Boltzmann constant $k_B$, as a function of temperature. These results indicate a transition temperature ($T_c$) of around $228$ K for the undistorted structure and around $68$ K for the CDW-distorted structure.}
    \label{vse2-mc}
\end{figure*}

From our DMC insight, we went on to estimate the bilinear isotropic exchange ($J$), anisotropic exchange ($\lambda$), and easy axis single ion anisotropy ($A$). The incorporation of spin-orbit within DMC is a relatively new development, with successful applications in band gap calculations \cite{PhysRevB.106.075127} and identifying band inversions in topological materials \cite{lopez2024identifyingbandinversionstopological}. For 1T-VSe$_2$, we were mainly focused on the spin-orbit-induced magnetic anisotropy. From our DFT calculations, we know that the magnetic anisotropy energy differences are on the order of fractions of an meV/f.u., while our DMC uncertainties for total energy are on the order of 10 meV/f.u. Since these spin-orbit-induced magnetic anisotropy energy differences are within our DMC uncertainty, we performed collinear DMC calculations (where spins can orient either up or down) for the undistorted FM and AFM orientations (depicted in Fig. S1 and Table \ref{dmc-table}). For the undistorted structure, we obtain a $J$ value of 71(10) meV (no spin-orbit contribution). Taking the insight from our DMC and DFT calculations, where we observed that SCAN quantitatively reproduced our DMC benchmarking results with closer agreement than r$^2$SCAN (see Table \ref{dmc-table}), we went on to perform spin-orbit SCAN calculations to explicitly obtain anisotropy parameters $\lambda$ and $A$ (see SI for more detailed information). From these calculations, we obtained a $\lambda$ value of -0.57 meV and a $A$ value of -0.77 meV. With regards to the distorted-CDW structure, we performed a DMC benchmark calculation for the FM orientation (see Fig. S1 and Table \ref{dmc-table}). Due to the reasonable agreement with DMC and SCAN and for consistency with the undistorted case, we went on to calculate exchange and anisotropy parameters for the distorted structure using spin-orbit SCAN calculations. In the case of SCAN (since the AFM-A CDW-distorted structure relaxed to the undistorted AFM orientation), the AFM-B2 configuration is the ground state AFM configuration. Using our CDW-FM and CDW-AFM-B2 noncollinear SCAN calculations, we obtained a obtained a $J$ value of 21 meV, a $\lambda$ value of -0.77 meV, and a $A$ value of 0.27 meV (see SI for more details). We went on to use these accurately computed magnetic exchange and anisotropy parameters for our spin Hamiltonian model shown in Eq. \ref{eq:hamiltonian}, where we performed classical Monte Carlo simulations to gain insight on the magnetic phase transitions of the undistorted and distorted structures.

Fig. \ref{vse2-mc} illustrates the magnetic phase transitions for both (a) undistorted and (b) CDW phases. In each case, the exchange interaction $J$ is predominant over other interactions. Notably, $J$ is over three times higher in the undistorted phase, measuring 71 meV, compared to 21 meV in the CDW phase. This pronounced disparity in interaction strength indicates that spins in the undistorted structure align more strongly with neighbors, leading to an extended correlation length and larger clusters of aligned spins, requiring more thermal energy to disrupt these interactions and disorder the aligned spins. This implies that there exists two potential magnetic phase transitions for freestanding (no substrate, strain or defects) 1T-VSe$_2$, one for the undistorted monolayer below room temperature ($\approx$ 100 K less than the transition temperature reported in Ref. \onlinecite{vse2-1stexp}) and one for the CDW-distorted (slightly below liquid N$_2$). These results indicate that the magnetic phase transition occurs below the CDW transition temperature, which has been reported to be above 121 K. This is expected since the CDW phase must exist before the CDW-FM phase can exist by further cooling. The small energy scale between the CDW-distorted phases and the undistorted phases (reported in our current work and previous work\cite{PhysRevB.106.085117}) poses a challenge in understanding which phase is energetically more favorable. This small energy scale opens the door to tune the energetic stability of various magnetic states (with and without CDW-distortion) through external mechanisms such as strain, defects, substrate engineering or temperature.

\begin{figure*}
    \centering
    \includegraphics[trim={0. 0cm 0 0cm},clip,width=0.95\textwidth]{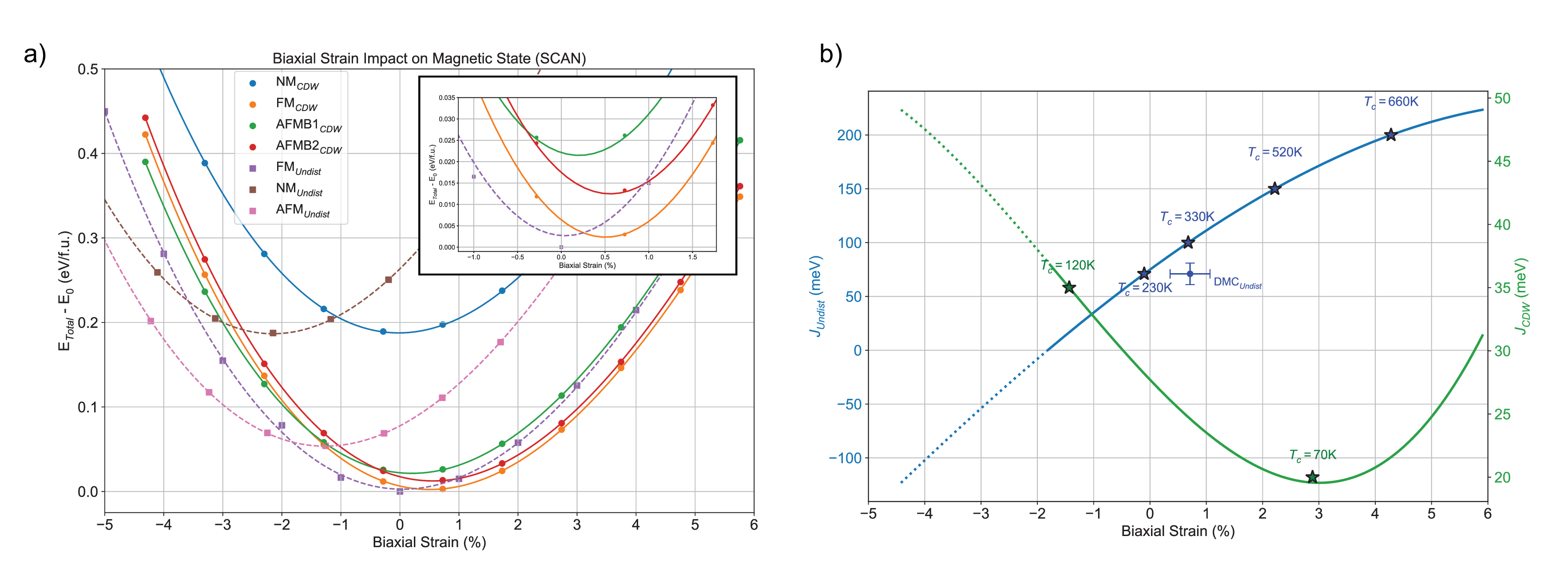}
    \caption{An illustration of the impact of strain on undistorted and CDW-distorted 1T-VSe$_2$: a) The total energy offset ($E_{Total} - E_0)$ in eV/f.u. (calculated with the SCAN functional) as a function of biaxial strain percentage (where the undistorted FM configuration is the reference point at 0 $\%$), demonstrating the impact of biaxial strain on magnetic state. The legend specifies the magnetic orientation of each set of data points, where the solid lines represent the distorted-CDW structures and the dashed lines represent the undistorted structures 
    . The inset depicts a zoomed in section of this curve, focusing on the lowest energy orientations. b) The isotropic exchange ($J$) for the undistorted (blue) and distorted-CDW structures (green) plotted on two separate axes. These values of $J$ were determined by the FM and AFM (undistorted) and the FM-CDW and AFMB2-CDW (distorted) structures. The dotted regions of these two curves indicate a transition to AFM favorability. The DMC results and uncertainties for $J$ and equilibrium lattice constant (for the undistorted FM structure) are also depicted. Finally, the star points on each respective curve represent the $T_c$ value computed under strain (with classical Monte Carlo) for that particular value of $J$.    }
    \label{strain}
\end{figure*}

For a deeper analysis of how structural parameters are related to the energetic favorability of different magnetic states, we performed biaxial strain calculations for the undistorted and distorted structures with the most reliable DFT functional (in our case, SCAN was most reliable for reproducing DMC geometries and quantitative DMC energy differences). These results are depicted in Fig \ref{strain}. It is important to note that for the CDW distorted structures, we kept the displaced atomic positions fixed while applying biaxial strain, and excluded the AFM-A orientation since it relaxed to the undistorted AFM (stripy) orientation with SCAN. We also kept the relative atomic positions fixed for the undistorted configurations while varying the lattice. 

From Fig. \ref{strain}a), we observe that the CDW-distorted structure in FM and AFM configurations is more sensitive to small amounts of strain in contrast to the undistorted structure. In other words, the energy differences between the strain curves are much smaller for the CDW-distorted structures (compared to undistorted). Notably, we observe that for positive values of strain (lattice expansion), the FM in the undistorted and CDW-distorted structures becomes more stable. For the undistorted structure, we see that the energy difference between the FM state and AFM state drastically increases with the application of positive (tensile) strain. We observe that with small amounts of compressive strain, the undistorted FM structure becomes more favorable than the CDW-distorted FM structure, which can be a promising route to stabilizing the FM in the undistorted structure. For larger amounts of compressive strain ($\approx$ $-2$ $\%$), AFM order begins to become more favorable for the undistorted and CDW-distorted (specifically AFM-B1) structures.  

To further understand this phenomena, we computed $J$ under strain for the energy differences depicted in Fig. \ref{strain}a) for the undistorted and CDW-distorted structures. We then went on to estimate $T_c$ under strain using these values of $J$ for additional Monte Carlo simulations (similar to those reported in Fig. \ref{vse2-mc}). It is important to note that for these Monte Carlo simulations under lattice strain, we did not include $A$ and $\lambda$ in the Hamiltonian (due to increased computational demand of spin-orbit SCAN calculations). Since the driving force behind the $T_c$ is $J$, we do not expect major fluctuations in $T_c$ from $A$ and $\lambda$ with applied strain. These results are depicted in Fig. \ref{strain}b). In addition to the strong competition between the undistorted FM and the CDW-FM structures (dotted purple  and orange curves in Fig. \ref{strain}a)), there is potential strain-tunability within each FM ground state (with and without CDW). In Fig. \ref{strain}b), we see that $J$ can be enhanced in the CDW-FM state by small amounts of biaxial expansion or compression, with the $T_c$ approaching values above 120 K. With regards to the undistorted FM structure, we find that tensile strain can enhance the $T_c$ to be above room temperature, but there is strong competition with the FM-CDW state for small values of tensile strain, with the CDW-FM becoming significantly more energetically favorable after $\approx$ $+2$ $\%$ strain. In addition to the tunability of magnetic states, it was revealed from anharmonic phonon calculations that the CDW order can be tuned with small amounts of lattice strain (as little as 1.5 $\%$), highlighting the strong substrate dependence \cite{vse2-anharm}. For example, common substrates for VSe$_2$ \cite{vse2-1stexp} such as MoS$_2$ (a = b = 3.16 \AA) and highly oriented pyrolytic graphite (a = b = 2.46 \AA) have significant lattice mismatch, which can induce strain (since the lattice constant of monolayer 1T-VSe$_2$ is estimated to be $\approx$ 3.41 \AA \cite{vse2-wines}). As an alternative to substrate engineering, substitutional doping can be a promising route to inducing tensile strain in 1T-VSe$_2$, which can stabilize and even enhance the magnetic properties in both the undistorted and CDW-distorted phases.

\begin{figure*}
    \centering
    \includegraphics[trim={0. 0cm 0 0cm},clip,width=0.6\textwidth]{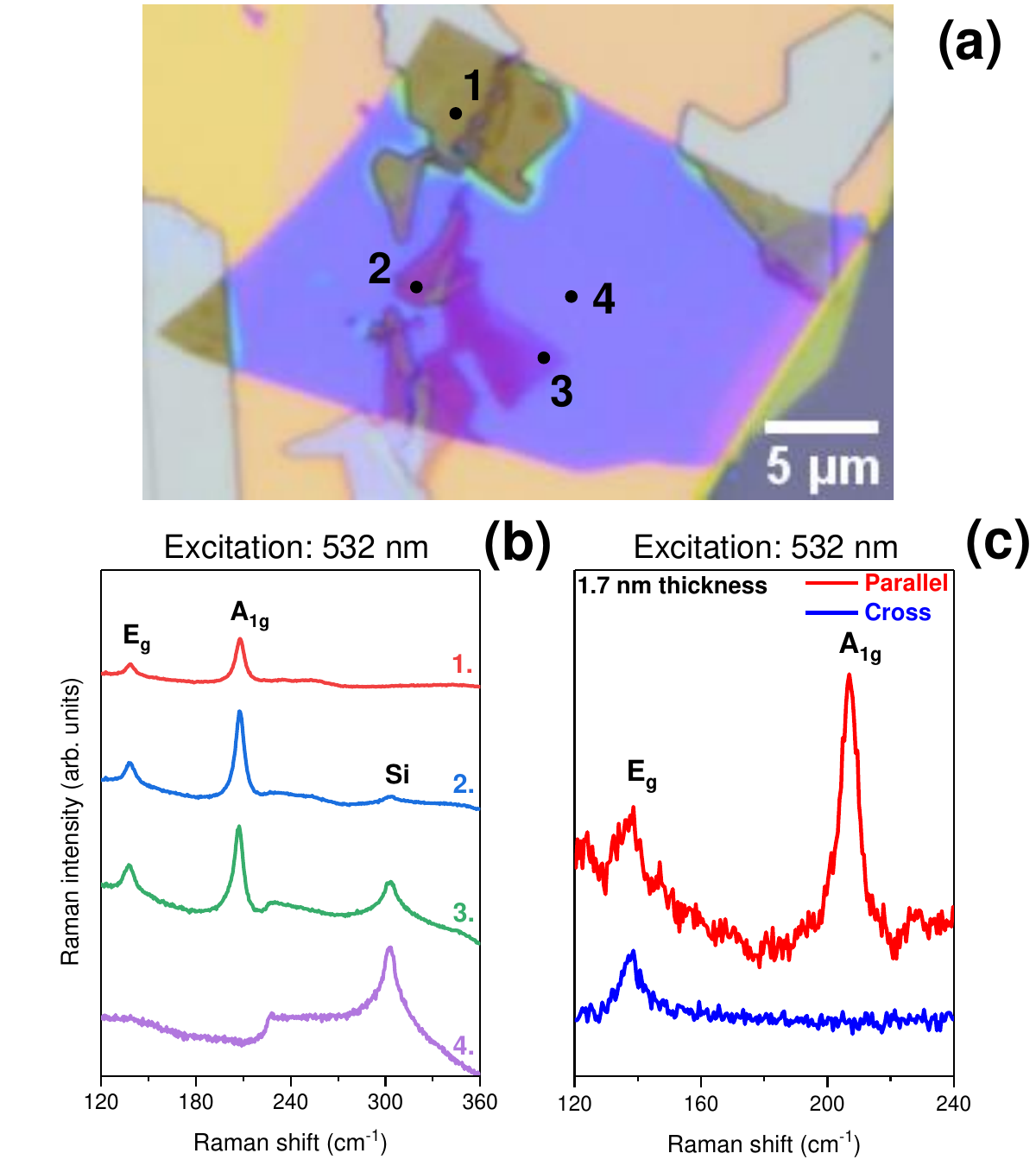}
    \caption{Optical image of h-BN encapsulated VSe$_2$ flakes with four labeled positions approximately identifying the locations from which Raman spectra were collected, position 1 is a bulk-like region and position 3 was measured by AFM to be about 1.7 nm thick (a). Raman spectra corresponding to the four labeled positions on the optical image (b), and polarization-dependent Raman spectra used to identify the A and E symmetry of the modes (c).}
    \label{exp-vse2}
\end{figure*}

\begin{table*}[]
\caption{DFT Raman mode results for VSe$_2$, for bulk and monolayer (ML) in FM, AFM and NM cases (in cm$^{-1}$) at 300 K. Experimental peak positions and uncertainties are given in the last column.}
 \begin{tabular}{c|c|c|c|c|c}
                     & Bulk FM                   & ML-FM                     & ML-AFM                    & ML-NM                        & Exp \\ \hline
A$_{1g}$                        & 199 & 197 & 197, 139              & 208 & 206.7(4) \\
E$_g$                         & 137   & 135   &                           & 130 & 137.5(3) \\
B$_g$                         &                           &                           & 133 &                           & \\ \hline
\hline
CDW-ML & NM    & FM    &                           &                           & \\ \hline
A$_{1g}$    & 204 & 205 &                           &                           &                          
\end{tabular}
\label{raman}
\end{table*}

For additional benchmarking purposes, we performed DFPT calculations to obtain the assignment and peak positions of the Raman modes of 1T-VSe$_2$ for the distorted and undistorted structures in various magnetic and nonmagnetic configurations at 300 K. Due to the higher computational cost and convergence issues of SCAN, we performed these DFPT calculations with PBE+$U$ ($U$ = 1), since this functional was able to match our DMC benchmark for lattice geometry. The logic behind using PBE+$U$ ($U$ = 1) stems from the fact that if it can correctly reproduce our lattice geometry, it can accurately capture vibrational properties. Table \ref{raman} depicts the Raman active modes calculated with DFT for the bulk and monolayer structures (with and without CDW distortions). For the undistorted structure, we performed calculations for the FM, AFM and NM orientations and for the CDW-distorted structure we considered the FM and orientations. As seen in Table \ref{raman}, each structure has a pronounced A$_{1g}$ peak ranging from (197 to 208) cm$^{-1}$. Due to a lack of symmetry in the CDW-distorted structure, we could only identify this A$_{1g}$, but it is entirely possible other Raman modes can appear experimentally. For the undistorted structure (bulk and monolayer), we observe an $E_g$ peak ranging from (130 to 137) cm$^{-1}$ for the FM and NM orientations. The most distinguishing features are apparent for the AFM orientation of the monolayer, where we observe an additional A$_{1g}$ peak at 139 cm$^{-1}$ and a B$_{g}$ peak at 133 cm$^{-1}$.



To validate our DMC-informed DFT calculations of the Raman modes, we grew a VSe$_{2}$ crystal by Chemical Vapor Transport (CVT) (see SI for details). The 1T phase of VSe$_{2}$ was confirmed by X-Ray diffraction (XRD) (see Fig. S2). Thin flakes of 1T-VSe$_{2}$ were exfoliated and encapsulated in h-BN (Fig. S3(a)) and subsequent Raman spectra measurements were performed. The Raman spectra were averaged (30 seconds, 20 accumulations) at four distinct positions on a 1T-VSe$_2$ flake capped with h-BN. Figure \ref{exp-vse2}(a) shows an optical image with the approximate locations where the spectra were measured and Fig. \ref{exp-vse2}(b) shows the corresponding spectra. The thinnest region measured to be about 1.7 nm by atomic force microscope Figure S3, position 3, was used for comparison to DFT predictions. To assign the phonon modes polarized Raman spectra were collected from the 1.7 nm thick region (position 3). Data is shown in both the parallel and cross configuration, i.e., the incoming laser excitation and the Raman scatter were either parallel to one another, or at 90-degrees (cross), confirming the respective A and E symmetry of the modes of the VSe$_2$. In the 1.7 nm thick region we identify two modes; the A$_{1g}$ at 206.7 cm$^{-1}$ $\pm$ 0.4 and the E$_g$ at 137.5 cm$^{-1}$ $\pm$ 0.3 in good agreement with the two modes predicted by DFT. The other features correspond to the Si/SiO$_2$ substrate, as can be seen when compared to measurement of the bare substrate (position 4). However, in the bulk-like region on the h-BN capped flake, position 1, there are additional observable features to the right of the A$_{1g}$ mode which differ from the edge-like feature observed in position 3 coming from the substrate. Previously a broad mode at about 257 cm$^{-1}$ has been attributed as an E$_g$ mode of the 1T phase in a bulk sample \cite{PhysRevResearch.2.033118}, however, we observe in some cases two distinct modes at 235.3 cm-1 and 253.2 cm$^{-1}$ in the bulk-like region in the h-BN capped flake and at 229.8 cm$^{-1}$ and 250.3 cm$^{-1}$ in the bulk crystal, shown in Figure S4. Since these are only observed in the bulk-like capped region/crystal, and are not predicted by the DFT results, these modes are likely due Se \cite{Se-ring,10.1063/1.4962315,PhysRevB.57.10414}. We conclude that the 1T phase of VSe$_2$ only has two phonon modes which we identify as the higher intensity A$_{1g}$ and the lower intensity E$_g$.

\section{Conclusion}

We have provided a comprehensive investigation of the magnetic properties and strain response of monolayer 1T-VSe$_2$ using DFT and highly accurate DMC methods. By extensively benchmarking various DFT functionals against our DMC results, we demonstrated the significant impact of the exchange-correlation functional on the accuracy of magnetic properties and energetic stability of various magnetic states of 2D 1T-VSe$_2$, with and without CDW. Our high-fidelity results aim to resolve previous discrepancies that exist in the theoretical and experimental literature for 1T-VSe$_2$. Through classical Monte Carlo simulations informed by DMC and DFT, we estimated the magnetic transition temperatures for the undistorted (228 K) and CDW-distorted (68 K) phases and our strain calculations indicated that small amounts of biaxial strain can enhance the transition temperature, which provides a viable route for engineering the magnetic properties. Furthermore, our Raman spectroscopy experiments on exfoliated flakes of 1T-VSe$_2$ validate our predictions. Our work underscores the important role of highly accurate many-body methods such as DMC in describing the electronic and magnetic properties of 2D materials. We hope our combined approach can pave the way for future explorations of correlated 2D magnetic materials.

\section{Methods}

Benchmarking DFT calculations for monolayer 1T-VSe$_2$ were performed with the Vienna Ab initio Simulation Package (VASP) code, using projector augmented wave (PAW) pseudopotentials \cite{PhysRevB.54.11169,PhysRevB.59.1758}. For testing purposes, we employed a variety of exchange-correlation functionals including the local density approximation (LDA)\cite{PhysRevB.23.5048,PhysRevLett.45.566}, Perdew-Burke-Ernzerhof (PBE)\cite{PhysRevLett.77.3865}, the strongly constrained and appropriately normed (SCAN)\cite{PhysRevLett.115.036402} meta-GGA functional and the r$^2$SCAN \cite{r2scan} meta-GGA functional with, and without the Hubbard correction $U$ \cite{PhysRevB.57.1505}, which was used to treat the on-site Coulomb interaction of $3d$ orbitals of V. A k-grid of $24 \times 24 \times 1$ was used and scaled proportionally to the supercell size, a plane-wave kinetic energy of 400 eV (increased to 800 eV for SCAN and r$^2$SCAN) was used and at least 20 \AA\space of vacuum was added between periodic layers. Our density functional perturbation theory (DFPT) \cite{RevModPhys.73.515} phonon calculations to obtain the peak positions of Raman modes were also performed with VASP along with the phonopy \cite{phonopy-phono3py-JPCM,phonopy-phono3py-JPSJ} package.

Our Quantum Monte Carlo (QMC) calculations for 1T-VSe$_2$ used the exact same settings as our previous work on 1T- and 2H-VSe$_2$ (see Ref. \onlinecite{vse2-wines} for more specific details regarding kinetic energy cutoff, k-point grid, finite-size, and timestep convergence). DFT-PBE was used to generate the trial wavefunction (nodal surface) for DMC using the Quantum Espresso (QE) \cite{Giannozzi_2009} code with a $U$ correction of 2 eV. This was due to the fact that PBE+$U$ = 2 eV resulted in a trial wavefunction with the lowest total energy calculated with DMC \cite{vse2-wines}. We used a k-point grid of $12 \times 12 \times 1$ and a kinetic energy cutoff of 4,080 eV (300 Ry) to generate the trial wavefunction. In terms of pseudopotentials, RRKJ (OPT) potentials were used for V \cite{PhysRevB.93.075143} and Burkatzki-Fillipi-Dolg (BFD) were used for Se \cite{doi:10.1063/1.2741534}. Variational Monte Carlo (VMC) and DMC \cite{RevModPhys.73.33} simulations were conducted using the QMCPACK \cite{Kim_2018,doi:10.1063/5.0004860} code and the Nexus \cite{nexus} workflow manager was used to automate the DFT-VMC-DMC calculations. Up to two-body Jastrow parameters \cite{PhysRev.34.1293,PhysRev.98.1479,PhysRevB.70.235119} were optimized using the linear method \cite{PhysRevLett.98.110201,PhysRevLett.94.150201} to minimize the variance and energy in VMC, with the goal of modeling electron correlation and reducing uncertainty in the DMC results \cite{PhysRevLett.94.150201,doi:10.1063/1.460849}. In order to compute the nonlocal part of the pseudopotentials, the locality approximation \cite{doi:10.1063/1.460849} was used in DMC. An optimal timestep of 0.01 Ha$^{-1}$ (0.27 eV$^{-1}$) \cite{vse2-wines} was used in DMC and to reduce finite-size errors, canonical twist averaging (with uniform weighting) \cite{ta} and extrapolating to larger supercells (up to 72 atoms) was performed. A more detailed explanation of the theory behind VMC and DMC can be found in Ref. \onlinecite{RevModPhys.73.33}. 

Energies of different spin configurations derived from first-principles calculations were mapped onto a classical spin Hamiltonian, characterizing three-dimensional spins arranged on a two-dimensional lattice at sites of V atom. Subsequently, this Hamiltonian served as the basis for executing classical Monte Carlo simulations aimed at estimating the transition temperature ($T_c$) for both the undistorted and CDW phases. The Hamiltonian, which has previously been used for other 2D magnetic systems \cite{Lado_2017,Torelli_2019,mno2-qmc}, includes the bilinear isotropic exchange ($J$), anisotropic exchange ($\lambda$), and easy axis single ion anisotropy ($A$),  

\begin{equation}
\label{eq:hamiltonian}
H = -\frac{J}{2} \sum_{i,j} \bar{S}_i \cdot \bar{S}_j - \frac{\lambda}{2} \sum_{i,j} S_i^z S_j^z - A \sum_i (S_i^{z})^2
\end{equation}
The indices $i$ and $j$ iterate over all magnetic sites and their corresponding first nearest neighbor (NN) magnetic sites, respectively. Only the first NN interactions are considered due to the strong localized magnetic moment on V. We have calculated $J_2/J_1$ for the CDW phase to be $0.068$, supporting these assumptions (using SCAN). According to the adopted sign convention, $J>0$ favors isotropic spin alignment, leading to a preferred ferromagnetic (FM) over an antiferromagnetic (AFM) phase. Similarly, $\lambda>0$ favors alignment of the spin $z$-components, and $A>0$ favors the out-of-plane direction as the easy axis. 
We employ a $40 ~a \times 23 \sqrt{3} a$ supercell, equivalent to a nearly $13.6 \times 13.6$ nm$^2$ square cell, for the Monte Carlo (MC) simulations. The spin configuration space is discretized into uniformly distributed points on the surface of a unit sphere with a fine resolution of $0.5$ degrees, facilitating detailed exploration of spin orientations. The MC simulation consists of numerous sweeps across all magnetic sites. During each sweep, the spin orientation at each site is updated randomly, one site at a time, and the new configuration is then either accepted or rejected based on the Metropolis algorithm. The convergence of both energy and magnetization is meticulously monitored across varying temperatures. Upon achieving equilibration, an ensemble of configurations is collected to calculate the magnetization and specific heat. We use an ensemble size of $20$ thousand for the CDW simulation and $60$ thousand for the undistorted simulation. Additional methodology details can be found in the SI.

 \section{Supporting Information}
Additional discussion of DFT benchmarking for undistorted and CDW structures, theoretical details of Heisenberg spin Hamiltonian and extraction of magnetic parameters, additional details of experimental synthesis and characterization

\section{Data Availability Statement}
The data from the present work is available on Figshare via \url{https://doi.org/10.6084/m9.figshare.28127477}.

\section{Code Availability Statement}
The code used to perform classical Monte Carlo simulations using the spin Hamiltonian is available at \url{https://github.com/UMBC-STEAM-LAB/SpinMCPack}.

 \section{Notes}
Please note that the use of commercial software (VASP) does not imply recommendation by the National Institute of Standards and Technology. Certain commercial equipment or materials are identified in this paper to adequately specify the experimental procedures.  In no case does the identification imply recommendation or endorsement by NIST, nor does it imply that the materials or equipment identified are necessarily the best available for the purpose. 

 \section{Competing interests}
The authors declare no competing interests.

\section{acknowledgments}
This work was supported by the National Science Foundation through the Division of Materials Research under NSF Grant No. DMR-2213398 and the Department of Energy (DOE) under Grant DE-SC0024236. The authors thank the National Institute of Standards and Technology for funding, computational, and data-management resources. Work by N.G. was supported by the National Institute of Standards and Technology Summer Undergraduate Research Fellowship program. Work by K.S. and J.T.K. (discussion, analysis of QMC calculations) was supported by the U.S. Department of Energy, Office of Science, Basic Energy Sciences, Materials Sciences and Engineering Division, as part of the Computational Materials Sciences Program and Center for Predictive Simulation of Functional Materials.  Work by T. B. (discussion, analysis of DFT+U calculations) was supported by the U.S. Department of Energy, Office of Science, Basic Energy Sciences, Materials Sciences and Engineering Division. F.A. and B.D. would like to acknowledge the support of Mr. Peter Morrison and the Office of Naval Research under Contract No. N00014-21-WX-01248. Research at the United States Naval Academy (M.E.J.) was supported by the Office of Naval Research under Contract No. N0001423WX02132.

Notice: This manuscript has been authored by UT-Battelle, LLC, under contract DE-AC05-00OR22725 with the US Department of Energy (DOE). The US government retains and the publisher, by accepting the article for publication, acknowledges that the US government retains a nonexclusive, paid-up, irrevocable, worldwide license to publish or reproduce the published form of this manuscript, or allow others to do so, for US government purposes. DOE will provide public access to these results of federally sponsored research in accordance with the DOE Public Access Plan (\url{https://www.energy.gov/doe-public-access-plan)}.

\section*{References}
\bibliography{Main}

\providecommand{\noopsort}[1]{}\providecommand{\singleletter}[1]{#1}%
\begin{thebibliography}{84}%
\makeatletter
\providecommand \@ifxundefined [1]{%
 \@ifx{#1\undefined}
}%
\providecommand \@ifnum [1]{%
 \ifnum #1\expandafter \@firstoftwo
 \else \expandafter \@secondoftwo
 \fi
}%
\providecommand \@ifx [1]{%
 \ifx #1\expandafter \@firstoftwo
 \else \expandafter \@secondoftwo
 \fi
}%
\providecommand \natexlab [1]{#1}%
\providecommand \enquote  [1]{``#1''}%
\providecommand \bibnamefont  [1]{#1}%
\providecommand \bibfnamefont [1]{#1}%
\providecommand \citenamefont [1]{#1}%
\providecommand \href@noop [0]{\@secondoftwo}%
\providecommand \href [0]{\begingroup \@sanitize@url \@href}%
\providecommand \@href[1]{\@@startlink{#1}\@@href}%
\providecommand \@@href[1]{\endgroup#1\@@endlink}%
\providecommand \@sanitize@url [0]{\catcode `\\12\catcode `\$12\catcode
  `\&12\catcode `\#12\catcode `\^12\catcode `\_12\catcode `\%12\relax}%
\providecommand \@@startlink[1]{}%
\providecommand \@@endlink[0]{}%
\providecommand \url  [0]{\begingroup\@sanitize@url \@url }%
\providecommand \@url [1]{\endgroup\@href {#1}{\urlprefix }}%
\providecommand \urlprefix  [0]{URL }%
\providecommand \Eprint [0]{\href }%
\providecommand \doibase [0]{https://doi.org/}%
\providecommand \selectlanguage [0]{\@gobble}%
\providecommand \bibinfo  [0]{\@secondoftwo}%
\providecommand \bibfield  [0]{\@secondoftwo}%
\providecommand \translation [1]{[#1]}%
\providecommand \BibitemOpen [0]{}%
\providecommand \bibitemStop [0]{}%
\providecommand \bibitemNoStop [0]{.\EOS\space}%
\providecommand \EOS [0]{\spacefactor3000\relax}%
\providecommand \BibitemShut  [1]{\csname bibitem#1\endcsname}%
\let\auto@bib@innerbib\@empty
\bibitem [{\citenamefont {Huang}\ \emph {et~al.}(2017)\citenamefont {Huang},
  \citenamefont {Clark}, \citenamefont {Navarro-Moratalla}, \citenamefont
  {Klein}, \citenamefont {Cheng}, \citenamefont {Seyler}, \citenamefont
  {Zhong}, \citenamefont {Schmidgall}, \citenamefont {McGuire}, \citenamefont
  {Cobden}, \citenamefont {Yao}, \citenamefont {Xiao}, \citenamefont
  {Jarillo-Herrero},\ and\ \citenamefont {Xu}}]{cri3}%
  \BibitemOpen
  \bibfield  {author} {\bibinfo {author} {\bibfnamefont {B.}~\bibnamefont
  {Huang}}, \bibinfo {author} {\bibfnamefont {G.}~\bibnamefont {Clark}},
  \bibinfo {author} {\bibfnamefont {E.}~\bibnamefont {Navarro-Moratalla}},
  \bibinfo {author} {\bibfnamefont {D.~R.}\ \bibnamefont {Klein}}, \bibinfo
  {author} {\bibfnamefont {R.}~\bibnamefont {Cheng}}, \bibinfo {author}
  {\bibfnamefont {K.~L.}\ \bibnamefont {Seyler}}, \bibinfo {author}
  {\bibfnamefont {D.}~\bibnamefont {Zhong}}, \bibinfo {author} {\bibfnamefont
  {E.}~\bibnamefont {Schmidgall}}, \bibinfo {author} {\bibfnamefont {M.~A.}\
  \bibnamefont {McGuire}}, \bibinfo {author} {\bibfnamefont {D.~H.}\
  \bibnamefont {Cobden}}, \bibinfo {author} {\bibfnamefont {W.}~\bibnamefont
  {Yao}}, \bibinfo {author} {\bibfnamefont {D.}~\bibnamefont {Xiao}}, \bibinfo
  {author} {\bibfnamefont {P.}~\bibnamefont {Jarillo-Herrero}},\ and\ \bibinfo
  {author} {\bibfnamefont {X.}~\bibnamefont {Xu}},\ }\bibfield  {title}
  {\enquote {\bibinfo {title} {Layer-dependent ferromagnetism in a van der
  waals crystal down to the monolayer limit},}\ }\href
  {https://doi.org/10.1038/nature22391} {\bibfield  {journal} {\bibinfo
  {journal} {Nature}\ }\textbf {\bibinfo {volume} {546}},\ \bibinfo {pages}
  {270--273} (\bibinfo {year} {2017})}\BibitemShut {NoStop}%
\bibitem [{\citenamefont {Gong}\ \emph {et~al.}(2017)\citenamefont {Gong},
  \citenamefont {Li}, \citenamefont {Li}, \citenamefont {Ji}, \citenamefont
  {Stern}, \citenamefont {Xia}, \citenamefont {Cao}, \citenamefont {Bao},
  \citenamefont {Wang}, \citenamefont {Wang}, \citenamefont {Qiu},
  \citenamefont {Cava}, \citenamefont {Louie}, \citenamefont {Xia},\ and\
  \citenamefont {Zhang}}]{cr2ge2te6}%
  \BibitemOpen
  \bibfield  {author} {\bibinfo {author} {\bibfnamefont {C.}~\bibnamefont
  {Gong}}, \bibinfo {author} {\bibfnamefont {L.}~\bibnamefont {Li}}, \bibinfo
  {author} {\bibfnamefont {Z.}~\bibnamefont {Li}}, \bibinfo {author}
  {\bibfnamefont {H.}~\bibnamefont {Ji}}, \bibinfo {author} {\bibfnamefont
  {A.}~\bibnamefont {Stern}}, \bibinfo {author} {\bibfnamefont
  {Y.}~\bibnamefont {Xia}}, \bibinfo {author} {\bibfnamefont {T.}~\bibnamefont
  {Cao}}, \bibinfo {author} {\bibfnamefont {W.}~\bibnamefont {Bao}}, \bibinfo
  {author} {\bibfnamefont {C.}~\bibnamefont {Wang}}, \bibinfo {author}
  {\bibfnamefont {Y.}~\bibnamefont {Wang}}, \bibinfo {author} {\bibfnamefont
  {Z.~Q.}\ \bibnamefont {Qiu}}, \bibinfo {author} {\bibfnamefont {R.~J.}\
  \bibnamefont {Cava}}, \bibinfo {author} {\bibfnamefont {S.~G.}\ \bibnamefont
  {Louie}}, \bibinfo {author} {\bibfnamefont {J.}~\bibnamefont {Xia}},\ and\
  \bibinfo {author} {\bibfnamefont {X.}~\bibnamefont {Zhang}},\ }\bibfield
  {title} {\enquote {\bibinfo {title} {Discovery of intrinsic ferromagnetism in
  two-dimensional van der waals crystals},}\ }\href
  {https://doi.org/10.1038/nature22060} {\bibfield  {journal} {\bibinfo
  {journal} {Nature}\ }\textbf {\bibinfo {volume} {546}},\ \bibinfo {pages}
  {265--269} (\bibinfo {year} {2017})}\BibitemShut {NoStop}%
\bibitem [{\citenamefont {Ataca}, \citenamefont {{\c S}ahin},\ and\
  \citenamefont {Ciraci}(2012)}]{ataca-mx2}%
  \BibitemOpen
  \bibfield  {author} {\bibinfo {author} {\bibfnamefont {C.}~\bibnamefont
  {Ataca}}, \bibinfo {author} {\bibfnamefont {H.}~\bibnamefont {{\c S}ahin}},\
  and\ \bibinfo {author} {\bibfnamefont {S.}~\bibnamefont {Ciraci}},\
  }\bibfield  {title} {\enquote {\bibinfo {title} {Stable, single-layer mx2
  transition-metal oxides and dichalcogenides in a honeycomb-like structure},}\
  }\href {https://doi.org/10.1021/jp212558p} {\bibfield  {journal} {\bibinfo
  {journal} {The Journal of Physical Chemistry C}\ }\textbf {\bibinfo {volume}
  {116}},\ \bibinfo {pages} {8983--8999} (\bibinfo {year} {2012})}\BibitemShut
  {NoStop}%
\bibitem [{\citenamefont {Li}\ \emph {et~al.}(2020)\citenamefont {Li},
  \citenamefont {Wang}, \citenamefont {Kan}, \citenamefont {He}, \citenamefont
  {Li}, \citenamefont {Hao}, \citenamefont {Zhao}, \citenamefont {Wu},
  \citenamefont {Jin},\ and\ \citenamefont {Cui}}]{struc-phase}%
  \BibitemOpen
  \bibfield  {author} {\bibinfo {author} {\bibfnamefont {D.}~\bibnamefont
  {Li}}, \bibinfo {author} {\bibfnamefont {X.}~\bibnamefont {Wang}}, \bibinfo
  {author} {\bibfnamefont {C.-m.}\ \bibnamefont {Kan}}, \bibinfo {author}
  {\bibfnamefont {D.}~\bibnamefont {He}}, \bibinfo {author} {\bibfnamefont
  {Z.}~\bibnamefont {Li}}, \bibinfo {author} {\bibfnamefont {Q.}~\bibnamefont
  {Hao}}, \bibinfo {author} {\bibfnamefont {H.}~\bibnamefont {Zhao}}, \bibinfo
  {author} {\bibfnamefont {C.}~\bibnamefont {Wu}}, \bibinfo {author}
  {\bibfnamefont {C.}~\bibnamefont {Jin}},\ and\ \bibinfo {author}
  {\bibfnamefont {X.}~\bibnamefont {Cui}},\ }\bibfield  {title} {\enquote
  {\bibinfo {title} {Structural phase transition of multilayer vse$_2$},}\
  }\href {https://doi.org/10.1021/acsami.0c04449} {\bibfield  {journal}
  {\bibinfo  {journal} {ACS Applied Materials \& Interfaces}\ }\textbf
  {\bibinfo {volume} {12}},\ \bibinfo {pages} {25143--25149} (\bibinfo {year}
  {2020})}\BibitemShut {NoStop}%
\bibitem [{\citenamefont {Pushkarev}\ \emph {et~al.}(2019)\citenamefont
  {Pushkarev}, \citenamefont {Mazurenko}, \citenamefont {Mazurenko},\ and\
  \citenamefont {Boukhvalov}}]{C9CP03726H}%
  \BibitemOpen
  \bibfield  {author} {\bibinfo {author} {\bibfnamefont {G.~V.}\ \bibnamefont
  {Pushkarev}}, \bibinfo {author} {\bibfnamefont {V.~G.}\ \bibnamefont
  {Mazurenko}}, \bibinfo {author} {\bibfnamefont {V.~V.}\ \bibnamefont
  {Mazurenko}},\ and\ \bibinfo {author} {\bibfnamefont {D.~W.}\ \bibnamefont
  {Boukhvalov}},\ }\bibfield  {title} {\enquote {\bibinfo {title} {Structural
  phase transitions in vse$_2$: Energetics{,} electronic structure and
  magnetism},}\ }\href {https://doi.org/10.1039/C9CP03726H} {\bibfield
  {journal} {\bibinfo  {journal} {Phys. Chem. Chem. Phys.}\ }\textbf {\bibinfo
  {volume} {21}},\ \bibinfo {pages} {22647--22653} (\bibinfo {year}
  {2019})}\BibitemShut {NoStop}%
\bibitem [{\citenamefont {Popov}\ \emph {et~al.}(2016)\citenamefont {Popov},
  \citenamefont {Mikhaleva}, \citenamefont {Visotin}, \citenamefont {Kuzubov},
  \citenamefont {Entani}, \citenamefont {Naramoto}, \citenamefont {Sakai},
  \citenamefont {Sorokin},\ and\ \citenamefont {Avramov}}]{C6CP06732H}%
  \BibitemOpen
  \bibfield  {author} {\bibinfo {author} {\bibfnamefont {Z.~I.}\ \bibnamefont
  {Popov}}, \bibinfo {author} {\bibfnamefont {N.~S.}\ \bibnamefont
  {Mikhaleva}}, \bibinfo {author} {\bibfnamefont {M.~A.}\ \bibnamefont
  {Visotin}}, \bibinfo {author} {\bibfnamefont {A.~A.}\ \bibnamefont
  {Kuzubov}}, \bibinfo {author} {\bibfnamefont {S.}~\bibnamefont {Entani}},
  \bibinfo {author} {\bibfnamefont {H.}~\bibnamefont {Naramoto}}, \bibinfo
  {author} {\bibfnamefont {S.}~\bibnamefont {Sakai}}, \bibinfo {author}
  {\bibfnamefont {P.~B.}\ \bibnamefont {Sorokin}},\ and\ \bibinfo {author}
  {\bibfnamefont {P.~V.}\ \bibnamefont {Avramov}},\ }\bibfield  {title}
  {\enquote {\bibinfo {title} {The electronic structure and spin states of 2d
  graphene/vx$_2$ (x = s{,} se) heterostructures},}\ }\href
  {https://doi.org/10.1039/C6CP06732H} {\bibfield  {journal} {\bibinfo
  {journal} {Phys. Chem. Chem. Phys.}\ }\textbf {\bibinfo {volume} {18}},\
  \bibinfo {pages} {33047--33052} (\bibinfo {year} {2016})}\BibitemShut
  {NoStop}%
\bibitem [{\citenamefont {Wang}\ \emph {et~al.}(2021)\citenamefont {Wang},
  \citenamefont {Li}, \citenamefont {Li}, \citenamefont {Wu}, \citenamefont
  {Che}, \citenamefont {Chen},\ and\ \citenamefont {Cui}}]{vse2-exp}%
  \BibitemOpen
  \bibfield  {author} {\bibinfo {author} {\bibfnamefont {X.}~\bibnamefont
  {Wang}}, \bibinfo {author} {\bibfnamefont {D.}~\bibnamefont {Li}}, \bibinfo
  {author} {\bibfnamefont {Z.}~\bibnamefont {Li}}, \bibinfo {author}
  {\bibfnamefont {C.}~\bibnamefont {Wu}}, \bibinfo {author} {\bibfnamefont
  {C.-M.}\ \bibnamefont {Che}}, \bibinfo {author} {\bibfnamefont
  {G.}~\bibnamefont {Chen}},\ and\ \bibinfo {author} {\bibfnamefont
  {X.}~\bibnamefont {Cui}},\ }\bibfield  {title} {\enquote {\bibinfo {title}
  {Ferromagnetism in 2d vanadium diselenide},}\ }\bibfield  {booktitle} {\emph
  {\bibinfo {booktitle} {ACS Nano}},\ }\href
  {https://doi.org/10.1021/acsnano.1c05232} {\bibfield  {journal} {\bibinfo
  {journal} {ACS Nano}\ }\textbf {\bibinfo {volume} {15}},\ \bibinfo {pages}
  {16236--16241} (\bibinfo {year} {2021})}\BibitemShut {NoStop}%
\bibitem [{\citenamefont {Wines}\ \emph {et~al.}(2023)\citenamefont {Wines},
  \citenamefont {Tiihonen}, \citenamefont {Saritas}, \citenamefont {Krogel},\
  and\ \citenamefont {Ataca}}]{vse2-wines}%
  \BibitemOpen
  \bibfield  {author} {\bibinfo {author} {\bibfnamefont {D.}~\bibnamefont
  {Wines}}, \bibinfo {author} {\bibfnamefont {J.}~\bibnamefont {Tiihonen}},
  \bibinfo {author} {\bibfnamefont {K.}~\bibnamefont {Saritas}}, \bibinfo
  {author} {\bibfnamefont {J.~T.}\ \bibnamefont {Krogel}},\ and\ \bibinfo
  {author} {\bibfnamefont {C.}~\bibnamefont {Ataca}},\ }\bibfield  {title}
  {\enquote {\bibinfo {title} {A quantum monte carlo study of the structural,
  energetic, and magnetic properties of two-dimensional h and t phase vse2},}\
  }\href {https://doi.org/10.1021/acs.jpclett.3c00497} {\bibfield  {journal}
  {\bibinfo  {journal} {The Journal of Physical Chemistry Letters}\ }\textbf
  {\bibinfo {volume} {14}},\ \bibinfo {pages} {3553--3560} (\bibinfo {year}
  {2023})}\BibitemShut {NoStop}%
\bibitem [{\citenamefont {Bonilla}\ \emph {et~al.}(2018)\citenamefont
  {Bonilla}, \citenamefont {Kolekar}, \citenamefont {Ma}, \citenamefont {Diaz},
  \citenamefont {Kalappattil}, \citenamefont {Das}, \citenamefont {Eggers},
  \citenamefont {Gutierrez}, \citenamefont {Phan},\ and\ \citenamefont
  {Batzill}}]{vse2-1stexp}%
  \BibitemOpen
  \bibfield  {author} {\bibinfo {author} {\bibfnamefont {M.}~\bibnamefont
  {Bonilla}}, \bibinfo {author} {\bibfnamefont {S.}~\bibnamefont {Kolekar}},
  \bibinfo {author} {\bibfnamefont {Y.}~\bibnamefont {Ma}}, \bibinfo {author}
  {\bibfnamefont {H.~C.}\ \bibnamefont {Diaz}}, \bibinfo {author}
  {\bibfnamefont {V.}~\bibnamefont {Kalappattil}}, \bibinfo {author}
  {\bibfnamefont {R.}~\bibnamefont {Das}}, \bibinfo {author} {\bibfnamefont
  {T.}~\bibnamefont {Eggers}}, \bibinfo {author} {\bibfnamefont {H.~R.}\
  \bibnamefont {Gutierrez}}, \bibinfo {author} {\bibfnamefont {M.-H.}\
  \bibnamefont {Phan}},\ and\ \bibinfo {author} {\bibfnamefont
  {M.}~\bibnamefont {Batzill}},\ }\bibfield  {title} {\enquote {\bibinfo
  {title} {Strong room-temperature ferromagnetism in vse2 monolayers on van der
  waals substrates},}\ }\href {https://doi.org/10.1038/s41565-018-0063-9}
  {\bibfield  {journal} {\bibinfo  {journal} {Nature Nanotechnology}\ }\textbf
  {\bibinfo {volume} {13}},\ \bibinfo {pages} {289--293} (\bibinfo {year}
  {2018})}\BibitemShut {NoStop}%
\bibitem [{\citenamefont {Yu}\ \emph {et~al.}(2019)\citenamefont {Yu},
  \citenamefont {Li}, \citenamefont {Herng}, \citenamefont {Wang},
  \citenamefont {Zhao}, \citenamefont {Chi}, \citenamefont {Fu}, \citenamefont
  {Abdelwahab}, \citenamefont {Zhou}, \citenamefont {Dan}, \citenamefont
  {Chen}, \citenamefont {Chen}, \citenamefont {Li}, \citenamefont {Lu},
  \citenamefont {Pennycook}, \citenamefont {Feng}, \citenamefont {Ding},\ and\
  \citenamefont {Loh}}]{https://doi.org/10.1002/adma.201903779}%
  \BibitemOpen
  \bibfield  {author} {\bibinfo {author} {\bibfnamefont {W.}~\bibnamefont
  {Yu}}, \bibinfo {author} {\bibfnamefont {J.}~\bibnamefont {Li}}, \bibinfo
  {author} {\bibfnamefont {T.~S.}\ \bibnamefont {Herng}}, \bibinfo {author}
  {\bibfnamefont {Z.}~\bibnamefont {Wang}}, \bibinfo {author} {\bibfnamefont
  {X.}~\bibnamefont {Zhao}}, \bibinfo {author} {\bibfnamefont {X.}~\bibnamefont
  {Chi}}, \bibinfo {author} {\bibfnamefont {W.}~\bibnamefont {Fu}}, \bibinfo
  {author} {\bibfnamefont {I.}~\bibnamefont {Abdelwahab}}, \bibinfo {author}
  {\bibfnamefont {J.}~\bibnamefont {Zhou}}, \bibinfo {author} {\bibfnamefont
  {J.}~\bibnamefont {Dan}}, \bibinfo {author} {\bibfnamefont {Z.}~\bibnamefont
  {Chen}}, \bibinfo {author} {\bibfnamefont {Z.}~\bibnamefont {Chen}}, \bibinfo
  {author} {\bibfnamefont {Z.}~\bibnamefont {Li}}, \bibinfo {author}
  {\bibfnamefont {J.}~\bibnamefont {Lu}}, \bibinfo {author} {\bibfnamefont
  {S.~J.}\ \bibnamefont {Pennycook}}, \bibinfo {author} {\bibfnamefont {Y.~P.}\
  \bibnamefont {Feng}}, \bibinfo {author} {\bibfnamefont {J.}~\bibnamefont
  {Ding}},\ and\ \bibinfo {author} {\bibfnamefont {K.~P.}\ \bibnamefont
  {Loh}},\ }\bibfield  {title} {\enquote {\bibinfo {title} {Chemically
  exfoliated vse2 monolayers with room-temperature ferromagnetism},}\ }\href
  {https://doi.org/https://doi.org/10.1002/adma.201903779} {\bibfield
  {journal} {\bibinfo  {journal} {Advanced Materials}\ }\textbf {\bibinfo
  {volume} {31}},\ \bibinfo {pages} {1903779} (\bibinfo {year} {2019})},\
  \Eprint
  {https://arxiv.org/abs/https://onlinelibrary.wiley.com/doi/pdf/10.1002/adma.201903779}
  {https://onlinelibrary.wiley.com/doi/pdf/10.1002/adma.201903779} \BibitemShut
  {NoStop}%
\bibitem [{\citenamefont {Sumida}\ \emph {et~al.}(2022)\citenamefont {Sumida},
  \citenamefont {Takeda}, \citenamefont {Kusaka}, \citenamefont {Kobayashi},\
  and\ \citenamefont {Hirahara}}]{PhysRevMaterials.6.014006}%
  \BibitemOpen
  \bibfield  {author} {\bibinfo {author} {\bibfnamefont {K.}~\bibnamefont
  {Sumida}}, \bibinfo {author} {\bibfnamefont {Y.}~\bibnamefont {Takeda}},
  \bibinfo {author} {\bibfnamefont {S.}~\bibnamefont {Kusaka}}, \bibinfo
  {author} {\bibfnamefont {K.}~\bibnamefont {Kobayashi}},\ and\ \bibinfo
  {author} {\bibfnamefont {T.}~\bibnamefont {Hirahara}},\ }\bibfield  {title}
  {\enquote {\bibinfo {title} {Short-range magnetic interaction in a monolayer
  $1t\text{\ensuremath{-}}{\mathrm{vse}}_{2}$ film revealed by element-specific
  x-ray magnetic circular dichroism},}\ }\href
  {https://doi.org/10.1103/PhysRevMaterials.6.014006} {\bibfield  {journal}
  {\bibinfo  {journal} {Phys. Rev. Mater.}\ }\textbf {\bibinfo {volume} {6}},\
  \bibinfo {pages} {014006} (\bibinfo {year} {2022})}\BibitemShut {NoStop}%
\bibitem [{\citenamefont {Duvjir}\ \emph {et~al.}(2018)\citenamefont {Duvjir},
  \citenamefont {Choi}, \citenamefont {Jang}, \citenamefont {Ulstrup},
  \citenamefont {Kang}, \citenamefont {Thi~Ly}, \citenamefont {Kim},
  \citenamefont {Choi}, \citenamefont {Jozwiak}, \citenamefont {Bostwick},
  \citenamefont {Rotenberg}, \citenamefont {Park}, \citenamefont {Sankar},
  \citenamefont {Kim}, \citenamefont {Kim},\ and\ \citenamefont
  {Chang}}]{vse2-moment-exp}%
  \BibitemOpen
  \bibfield  {author} {\bibinfo {author} {\bibfnamefont {G.}~\bibnamefont
  {Duvjir}}, \bibinfo {author} {\bibfnamefont {B.~K.}\ \bibnamefont {Choi}},
  \bibinfo {author} {\bibfnamefont {I.}~\bibnamefont {Jang}}, \bibinfo {author}
  {\bibfnamefont {S.}~\bibnamefont {Ulstrup}}, \bibinfo {author} {\bibfnamefont
  {S.}~\bibnamefont {Kang}}, \bibinfo {author} {\bibfnamefont {T.}~\bibnamefont
  {Thi~Ly}}, \bibinfo {author} {\bibfnamefont {S.}~\bibnamefont {Kim}},
  \bibinfo {author} {\bibfnamefont {Y.~H.}\ \bibnamefont {Choi}}, \bibinfo
  {author} {\bibfnamefont {C.}~\bibnamefont {Jozwiak}}, \bibinfo {author}
  {\bibfnamefont {A.}~\bibnamefont {Bostwick}}, \bibinfo {author}
  {\bibfnamefont {E.}~\bibnamefont {Rotenberg}}, \bibinfo {author}
  {\bibfnamefont {J.-G.}\ \bibnamefont {Park}}, \bibinfo {author}
  {\bibfnamefont {R.}~\bibnamefont {Sankar}}, \bibinfo {author} {\bibfnamefont
  {K.-S.}\ \bibnamefont {Kim}}, \bibinfo {author} {\bibfnamefont
  {J.}~\bibnamefont {Kim}},\ and\ \bibinfo {author} {\bibfnamefont {Y.~J.}\
  \bibnamefont {Chang}},\ }\bibfield  {title} {\enquote {\bibinfo {title}
  {Emergence of a metal--insulator transition and high-temperature
  charge-density waves in vse$_2$ at the monolayer limit},}\ }\bibfield
  {booktitle} {\emph {\bibinfo {booktitle} {Nano Lett.}},\ }\href
  {https://doi.org/10.1021/acs.nanolett.8b01764} {\bibfield  {journal}
  {\bibinfo  {journal} {Nano Lett.}\ }\textbf {\bibinfo {volume} {18}},\
  \bibinfo {pages} {5432--5438} (\bibinfo {year} {2018})}\BibitemShut {NoStop}%
\bibitem [{\citenamefont {Feng}\ \emph {et~al.}(2018)\citenamefont {Feng},
  \citenamefont {Biswas}, \citenamefont {Rajan}, \citenamefont {Watson},
  \citenamefont {Mazzola}, \citenamefont {Clark}, \citenamefont {Underwood},
  \citenamefont {Markovi{\'c}}, \citenamefont {McLaren}, \citenamefont
  {Hunter}, \citenamefont {Burn}, \citenamefont {Duffy}, \citenamefont {Barua},
  \citenamefont {Balakrishnan}, \citenamefont {Bertran}, \citenamefont
  {Le~F{\`e}vre}, \citenamefont {Kim}, \citenamefont {van~der Laan},
  \citenamefont {Hesjedal}, \citenamefont {Wahl},\ and\ \citenamefont
  {King}}]{cdw}%
  \BibitemOpen
  \bibfield  {author} {\bibinfo {author} {\bibfnamefont {J.}~\bibnamefont
  {Feng}}, \bibinfo {author} {\bibfnamefont {D.}~\bibnamefont {Biswas}},
  \bibinfo {author} {\bibfnamefont {A.}~\bibnamefont {Rajan}}, \bibinfo
  {author} {\bibfnamefont {M.~D.}\ \bibnamefont {Watson}}, \bibinfo {author}
  {\bibfnamefont {F.}~\bibnamefont {Mazzola}}, \bibinfo {author} {\bibfnamefont
  {O.~J.}\ \bibnamefont {Clark}}, \bibinfo {author} {\bibfnamefont
  {K.}~\bibnamefont {Underwood}}, \bibinfo {author} {\bibfnamefont
  {I.}~\bibnamefont {Markovi{\'c}}}, \bibinfo {author} {\bibfnamefont
  {M.}~\bibnamefont {McLaren}}, \bibinfo {author} {\bibfnamefont
  {A.}~\bibnamefont {Hunter}}, \bibinfo {author} {\bibfnamefont {D.~M.}\
  \bibnamefont {Burn}}, \bibinfo {author} {\bibfnamefont {L.~B.}\ \bibnamefont
  {Duffy}}, \bibinfo {author} {\bibfnamefont {S.}~\bibnamefont {Barua}},
  \bibinfo {author} {\bibfnamefont {G.}~\bibnamefont {Balakrishnan}}, \bibinfo
  {author} {\bibfnamefont {F.}~\bibnamefont {Bertran}}, \bibinfo {author}
  {\bibfnamefont {P.}~\bibnamefont {Le~F{\`e}vre}}, \bibinfo {author}
  {\bibfnamefont {T.~K.}\ \bibnamefont {Kim}}, \bibinfo {author} {\bibfnamefont
  {G.}~\bibnamefont {van~der Laan}}, \bibinfo {author} {\bibfnamefont
  {T.}~\bibnamefont {Hesjedal}}, \bibinfo {author} {\bibfnamefont
  {P.}~\bibnamefont {Wahl}},\ and\ \bibinfo {author} {\bibfnamefont {P.~D.~C.}\
  \bibnamefont {King}},\ }\bibfield  {title} {\enquote {\bibinfo {title}
  {Electronic structure and enhanced charge-density wave order of monolayer
  vse2},}\ }\href {https://doi.org/10.1021/acs.nanolett.8b01649} {\bibfield
  {journal} {\bibinfo  {journal} {Nano Letters}\ }\textbf {\bibinfo {volume}
  {18}},\ \bibinfo {pages} {4493--4499} (\bibinfo {year} {2018})}\BibitemShut
  {NoStop}%
\bibitem [{\citenamefont {Fumega}\ \emph {et~al.}(2019)\citenamefont {Fumega},
  \citenamefont {Gobbi}, \citenamefont {Dreher}, \citenamefont {Wan},
  \citenamefont {Gonz{\'a}lez-Orellana}, \citenamefont {Pe{\~n}a-D{\'\i}az},
  \citenamefont {Rogero}, \citenamefont {Herrero-Mart{\'\i}n}, \citenamefont
  {Gargiani}, \citenamefont {Ilyn}, \citenamefont {Ugeda}, \citenamefont
  {Pardo},\ and\ \citenamefont {Blanco-Canosa}}]{cdw2}%
  \BibitemOpen
  \bibfield  {author} {\bibinfo {author} {\bibfnamefont {A.~O.}\ \bibnamefont
  {Fumega}}, \bibinfo {author} {\bibfnamefont {M.}~\bibnamefont {Gobbi}},
  \bibinfo {author} {\bibfnamefont {P.}~\bibnamefont {Dreher}}, \bibinfo
  {author} {\bibfnamefont {W.}~\bibnamefont {Wan}}, \bibinfo {author}
  {\bibfnamefont {C.}~\bibnamefont {Gonz{\'a}lez-Orellana}}, \bibinfo {author}
  {\bibfnamefont {M.}~\bibnamefont {Pe{\~n}a-D{\'\i}az}}, \bibinfo {author}
  {\bibfnamefont {C.}~\bibnamefont {Rogero}}, \bibinfo {author} {\bibfnamefont
  {J.}~\bibnamefont {Herrero-Mart{\'\i}n}}, \bibinfo {author} {\bibfnamefont
  {P.}~\bibnamefont {Gargiani}}, \bibinfo {author} {\bibfnamefont
  {M.}~\bibnamefont {Ilyn}}, \bibinfo {author} {\bibfnamefont {M.~M.}\
  \bibnamefont {Ugeda}}, \bibinfo {author} {\bibfnamefont {V.}~\bibnamefont
  {Pardo}},\ and\ \bibinfo {author} {\bibfnamefont {S.}~\bibnamefont
  {Blanco-Canosa}},\ }\bibfield  {title} {\enquote {\bibinfo {title} {Absence
  of ferromagnetism in vse2 caused by its charge density wave phase},}\ }\href
  {https://doi.org/10.1021/acs.jpcc.9b08868} {\bibfield  {journal} {\bibinfo
  {journal} {The Journal of Physical Chemistry C}\ }\textbf {\bibinfo {volume}
  {123}},\ \bibinfo {pages} {27802--27810} (\bibinfo {year}
  {2019})}\BibitemShut {NoStop}%
\bibitem [{\citenamefont {Coelho}\ \emph {et~al.}(2019)\citenamefont {Coelho},
  \citenamefont {Nguyen~Cong}, \citenamefont {Bonilla}, \citenamefont
  {Kolekar}, \citenamefont {Phan}, \citenamefont {Avila}, \citenamefont
  {Asensio}, \citenamefont {Oleynik},\ and\ \citenamefont {Batzill}}]{cdw3}%
  \BibitemOpen
  \bibfield  {author} {\bibinfo {author} {\bibfnamefont {P.~M.}\ \bibnamefont
  {Coelho}}, \bibinfo {author} {\bibfnamefont {K.}~\bibnamefont {Nguyen~Cong}},
  \bibinfo {author} {\bibfnamefont {M.}~\bibnamefont {Bonilla}}, \bibinfo
  {author} {\bibfnamefont {S.}~\bibnamefont {Kolekar}}, \bibinfo {author}
  {\bibfnamefont {M.-H.}\ \bibnamefont {Phan}}, \bibinfo {author}
  {\bibfnamefont {J.}~\bibnamefont {Avila}}, \bibinfo {author} {\bibfnamefont
  {M.~C.}\ \bibnamefont {Asensio}}, \bibinfo {author} {\bibfnamefont {I.~I.}\
  \bibnamefont {Oleynik}},\ and\ \bibinfo {author} {\bibfnamefont
  {M.}~\bibnamefont {Batzill}},\ }\bibfield  {title} {\enquote {\bibinfo
  {title} {Charge density wave state suppresses ferromagnetic ordering in vse2
  monolayers},}\ }\href {https://doi.org/10.1021/acs.jpcc.9b04281} {\bibfield
  {journal} {\bibinfo  {journal} {The Journal of Physical Chemistry C}\
  }\textbf {\bibinfo {volume} {123}},\ \bibinfo {pages} {14089--14096}
  (\bibinfo {year} {2019})}\BibitemShut {NoStop}%
\bibitem [{\citenamefont {Chua}\ \emph {et~al.}(2022)\citenamefont {Chua},
  \citenamefont {Henke}, \citenamefont {Saha}, \citenamefont {Huang},
  \citenamefont {Gou}, \citenamefont {He}, \citenamefont {Das}, \citenamefont
  {van Wezel}, \citenamefont {Soumyanarayanan},\ and\ \citenamefont
  {Wee}}]{cdw4}%
  \BibitemOpen
  \bibfield  {author} {\bibinfo {author} {\bibfnamefont {R.}~\bibnamefont
  {Chua}}, \bibinfo {author} {\bibfnamefont {J.}~\bibnamefont {Henke}},
  \bibinfo {author} {\bibfnamefont {S.}~\bibnamefont {Saha}}, \bibinfo {author}
  {\bibfnamefont {Y.}~\bibnamefont {Huang}}, \bibinfo {author} {\bibfnamefont
  {J.}~\bibnamefont {Gou}}, \bibinfo {author} {\bibfnamefont {X.}~\bibnamefont
  {He}}, \bibinfo {author} {\bibfnamefont {T.}~\bibnamefont {Das}}, \bibinfo
  {author} {\bibfnamefont {J.}~\bibnamefont {van Wezel}}, \bibinfo {author}
  {\bibfnamefont {A.}~\bibnamefont {Soumyanarayanan}},\ and\ \bibinfo {author}
  {\bibfnamefont {A.~T.~S.}\ \bibnamefont {Wee}},\ }\bibfield  {title}
  {\enquote {\bibinfo {title} {Coexisting charge-ordered states with distinct
  driving mechanisms in monolayer vse2},}\ }\href
  {https://doi.org/10.1021/acsnano.1c08304} {\bibfield  {journal} {\bibinfo
  {journal} {ACS Nano}\ }\textbf {\bibinfo {volume} {16}},\ \bibinfo {pages}
  {783--791} (\bibinfo {year} {2022})}\BibitemShut {NoStop}%
\bibitem [{\citenamefont {Chen}\ \emph {et~al.}(2018)\citenamefont {Chen},
  \citenamefont {Pai}, \citenamefont {Chan}, \citenamefont {Madhavan},
  \citenamefont {Chou}, \citenamefont {Mo}, \citenamefont {Fedorov},\ and\
  \citenamefont {Chiang}}]{PhysRevLett.121.196402}%
  \BibitemOpen
  \bibfield  {author} {\bibinfo {author} {\bibfnamefont {P.}~\bibnamefont
  {Chen}}, \bibinfo {author} {\bibfnamefont {W.~W.}\ \bibnamefont {Pai}},
  \bibinfo {author} {\bibfnamefont {Y.-H.}\ \bibnamefont {Chan}}, \bibinfo
  {author} {\bibfnamefont {V.}~\bibnamefont {Madhavan}}, \bibinfo {author}
  {\bibfnamefont {M.~Y.}\ \bibnamefont {Chou}}, \bibinfo {author}
  {\bibfnamefont {S.-K.}\ \bibnamefont {Mo}}, \bibinfo {author} {\bibfnamefont
  {A.-V.}\ \bibnamefont {Fedorov}},\ and\ \bibinfo {author} {\bibfnamefont
  {T.-C.}\ \bibnamefont {Chiang}},\ }\bibfield  {title} {\enquote {\bibinfo
  {title} {Unique gap structure and symmetry of the charge density wave in
  single-layer ${\mathrm{vse}}_{2}$},}\ }\href
  {https://doi.org/10.1103/PhysRevLett.121.196402} {\bibfield  {journal}
  {\bibinfo  {journal} {Phys. Rev. Lett.}\ }\textbf {\bibinfo {volume} {121}},\
  \bibinfo {pages} {196402} (\bibinfo {year} {2018})}\BibitemShut {NoStop}%
\bibitem [{\citenamefont {Sahoo}\ \emph {et~al.}(2020)\citenamefont {Sahoo},
  \citenamefont {Dutta}, \citenamefont {Harnagea}, \citenamefont {Sood},\ and\
  \citenamefont {Karmakar}}]{PhysRevB.101.014514}%
  \BibitemOpen
  \bibfield  {author} {\bibinfo {author} {\bibfnamefont {S.}~\bibnamefont
  {Sahoo}}, \bibinfo {author} {\bibfnamefont {U.}~\bibnamefont {Dutta}},
  \bibinfo {author} {\bibfnamefont {L.}~\bibnamefont {Harnagea}}, \bibinfo
  {author} {\bibfnamefont {A.~K.}\ \bibnamefont {Sood}},\ and\ \bibinfo
  {author} {\bibfnamefont {S.}~\bibnamefont {Karmakar}},\ }\bibfield  {title}
  {\enquote {\bibinfo {title} {Pressure-induced suppression of charge density
  wave and emergence of superconductivity in
  $1t\ensuremath{-}{\mathrm{vse}}_{2}$},}\ }\href
  {https://doi.org/10.1103/PhysRevB.101.014514} {\bibfield  {journal} {\bibinfo
   {journal} {Phys. Rev. B}\ }\textbf {\bibinfo {volume} {101}},\ \bibinfo
  {pages} {014514} (\bibinfo {year} {2020})}\BibitemShut {NoStop}%
\bibitem [{\citenamefont {Fuh}\ \emph {et~al.}(2016)\citenamefont {Fuh},
  \citenamefont {Chang}, \citenamefont {Wang}, \citenamefont {Evans},
  \citenamefont {Chantrell},\ and\ \citenamefont {Jeng}}]{original-vse2}%
  \BibitemOpen
  \bibfield  {author} {\bibinfo {author} {\bibfnamefont {H.-R.}\ \bibnamefont
  {Fuh}}, \bibinfo {author} {\bibfnamefont {C.-R.}\ \bibnamefont {Chang}},
  \bibinfo {author} {\bibfnamefont {Y.-K.}\ \bibnamefont {Wang}}, \bibinfo
  {author} {\bibfnamefont {R.~F.~L.}\ \bibnamefont {Evans}}, \bibinfo {author}
  {\bibfnamefont {R.~W.}\ \bibnamefont {Chantrell}},\ and\ \bibinfo {author}
  {\bibfnamefont {H.-T.}\ \bibnamefont {Jeng}},\ }\bibfield  {title} {\enquote
  {\bibinfo {title} {Newtype single-layer magnetic semiconductor in
  transition-metal dichalcogenides vx2 (x = s, se and te)},}\ }\href
  {https://doi.org/10.1038/srep32625} {\bibfield  {journal} {\bibinfo
  {journal} {Scientific Reports}\ }\textbf {\bibinfo {volume} {6}},\ \bibinfo
  {pages} {32625} (\bibinfo {year} {2016})}\BibitemShut {NoStop}%
\bibitem [{\citenamefont {He}, \citenamefont {Xie},\ and\ \citenamefont
  {Xu}(2021)}]{He_2021}%
  \BibitemOpen
  \bibfield  {author} {\bibinfo {author} {\bibfnamefont {J.}~\bibnamefont
  {He}}, \bibinfo {author} {\bibfnamefont {Q.}~\bibnamefont {Xie}},\ and\
  \bibinfo {author} {\bibfnamefont {G.}~\bibnamefont {Xu}},\ }\bibfield
  {title} {\enquote {\bibinfo {title} {Confinement effect enhanced stoner
  ferromagnetic instability in monolayer 1t-vse2},}\ }\href
  {https://doi.org/10.1088/1367-2630/abdfef} {\bibfield  {journal} {\bibinfo
  {journal} {New Journal of Physics}\ }\textbf {\bibinfo {volume} {23}},\
  \bibinfo {pages} {023027} (\bibinfo {year} {2021})}\BibitemShut {NoStop}%
\bibitem [{\citenamefont {Esters}, \citenamefont {Hennig},\ and\ \citenamefont
  {Johnson}(2017)}]{PhysRevB.96.235147}%
  \BibitemOpen
  \bibfield  {author} {\bibinfo {author} {\bibfnamefont {M.}~\bibnamefont
  {Esters}}, \bibinfo {author} {\bibfnamefont {R.~G.}\ \bibnamefont {Hennig}},\
  and\ \bibinfo {author} {\bibfnamefont {D.~C.}\ \bibnamefont {Johnson}},\
  }\bibfield  {title} {\enquote {\bibinfo {title} {Dynamic instabilities in
  strongly correlated ${\text{vse}}_{2}$ monolayers and bilayers},}\ }\href
  {https://doi.org/10.1103/PhysRevB.96.235147} {\bibfield  {journal} {\bibinfo
  {journal} {Phys. Rev. B}\ }\textbf {\bibinfo {volume} {96}},\ \bibinfo
  {pages} {235147} (\bibinfo {year} {2017})}\BibitemShut {NoStop}%
\bibitem [{\citenamefont {Yilmaz}\ \emph {et~al.}(2024)\citenamefont {Yilmaz},
  \citenamefont {Tong}, \citenamefont {Sadowski}, \citenamefont {Hwang},
  \citenamefont {Evans-Lutterodt}, \citenamefont {Kisslinger},\ and\
  \citenamefont {Vescovo}}]{yilmaz2024evolutionfermisurface1tvse2}%
  \BibitemOpen
  \bibfield  {author} {\bibinfo {author} {\bibfnamefont {T.}~\bibnamefont
  {Yilmaz}}, \bibinfo {author} {\bibfnamefont {X.}~\bibnamefont {Tong}},
  \bibinfo {author} {\bibfnamefont {J.~T.}\ \bibnamefont {Sadowski}}, \bibinfo
  {author} {\bibfnamefont {S.}~\bibnamefont {Hwang}}, \bibinfo {author}
  {\bibfnamefont {K.}~\bibnamefont {Evans-Lutterodt}}, \bibinfo {author}
  {\bibfnamefont {K.}~\bibnamefont {Kisslinger}},\ and\ \bibinfo {author}
  {\bibfnamefont {E.}~\bibnamefont {Vescovo}},\ }\href
  {https://arxiv.org/abs/2408.05930} {\enquote {\bibinfo {title} {Evolution of
  the fermi surface of 1t-vse$_2$ across a structural phase transition},}\ }
  (\bibinfo {year} {2024}),\ \Eprint {https://arxiv.org/abs/2408.05930}
  {arXiv:2408.05930 [cond-mat.mtrl-sci]} \BibitemShut {NoStop}%
\bibitem [{\citenamefont {Pan}(2014)}]{hydro}%
  \BibitemOpen
  \bibfield  {author} {\bibinfo {author} {\bibfnamefont {H.}~\bibnamefont
  {Pan}},\ }\bibfield  {title} {\enquote {\bibinfo {title} {Electronic and
  magnetic properties of vanadium dichalcogenides monolayers tuned by
  hydrogenation},}\ }\href {https://doi.org/10.1021/jp503030b} {\bibfield
  {journal} {\bibinfo  {journal} {The Journal of Physical Chemistry C}\
  }\textbf {\bibinfo {volume} {118}},\ \bibinfo {pages} {13248--13253}
  (\bibinfo {year} {2014})}\BibitemShut {NoStop}%
\bibitem [{\citenamefont {Zhang}\ \emph {et~al.}(2019)\citenamefont {Zhang},
  \citenamefont {Zhang}, \citenamefont {Wong}, \citenamefont {Yuan},
  \citenamefont {Vinai}, \citenamefont {Torelli}, \citenamefont {van~der Laan},
  \citenamefont {Feng},\ and\ \citenamefont {Wee}}]{interface-vse2}%
  \BibitemOpen
  \bibfield  {author} {\bibinfo {author} {\bibfnamefont {W.}~\bibnamefont
  {Zhang}}, \bibinfo {author} {\bibfnamefont {L.}~\bibnamefont {Zhang}},
  \bibinfo {author} {\bibfnamefont {P.~K.~J.}\ \bibnamefont {Wong}}, \bibinfo
  {author} {\bibfnamefont {J.}~\bibnamefont {Yuan}}, \bibinfo {author}
  {\bibfnamefont {G.}~\bibnamefont {Vinai}}, \bibinfo {author} {\bibfnamefont
  {P.}~\bibnamefont {Torelli}}, \bibinfo {author} {\bibfnamefont
  {G.}~\bibnamefont {van~der Laan}}, \bibinfo {author} {\bibfnamefont {Y.~P.}\
  \bibnamefont {Feng}},\ and\ \bibinfo {author} {\bibfnamefont {A.~T.~S.}\
  \bibnamefont {Wee}},\ }\bibfield  {title} {\enquote {\bibinfo {title}
  {Magnetic transition in monolayer vse2 via interface hybridization},}\ }\href
  {https://doi.org/10.1021/acsnano.9b02996} {\bibfield  {journal} {\bibinfo
  {journal} {ACS Nano}\ }\textbf {\bibinfo {volume} {13}},\ \bibinfo {pages}
  {8997--9004} (\bibinfo {year} {2019})}\BibitemShut {NoStop}%
\bibitem [{\citenamefont {Ma}\ \emph {et~al.}(2012)\citenamefont {Ma},
  \citenamefont {Dai}, \citenamefont {Guo}, \citenamefont {Niu}, \citenamefont
  {Zhu},\ and\ \citenamefont {Huang}}]{strain-induced}%
  \BibitemOpen
  \bibfield  {author} {\bibinfo {author} {\bibfnamefont {Y.}~\bibnamefont
  {Ma}}, \bibinfo {author} {\bibfnamefont {Y.}~\bibnamefont {Dai}}, \bibinfo
  {author} {\bibfnamefont {M.}~\bibnamefont {Guo}}, \bibinfo {author}
  {\bibfnamefont {C.}~\bibnamefont {Niu}}, \bibinfo {author} {\bibfnamefont
  {Y.}~\bibnamefont {Zhu}},\ and\ \bibinfo {author} {\bibfnamefont
  {B.}~\bibnamefont {Huang}},\ }\bibfield  {title} {\enquote {\bibinfo {title}
  {Evidence of the existence of magnetism in pristine vx2 monolayers (x = s,
  se) and their strain-induced tunable magnetic properties},}\ }\href
  {https://doi.org/10.1021/nn204667z} {\bibfield  {journal} {\bibinfo
  {journal} {ACS Nano}\ }\textbf {\bibinfo {volume} {6}},\ \bibinfo {pages}
  {1695--1701} (\bibinfo {year} {2012})}\BibitemShut {NoStop}%
\bibitem [{\citenamefont {Yin}\ \emph {et~al.}(2022)\citenamefont {Yin},
  \citenamefont {Berlijn}, \citenamefont {Juneja}, \citenamefont {Lindsay},\
  and\ \citenamefont {Parker}}]{PhysRevB.106.085117}%
  \BibitemOpen
  \bibfield  {author} {\bibinfo {author} {\bibfnamefont {L.}~\bibnamefont
  {Yin}}, \bibinfo {author} {\bibfnamefont {T.}~\bibnamefont {Berlijn}},
  \bibinfo {author} {\bibfnamefont {R.}~\bibnamefont {Juneja}}, \bibinfo
  {author} {\bibfnamefont {L.}~\bibnamefont {Lindsay}},\ and\ \bibinfo {author}
  {\bibfnamefont {D.~S.}\ \bibnamefont {Parker}},\ }\bibfield  {title}
  {\enquote {\bibinfo {title} {Competing magnetic and nonmagnetic states in
  monolayer ${\mathrm{vse}}_{2}$ with charge density wave},}\ }\href
  {https://doi.org/10.1103/PhysRevB.106.085117} {\bibfield  {journal} {\bibinfo
   {journal} {Phys. Rev. B}\ }\textbf {\bibinfo {volume} {106}},\ \bibinfo
  {pages} {085117} (\bibinfo {year} {2022})}\BibitemShut {NoStop}%
\bibitem [{\citenamefont {Wei}\ \emph {et~al.}(2023)\citenamefont {Wei},
  \citenamefont {Ma}, \citenamefont {Ye}, \citenamefont {Wang},\ and\
  \citenamefont {Bai}}]{WEI2023170683}%
  \BibitemOpen
  \bibfield  {author} {\bibinfo {author} {\bibfnamefont {M.}~\bibnamefont
  {Wei}}, \bibinfo {author} {\bibfnamefont {H.}~\bibnamefont {Ma}}, \bibinfo
  {author} {\bibfnamefont {H.}~\bibnamefont {Ye}}, \bibinfo {author}
  {\bibfnamefont {J.}~\bibnamefont {Wang}},\ and\ \bibinfo {author}
  {\bibfnamefont {D.}~\bibnamefont {Bai}},\ }\bibfield  {title} {\enquote
  {\bibinfo {title} {Magnetic and transport properties of two-dimensional
  ferromagnet vse2 with se vacancies},}\ }\href
  {https://doi.org/https://doi.org/10.1016/j.jmmm.2023.170683} {\bibfield
  {journal} {\bibinfo  {journal} {Journal of Magnetism and Magnetic Materials}\
  }\textbf {\bibinfo {volume} {574}},\ \bibinfo {pages} {170683} (\bibinfo
  {year} {2023})}\BibitemShut {NoStop}%
\bibitem [{\citenamefont {Zhu}\ \emph {et~al.}(2022)\citenamefont {Zhu},
  \citenamefont {Gao}, \citenamefont {Jiang},\ and\ \citenamefont
  {Zhao}}]{D2CP01537D}%
  \BibitemOpen
  \bibfield  {author} {\bibinfo {author} {\bibfnamefont {Y.}~\bibnamefont
  {Zhu}}, \bibinfo {author} {\bibfnamefont {Y.}~\bibnamefont {Gao}}, \bibinfo
  {author} {\bibfnamefont {X.}~\bibnamefont {Jiang}},\ and\ \bibinfo {author}
  {\bibfnamefont {J.}~\bibnamefont {Zhao}},\ }\bibfield  {title} {\enquote
  {\bibinfo {title} {Effects of vacancy defects on the magnetic properties of
  vanadium diselenide monolayers: a first principle investigation},}\ }\href
  {https://doi.org/10.1039/D2CP01537D} {\bibfield  {journal} {\bibinfo
  {journal} {Phys. Chem. Chem. Phys.}\ }\textbf {\bibinfo {volume} {24}},\
  \bibinfo {pages} {17615--17622} (\bibinfo {year} {2022})}\BibitemShut
  {NoStop}%
\bibitem [{\citenamefont {Jiang}, \citenamefont {Li},\ and\ \citenamefont
  {Mi}(2022)}]{D2MH00888B}%
  \BibitemOpen
  \bibfield  {author} {\bibinfo {author} {\bibfnamefont {J.}~\bibnamefont
  {Jiang}}, \bibinfo {author} {\bibfnamefont {R.}~\bibnamefont {Li}},\ and\
  \bibinfo {author} {\bibfnamefont {W.}~\bibnamefont {Mi}},\ }\bibfield
  {title} {\enquote {\bibinfo {title} {Exchange interactions in the 1t-vse2
  monolayer and their modulation via electron doping using alkali metal
  adsorption and the electride substrate},}\ }\href
  {https://doi.org/10.1039/D2MH00888B} {\bibfield  {journal} {\bibinfo
  {journal} {Mater. Horiz.}\ }\textbf {\bibinfo {volume} {9}},\ \bibinfo
  {pages} {2785--2796} (\bibinfo {year} {2022})}\BibitemShut {NoStop}%
\bibitem [{\citenamefont {Boukhvalov}\ and\ \citenamefont
  {Politano}(2020)}]{D0NR04663A}%
  \BibitemOpen
  \bibfield  {author} {\bibinfo {author} {\bibfnamefont {D.~W.}\ \bibnamefont
  {Boukhvalov}}\ and\ \bibinfo {author} {\bibfnamefont {A.}~\bibnamefont
  {Politano}},\ }\bibfield  {title} {\enquote {\bibinfo {title} {Unveiling the
  origin of room-temperature ferromagnetism in monolayer vse2: the role of
  extrinsic effects},}\ }\href {https://doi.org/10.1039/D0NR04663A} {\bibfield
  {journal} {\bibinfo  {journal} {Nanoscale}\ }\textbf {\bibinfo {volume}
  {12}},\ \bibinfo {pages} {20875--20882} (\bibinfo {year} {2020})}\BibitemShut
  {NoStop}%
\bibitem [{\citenamefont {Wines}\ \emph {et~al.}(2024)\citenamefont {Wines},
  \citenamefont {Ahn}, \citenamefont {Benali}, \citenamefont {Kent},
  \citenamefont {Krogel}, \citenamefont {Kwon}, \citenamefont {Mitas},
  \citenamefont {Reboredo}, \citenamefont {Rubenstein}, \citenamefont
  {Saritas}, \citenamefont {Shin}, \citenamefont {Štich},\ and\ \citenamefont
  {Ataca}}]{wines2024improvedpropertypredictiontwodimensional}%
  \BibitemOpen
  \bibfield  {author} {\bibinfo {author} {\bibfnamefont {D.}~\bibnamefont
  {Wines}}, \bibinfo {author} {\bibfnamefont {J.}~\bibnamefont {Ahn}}, \bibinfo
  {author} {\bibfnamefont {A.}~\bibnamefont {Benali}}, \bibinfo {author}
  {\bibfnamefont {P.~R.~C.}\ \bibnamefont {Kent}}, \bibinfo {author}
  {\bibfnamefont {J.~T.}\ \bibnamefont {Krogel}}, \bibinfo {author}
  {\bibfnamefont {Y.}~\bibnamefont {Kwon}}, \bibinfo {author} {\bibfnamefont
  {L.}~\bibnamefont {Mitas}}, \bibinfo {author} {\bibfnamefont {F.~A.}\
  \bibnamefont {Reboredo}}, \bibinfo {author} {\bibfnamefont {B.}~\bibnamefont
  {Rubenstein}}, \bibinfo {author} {\bibfnamefont {K.}~\bibnamefont {Saritas}},
  \bibinfo {author} {\bibfnamefont {H.}~\bibnamefont {Shin}}, \bibinfo {author}
  {\bibfnamefont {I.}~\bibnamefont {Štich}},\ and\ \bibinfo {author}
  {\bibfnamefont {C.}~\bibnamefont {Ataca}},\ }\href
  {https://arxiv.org/abs/2406.02753} {\enquote {\bibinfo {title} {Towards
  improved property prediction of two-dimensional (2d) materials using
  many-body quantum monte carlo methods},}\ } (\bibinfo {year} {2024}),\
  \Eprint {https://arxiv.org/abs/2406.02753} {arXiv:2406.02753
  [cond-mat.mtrl-sci]} \BibitemShut {NoStop}%
\bibitem [{\citenamefont {Foulkes}\ \emph {et~al.}(2001)\citenamefont
  {Foulkes}, \citenamefont {Mitas}, \citenamefont {Needs},\ and\ \citenamefont
  {Rajagopal}}]{RevModPhys.73.33}%
  \BibitemOpen
  \bibfield  {author} {\bibinfo {author} {\bibfnamefont {W.~M.~C.}\
  \bibnamefont {Foulkes}}, \bibinfo {author} {\bibfnamefont {L.}~\bibnamefont
  {Mitas}}, \bibinfo {author} {\bibfnamefont {R.~J.}\ \bibnamefont {Needs}},\
  and\ \bibinfo {author} {\bibfnamefont {G.}~\bibnamefont {Rajagopal}},\
  }\bibfield  {title} {\enquote {\bibinfo {title} {{Quantum Monte Carlo
  Simulations of Solids}},}\ }\href {https://doi.org/10.1103/RevModPhys.73.33}
  {\bibfield  {journal} {\bibinfo  {journal} {Rev. Mod. Phys.}\ }\textbf
  {\bibinfo {volume} {73}},\ \bibinfo {pages} {33--83} (\bibinfo {year}
  {2001})}\BibitemShut {NoStop}%
\bibitem [{\citenamefont {Dudarev}\ \emph {et~al.}(1998)\citenamefont
  {Dudarev}, \citenamefont {Botton}, \citenamefont {Savrasov}, \citenamefont
  {Humphreys},\ and\ \citenamefont {Sutton}}]{PhysRevB.57.1505}%
  \BibitemOpen
  \bibfield  {author} {\bibinfo {author} {\bibfnamefont {S.~L.}\ \bibnamefont
  {Dudarev}}, \bibinfo {author} {\bibfnamefont {G.~A.}\ \bibnamefont {Botton}},
  \bibinfo {author} {\bibfnamefont {S.~Y.}\ \bibnamefont {Savrasov}}, \bibinfo
  {author} {\bibfnamefont {C.~J.}\ \bibnamefont {Humphreys}},\ and\ \bibinfo
  {author} {\bibfnamefont {A.~P.}\ \bibnamefont {Sutton}},\ }\bibfield  {title}
  {\enquote {\bibinfo {title} {Electron-energy-loss spectra and the structural
  stability of nickel oxide: An {LSDA+U} study},}\ }\href
  {https://doi.org/10.1103/PhysRevB.57.1505} {\bibfield  {journal} {\bibinfo
  {journal} {Phys. Rev. B}\ }\textbf {\bibinfo {volume} {57}},\ \bibinfo
  {pages} {1505--1509} (\bibinfo {year} {1998})}\BibitemShut {NoStop}%
\bibitem [{\citenamefont {Foyevtsova}\ \emph {et~al.}(2014)\citenamefont
  {Foyevtsova}, \citenamefont {Krogel}, \citenamefont {Kim}, \citenamefont
  {Kent}, \citenamefont {Dagotto},\ and\ \citenamefont
  {Reboredo}}]{PhysRevX.4.031003}%
  \BibitemOpen
  \bibfield  {author} {\bibinfo {author} {\bibfnamefont {K.}~\bibnamefont
  {Foyevtsova}}, \bibinfo {author} {\bibfnamefont {J.~T.}\ \bibnamefont
  {Krogel}}, \bibinfo {author} {\bibfnamefont {J.}~\bibnamefont {Kim}},
  \bibinfo {author} {\bibfnamefont {P.~R.~C.}\ \bibnamefont {Kent}}, \bibinfo
  {author} {\bibfnamefont {E.}~\bibnamefont {Dagotto}},\ and\ \bibinfo {author}
  {\bibfnamefont {F.~A.}\ \bibnamefont {Reboredo}},\ }\bibfield  {title}
  {\enquote {\bibinfo {title} {Ab initio {Quantum Monte Carlo} calculations of
  spin superexchange in cuprates: The benchmarking case of
  {${\mathrm{Ca}}_{2}{\mathrm{CuO}}_{3}$}},}\ }\href
  {https://doi.org/10.1103/PhysRevX.4.031003} {\bibfield  {journal} {\bibinfo
  {journal} {Phys. Rev. X}\ }\textbf {\bibinfo {volume} {4}},\ \bibinfo {pages}
  {031003} (\bibinfo {year} {2014})}\BibitemShut {NoStop}%
\bibitem [{\citenamefont {Wines}, \citenamefont {Saritas},\ and\ \citenamefont
  {Ataca}(2020)}]{doi:10.1063/5.0023223}%
  \BibitemOpen
  \bibfield  {author} {\bibinfo {author} {\bibfnamefont {D.}~\bibnamefont
  {Wines}}, \bibinfo {author} {\bibfnamefont {K.}~\bibnamefont {Saritas}},\
  and\ \bibinfo {author} {\bibfnamefont {C.}~\bibnamefont {Ataca}},\ }\bibfield
   {title} {\enquote {\bibinfo {title} {A first-principles {Quantum Monte
  Carlo} study of two-dimensional {(2D) GaSe}},}\ }\href
  {https://doi.org/10.1063/5.0023223} {\bibfield  {journal} {\bibinfo
  {journal} {J. Chem. Phys.}\ }\textbf {\bibinfo {volume} {153}},\ \bibinfo
  {pages} {154704} (\bibinfo {year} {2020})}\BibitemShut {NoStop}%
\bibitem [{\citenamefont {Wines}, \citenamefont {Saritas},\ and\ \citenamefont
  {Ataca}(2021)}]{wines2021pathway}%
  \BibitemOpen
  \bibfield  {author} {\bibinfo {author} {\bibfnamefont {D.}~\bibnamefont
  {Wines}}, \bibinfo {author} {\bibfnamefont {K.}~\bibnamefont {Saritas}},\
  and\ \bibinfo {author} {\bibfnamefont {C.}~\bibnamefont {Ataca}},\ }\bibfield
   {title} {\enquote {\bibinfo {title} {A pathway toward high-throughput
  quantum {Monte Carlo} simulations for alloys: A case study of two-dimensional
  (2d) {GaS}$_x${Se}$_{1-x}$},}\ }\href {https://doi.org/10.1063/5.0070423}
  {\bibfield  {journal} {\bibinfo  {journal} {J. Chem. Phys.}\ }\textbf
  {\bibinfo {volume} {155}},\ \bibinfo {pages} {194112} (\bibinfo {year}
  {2021})}\BibitemShut {NoStop}%
\bibitem [{\citenamefont {Wines}, \citenamefont {Saritas},\ and\ \citenamefont
  {Ataca}(2022)}]{mno2-qmc}%
  \BibitemOpen
  \bibfield  {author} {\bibinfo {author} {\bibfnamefont {D.}~\bibnamefont
  {Wines}}, \bibinfo {author} {\bibfnamefont {K.}~\bibnamefont {Saritas}},\
  and\ \bibinfo {author} {\bibfnamefont {C.}~\bibnamefont {Ataca}},\ }\bibfield
   {title} {\enquote {\bibinfo {title} {Intrinsic ferromagnetism of
  two-dimensional (2d) mno$_2$ revisited: A many-body quantum monte carlo and
  dft+u study},}\ }\href {https://doi.org/10.1021/acs.jpcc.1c10841} {\bibfield
  {journal} {\bibinfo  {journal} {J. Phys. Chem. C}\ }\textbf {\bibinfo
  {volume} {126}},\ \bibinfo {pages} {5813--5821} (\bibinfo {year}
  {2022})}\BibitemShut {NoStop}%
\bibitem [{\citenamefont {Wines}, \citenamefont {Choudhary},\ and\
  \citenamefont {Tavazza}(2023)}]{https://doi.org/10.48550/arxiv.2209.10379}%
  \BibitemOpen
  \bibfield  {author} {\bibinfo {author} {\bibfnamefont {D.}~\bibnamefont
  {Wines}}, \bibinfo {author} {\bibfnamefont {K.}~\bibnamefont {Choudhary}},\
  and\ \bibinfo {author} {\bibfnamefont {F.}~\bibnamefont {Tavazza}},\
  }\bibfield  {title} {\enquote {\bibinfo {title} {Systematic dft+u and quantum
  monte carlo benchmark of magnetic two-dimensional (2d) crx$_3$ (x = i, br,
  cl, f)},}\ }\href {https://doi.org/10.1021/acs.jpcc.2c06733} {\bibfield
  {journal} {\bibinfo  {journal} {J. Phys. Chem. C}\ }\textbf {\bibinfo
  {volume} {127}},\ \bibinfo {pages} {1176--1188} (\bibinfo {year}
  {2023})}\BibitemShut {NoStop}%
\bibitem [{\citenamefont {Shin}\ \emph {et~al.}(2021)\citenamefont {Shin},
  \citenamefont {Krogel}, \citenamefont {Gasperich}, \citenamefont {Kent},
  \citenamefont {Benali},\ and\ \citenamefont
  {Heinonen}}]{PhysRevMaterials.5.024002}%
  \BibitemOpen
  \bibfield  {author} {\bibinfo {author} {\bibfnamefont {H.}~\bibnamefont
  {Shin}}, \bibinfo {author} {\bibfnamefont {J.~T.}\ \bibnamefont {Krogel}},
  \bibinfo {author} {\bibfnamefont {K.}~\bibnamefont {Gasperich}}, \bibinfo
  {author} {\bibfnamefont {P.~R.~C.}\ \bibnamefont {Kent}}, \bibinfo {author}
  {\bibfnamefont {A.}~\bibnamefont {Benali}},\ and\ \bibinfo {author}
  {\bibfnamefont {O.}~\bibnamefont {Heinonen}},\ }\bibfield  {title} {\enquote
  {\bibinfo {title} {Optimized structure and electronic band gap of monolayer
  {GeSe} from {Quantum Monte Carlo} methods},}\ }\href
  {https://doi.org/10.1103/PhysRevMaterials.5.024002} {\bibfield  {journal}
  {\bibinfo  {journal} {Phys. Rev. Materials}\ }\textbf {\bibinfo {volume}
  {5}},\ \bibinfo {pages} {024002} (\bibinfo {year} {2021})}\BibitemShut
  {NoStop}%
\bibitem [{\citenamefont {Staros}\ \emph {et~al.}(2021)\citenamefont {Staros},
  \citenamefont {Hu}, \citenamefont {Tiihonen}, \citenamefont {Nanguneri},
  \citenamefont {Krogel}, \citenamefont {Bennett}, \citenamefont {Heinonen},
  \citenamefont {Ganesh},\ and\ \citenamefont {Rubenstein}}]{staros}%
  \BibitemOpen
  \bibfield  {author} {\bibinfo {author} {\bibfnamefont {D.}~\bibnamefont
  {Staros}}, \bibinfo {author} {\bibfnamefont {G.}~\bibnamefont {Hu}}, \bibinfo
  {author} {\bibfnamefont {J.}~\bibnamefont {Tiihonen}}, \bibinfo {author}
  {\bibfnamefont {R.}~\bibnamefont {Nanguneri}}, \bibinfo {author}
  {\bibfnamefont {J.}~\bibnamefont {Krogel}}, \bibinfo {author} {\bibfnamefont
  {M.~C.}\ \bibnamefont {Bennett}}, \bibinfo {author} {\bibfnamefont
  {O.}~\bibnamefont {Heinonen}}, \bibinfo {author} {\bibfnamefont
  {P.}~\bibnamefont {Ganesh}},\ and\ \bibinfo {author} {\bibfnamefont
  {B.}~\bibnamefont {Rubenstein}},\ }\bibfield  {title} {\enquote {\bibinfo
  {title} {A combined first principles study of the structural, magnetic, and
  phonon properties of monolayer cri$_3$},}\ }\bibfield  {booktitle} {\emph
  {\bibinfo {booktitle} {The Journal of Chemical Physics}},\ }\href
  {https://doi.org/10.1063/5.0074848} {\bibfield  {journal} {\bibinfo
  {journal} {J. Chem. Phys.}\ }\textbf {\bibinfo {volume} {156}},\ \bibinfo
  {pages} {014707} (\bibinfo {year} {2021})}\BibitemShut {NoStop}%
\bibitem [{\citenamefont {Annaberdiyev}\ \emph {et~al.}(2022)\citenamefont
  {Annaberdiyev}, \citenamefont {Melton}, \citenamefont {Wang},\ and\
  \citenamefont {Mitas}}]{PhysRevB.106.075127}%
  \BibitemOpen
  \bibfield  {author} {\bibinfo {author} {\bibfnamefont {A.}~\bibnamefont
  {Annaberdiyev}}, \bibinfo {author} {\bibfnamefont {C.~A.}\ \bibnamefont
  {Melton}}, \bibinfo {author} {\bibfnamefont {G.}~\bibnamefont {Wang}},\ and\
  \bibinfo {author} {\bibfnamefont {L.}~\bibnamefont {Mitas}},\ }\bibfield
  {title} {\enquote {\bibinfo {title} {Electronic structure of
  $\ensuremath{\alpha}\text{\ensuremath{-}}{\mathrm{rucl}}_{3}$ by fixed-node
  and fixed-phase diffusion monte carlo methods},}\ }\href
  {https://doi.org/10.1103/PhysRevB.106.075127} {\bibfield  {journal} {\bibinfo
   {journal} {Phys. Rev. B}\ }\textbf {\bibinfo {volume} {106}},\ \bibinfo
  {pages} {075127} (\bibinfo {year} {2022})}\BibitemShut {NoStop}%
\bibitem [{\citenamefont {Huang}\ \emph {et~al.}(2023)\citenamefont {Huang},
  \citenamefont {Faizan}, \citenamefont {Manzoor}, \citenamefont {Brndiar},
  \citenamefont {Mitas}, \citenamefont {Fabian},\ and\ \citenamefont
  {\ifmmode~\check{S}\else \v{S}\fi{}tich}}]{PhysRevResearch.5.033223}%
  \BibitemOpen
  \bibfield  {author} {\bibinfo {author} {\bibfnamefont {Y.}~\bibnamefont
  {Huang}}, \bibinfo {author} {\bibfnamefont {A.}~\bibnamefont {Faizan}},
  \bibinfo {author} {\bibfnamefont {M.}~\bibnamefont {Manzoor}}, \bibinfo
  {author} {\bibfnamefont {J.}~\bibnamefont {Brndiar}}, \bibinfo {author}
  {\bibfnamefont {L.}~\bibnamefont {Mitas}}, \bibinfo {author} {\bibfnamefont
  {J.}~\bibnamefont {Fabian}},\ and\ \bibinfo {author} {\bibfnamefont
  {I.}~\bibnamefont {\ifmmode~\check{S}\else \v{S}\fi{}tich}},\ }\bibfield
  {title} {\enquote {\bibinfo {title} {Colossal band gap response of
  single-layer phosphorene to strain predicted by quantum monte carlo},}\
  }\href {https://doi.org/10.1103/PhysRevResearch.5.033223} {\bibfield
  {journal} {\bibinfo  {journal} {Phys. Rev. Res.}\ }\textbf {\bibinfo {volume}
  {5}},\ \bibinfo {pages} {033223} (\bibinfo {year} {2023})}\BibitemShut
  {NoStop}%
\bibitem [{\citenamefont {Lee}\ \emph {et~al.}(2022)\citenamefont {Lee},
  \citenamefont {Hong}, \citenamefont {Ahn}, \citenamefont {Shin},
  \citenamefont {Benali},\ and\ \citenamefont {Kwon}}]{10.1063/5.0116092}%
  \BibitemOpen
  \bibfield  {author} {\bibinfo {author} {\bibfnamefont {G.}~\bibnamefont
  {Lee}}, \bibinfo {author} {\bibfnamefont {I.}~\bibnamefont {Hong}}, \bibinfo
  {author} {\bibfnamefont {J.}~\bibnamefont {Ahn}}, \bibinfo {author}
  {\bibfnamefont {H.}~\bibnamefont {Shin}}, \bibinfo {author} {\bibfnamefont
  {A.}~\bibnamefont {Benali}},\ and\ \bibinfo {author} {\bibfnamefont
  {Y.}~\bibnamefont {Kwon}},\ }\bibfield  {title} {\enquote {\bibinfo {title}
  {{Hydrogen separation with a graphenylene monolayer: Diffusion Monte Carlo
  study}},}\ }\href {https://doi.org/10.1063/5.0116092} {\bibfield  {journal}
  {\bibinfo  {journal} {The Journal of Chemical Physics}\ }\textbf {\bibinfo
  {volume} {157}},\ \bibinfo {pages} {144703} (\bibinfo {year} {2022})},\
  \Eprint
  {https://arxiv.org/abs/https://pubs.aip.org/aip/jcp/article-pdf/doi/10.1063/5.0116092/16550974/144703\_1\_online.pdf}
  {https://pubs.aip.org/aip/jcp/article-pdf/doi/10.1063/5.0116092/16550974/144703\_1\_online.pdf}
  \BibitemShut {NoStop}%
\bibitem [{\citenamefont {Huang}\ \emph {et~al.}(2024)\citenamefont {Huang},
  \citenamefont {Manzoor}, \citenamefont {Brndiar}, \citenamefont
  {Milivojevic},\ and\ \citenamefont {\ifmmode~\check{S}\else
  \v{S}\fi{}tich}}]{PhysRevResearch.6.013007}%
  \BibitemOpen
  \bibfield  {author} {\bibinfo {author} {\bibfnamefont {Y.}~\bibnamefont
  {Huang}}, \bibinfo {author} {\bibfnamefont {M.}~\bibnamefont {Manzoor}},
  \bibinfo {author} {\bibfnamefont {J.}~\bibnamefont {Brndiar}}, \bibinfo
  {author} {\bibfnamefont {M.}~\bibnamefont {Milivojevic}},\ and\ \bibinfo
  {author} {\bibfnamefont {I.}~\bibnamefont {\ifmmode~\check{S}\else
  \v{S}\fi{}tich}},\ }\bibfield  {title} {\enquote {\bibinfo {title}
  {Straintronics with single-layer ${\mathrm{mos}}_{2}$: A quantum monte carlo
  study},}\ }\href {https://doi.org/10.1103/PhysRevResearch.6.013007}
  {\bibfield  {journal} {\bibinfo  {journal} {Phys. Rev. Res.}\ }\textbf
  {\bibinfo {volume} {6}},\ \bibinfo {pages} {013007} (\bibinfo {year}
  {2024})}\BibitemShut {NoStop}%
\bibitem [{\citenamefont {Ahn}\ \emph {et~al.}(2018)\citenamefont {Ahn},
  \citenamefont {Hong}, \citenamefont {Kwon}, \citenamefont {Clay},
  \citenamefont {Shulenburger}, \citenamefont {Shin},\ and\ \citenamefont
  {Benali}}]{PhysRevB.98.085429}%
  \BibitemOpen
  \bibfield  {author} {\bibinfo {author} {\bibfnamefont {J.}~\bibnamefont
  {Ahn}}, \bibinfo {author} {\bibfnamefont {I.}~\bibnamefont {Hong}}, \bibinfo
  {author} {\bibfnamefont {Y.}~\bibnamefont {Kwon}}, \bibinfo {author}
  {\bibfnamefont {R.~C.}\ \bibnamefont {Clay}}, \bibinfo {author}
  {\bibfnamefont {L.}~\bibnamefont {Shulenburger}}, \bibinfo {author}
  {\bibfnamefont {H.}~\bibnamefont {Shin}},\ and\ \bibinfo {author}
  {\bibfnamefont {A.}~\bibnamefont {Benali}},\ }\bibfield  {title} {\enquote
  {\bibinfo {title} {Phase stability and interlayer interaction of blue
  phosphorene},}\ }\href {https://doi.org/10.1103/PhysRevB.98.085429}
  {\bibfield  {journal} {\bibinfo  {journal} {Phys. Rev. B}\ }\textbf {\bibinfo
  {volume} {98}},\ \bibinfo {pages} {085429} (\bibinfo {year}
  {2018})}\BibitemShut {NoStop}%
\bibitem [{\citenamefont {Ahn}\ \emph {et~al.}(2021{\natexlab{a}})\citenamefont
  {Ahn}, \citenamefont {Hong}, \citenamefont {Lee}, \citenamefont {Shin},
  \citenamefont {Benali},\ and\ \citenamefont {Kwon}}]{D1CP02473F}%
  \BibitemOpen
  \bibfield  {author} {\bibinfo {author} {\bibfnamefont {J.}~\bibnamefont
  {Ahn}}, \bibinfo {author} {\bibfnamefont {I.}~\bibnamefont {Hong}}, \bibinfo
  {author} {\bibfnamefont {G.}~\bibnamefont {Lee}}, \bibinfo {author}
  {\bibfnamefont {H.}~\bibnamefont {Shin}}, \bibinfo {author} {\bibfnamefont
  {A.}~\bibnamefont {Benali}},\ and\ \bibinfo {author} {\bibfnamefont
  {Y.}~\bibnamefont {Kwon}},\ }\bibfield  {title} {\enquote {\bibinfo {title}
  {Adsorption of a single pt atom on graphene: spin crossing between
  physisorbed triplet and chemisorbed singlet states},}\ }\href
  {https://doi.org/10.1039/D1CP02473F} {\bibfield  {journal} {\bibinfo
  {journal} {Phys. Chem. Chem. Phys.}\ }\textbf {\bibinfo {volume} {23}},\
  \bibinfo {pages} {22147--22154} (\bibinfo {year}
  {2021}{\natexlab{a}})}\BibitemShut {NoStop}%
\bibitem [{\citenamefont {Dubecký}\ \emph {et~al.}(2020)\citenamefont
  {Dubecký}, \citenamefont {Karlický}, \citenamefont {Minárik},\ and\
  \citenamefont {Mitas}}]{10.1063/5.0030952}%
  \BibitemOpen
  \bibfield  {author} {\bibinfo {author} {\bibfnamefont {M.}~\bibnamefont
  {Dubecký}}, \bibinfo {author} {\bibfnamefont {F.}~\bibnamefont {Karlický}},
  \bibinfo {author} {\bibfnamefont {S.}~\bibnamefont {Minárik}},\ and\
  \bibinfo {author} {\bibfnamefont {L.}~\bibnamefont {Mitas}},\ }\bibfield
  {title} {\enquote {\bibinfo {title} {{Fundamental gap of fluorographene by
  many-body GW and fixed-node diffusion Monte Carlo methods}},}\ }\href
  {https://doi.org/10.1063/5.0030952} {\bibfield  {journal} {\bibinfo
  {journal} {The Journal of Chemical Physics}\ }\textbf {\bibinfo {volume}
  {153}},\ \bibinfo {pages} {184706} (\bibinfo {year} {2020})},\ \Eprint
  {https://arxiv.org/abs/https://pubs.aip.org/aip/jcp/article-pdf/doi/10.1063/5.0030952/15583281/184706\_1\_online.pdf}
  {https://pubs.aip.org/aip/jcp/article-pdf/doi/10.1063/5.0030952/15583281/184706\_1\_online.pdf}
  \BibitemShut {NoStop}%
\bibitem [{\citenamefont {Ahn}\ \emph {et~al.}(2021{\natexlab{b}})\citenamefont
  {Ahn}, \citenamefont {Hong}, \citenamefont {Lee}, \citenamefont {Shin},
  \citenamefont {Benali},\ and\ \citenamefont {Kwon}}]{bluephos}%
  \BibitemOpen
  \bibfield  {author} {\bibinfo {author} {\bibfnamefont {J.}~\bibnamefont
  {Ahn}}, \bibinfo {author} {\bibfnamefont {I.}~\bibnamefont {Hong}}, \bibinfo
  {author} {\bibfnamefont {G.}~\bibnamefont {Lee}}, \bibinfo {author}
  {\bibfnamefont {H.}~\bibnamefont {Shin}}, \bibinfo {author} {\bibfnamefont
  {A.}~\bibnamefont {Benali}},\ and\ \bibinfo {author} {\bibfnamefont
  {Y.}~\bibnamefont {Kwon}},\ }\bibfield  {title} {\enquote {\bibinfo {title}
  {Metastable metallic phase of a bilayer blue phosphorene induced by
  interlayer bonding and intralayer charge redistributions},}\ }\href
  {https://doi.org/10.1021/acs.jpclett.1c03045} {\bibfield  {journal} {\bibinfo
   {journal} {The Journal of Physical Chemistry Letters}\ }\textbf {\bibinfo
  {volume} {12}},\ \bibinfo {pages} {10981--10986} (\bibinfo {year}
  {2021}{\natexlab{b}})}\BibitemShut {NoStop}%
\bibitem [{\citenamefont {Annaberdiyev}\ \emph {et~al.}(2023)\citenamefont
  {Annaberdiyev}, \citenamefont {Mandal}, \citenamefont {Mitas}, \citenamefont
  {Krogel},\ and\ \citenamefont {Ganesh}}]{chern}%
  \BibitemOpen
  \bibfield  {author} {\bibinfo {author} {\bibfnamefont {A.}~\bibnamefont
  {Annaberdiyev}}, \bibinfo {author} {\bibfnamefont {S.}~\bibnamefont
  {Mandal}}, \bibinfo {author} {\bibfnamefont {L.}~\bibnamefont {Mitas}},
  \bibinfo {author} {\bibfnamefont {J.~T.}\ \bibnamefont {Krogel}},\ and\
  \bibinfo {author} {\bibfnamefont {P.}~\bibnamefont {Ganesh}},\ }\bibfield
  {title} {\enquote {\bibinfo {title} {The role of electron correlations in the
  electronic structure of putative chern magnet tbmn6sn6},}\ }\href
  {https://doi.org/10.1038/s41535-023-00583-6} {\bibfield  {journal} {\bibinfo
  {journal} {npj Quantum Materials}\ }\textbf {\bibinfo {volume} {8}},\
  \bibinfo {pages} {50} (\bibinfo {year} {2023})}\BibitemShut {NoStop}%
\bibitem [{\citenamefont {Furness}\ \emph {et~al.}(2020)\citenamefont
  {Furness}, \citenamefont {Kaplan}, \citenamefont {Ning}, \citenamefont
  {Perdew},\ and\ \citenamefont {Sun}}]{r2scan}%
  \BibitemOpen
  \bibfield  {author} {\bibinfo {author} {\bibfnamefont {J.~W.}\ \bibnamefont
  {Furness}}, \bibinfo {author} {\bibfnamefont {A.~D.}\ \bibnamefont {Kaplan}},
  \bibinfo {author} {\bibfnamefont {J.}~\bibnamefont {Ning}}, \bibinfo {author}
  {\bibfnamefont {J.~P.}\ \bibnamefont {Perdew}},\ and\ \bibinfo {author}
  {\bibfnamefont {J.}~\bibnamefont {Sun}},\ }\bibfield  {title} {\enquote
  {\bibinfo {title} {Accurate and numerically efficient r2scan meta-generalized
  gradient approximation},}\ }\href
  {https://doi.org/10.1021/acs.jpclett.0c02405} {\bibfield  {journal} {\bibinfo
   {journal} {The Journal of Physical Chemistry Letters}\ }\textbf {\bibinfo
  {volume} {11}},\ \bibinfo {pages} {8208--8215} (\bibinfo {year}
  {2020})}\BibitemShut {NoStop}%
\bibitem [{\citenamefont {Sokolovskiy}\ \emph {et~al.}(2023)\citenamefont
  {Sokolovskiy}, \citenamefont {Baigutlin}, \citenamefont {Miroshkina},\ and\
  \citenamefont {Buchelnikov}}]{met13040728}%
  \BibitemOpen
  \bibfield  {author} {\bibinfo {author} {\bibfnamefont {V.}~\bibnamefont
  {Sokolovskiy}}, \bibinfo {author} {\bibfnamefont {D.}~\bibnamefont
  {Baigutlin}}, \bibinfo {author} {\bibfnamefont {O.}~\bibnamefont
  {Miroshkina}},\ and\ \bibinfo {author} {\bibfnamefont {V.}~\bibnamefont
  {Buchelnikov}},\ }\bibfield  {title} {\enquote {\bibinfo {title} {Meta-gga
  scan functional in the prediction of ground state properties of magnetic
  materials: Review of the current state},}\ }\href
  {https://doi.org/10.3390/met13040728} {\bibfield  {journal} {\bibinfo
  {journal} {Metals}\ }\textbf {\bibinfo {volume} {13}} (\bibinfo {year}
  {2023}),\ 10.3390/met13040728}\BibitemShut {NoStop}%
\bibitem [{\citenamefont {Swathilakshmi}, \citenamefont {Devi},\ and\
  \citenamefont {Sai~Gautam}(2023)}]{r2scan-mag}%
  \BibitemOpen
  \bibfield  {author} {\bibinfo {author} {\bibfnamefont {S.}~\bibnamefont
  {Swathilakshmi}}, \bibinfo {author} {\bibfnamefont {R.}~\bibnamefont
  {Devi}},\ and\ \bibinfo {author} {\bibfnamefont {G.}~\bibnamefont
  {Sai~Gautam}},\ }\bibfield  {title} {\enquote {\bibinfo {title} {Performance
  of the r2scan functional in transition metal oxides},}\ }\href
  {https://doi.org/10.1021/acs.jctc.3c00030} {\bibfield  {journal} {\bibinfo
  {journal} {Journal of Chemical Theory and Computation}\ }\textbf {\bibinfo
  {volume} {19}},\ \bibinfo {pages} {4202--4215} (\bibinfo {year}
  {2023})}\BibitemShut {NoStop}%
\bibitem [{\citenamefont {Ekholm}\ \emph {et~al.}(2018)\citenamefont {Ekholm},
  \citenamefont {Gambino}, \citenamefont {J\"onsson}, \citenamefont
  {Tasn\'adi}, \citenamefont {Alling},\ and\ \citenamefont
  {Abrikosov}}]{PhysRevB.98.094413}%
  \BibitemOpen
  \bibfield  {author} {\bibinfo {author} {\bibfnamefont {M.}~\bibnamefont
  {Ekholm}}, \bibinfo {author} {\bibfnamefont {D.}~\bibnamefont {Gambino}},
  \bibinfo {author} {\bibfnamefont {H.~J.~M.}\ \bibnamefont {J\"onsson}},
  \bibinfo {author} {\bibfnamefont {F.}~\bibnamefont {Tasn\'adi}}, \bibinfo
  {author} {\bibfnamefont {B.}~\bibnamefont {Alling}},\ and\ \bibinfo {author}
  {\bibfnamefont {I.~A.}\ \bibnamefont {Abrikosov}},\ }\bibfield  {title}
  {\enquote {\bibinfo {title} {Assessing the scan functional for itinerant
  electron ferromagnets},}\ }\href {https://doi.org/10.1103/PhysRevB.98.094413}
  {\bibfield  {journal} {\bibinfo  {journal} {Phys. Rev. B}\ }\textbf {\bibinfo
  {volume} {98}},\ \bibinfo {pages} {094413} (\bibinfo {year}
  {2018})}\BibitemShut {NoStop}%
\bibitem [{\citenamefont {Fu}\ and\ \citenamefont
  {Singh}(2019)}]{PhysRevB.100.045126}%
  \BibitemOpen
  \bibfield  {author} {\bibinfo {author} {\bibfnamefont {Y.}~\bibnamefont
  {Fu}}\ and\ \bibinfo {author} {\bibfnamefont {D.~J.}\ \bibnamefont {Singh}},\
  }\bibfield  {title} {\enquote {\bibinfo {title} {Density functional methods
  for the magnetism of transition metals: Scan in relation to other
  functionals},}\ }\href {https://doi.org/10.1103/PhysRevB.100.045126}
  {\bibfield  {journal} {\bibinfo  {journal} {Phys. Rev. B}\ }\textbf {\bibinfo
  {volume} {100}},\ \bibinfo {pages} {045126} (\bibinfo {year}
  {2019})}\BibitemShut {NoStop}%
\bibitem [{\citenamefont {Lopez}\ \emph {et~al.}(2024)\citenamefont {Lopez},
  \citenamefont {Melton}, \citenamefont {Ahn}, \citenamefont {Rubenstein},\
  and\ \citenamefont {Krogel}}]{lopez2024identifyingbandinversionstopological}%
  \BibitemOpen
  \bibfield  {author} {\bibinfo {author} {\bibfnamefont {A.}~\bibnamefont
  {Lopez}}, \bibinfo {author} {\bibfnamefont {C.~A.}\ \bibnamefont {Melton}},
  \bibinfo {author} {\bibfnamefont {J.}~\bibnamefont {Ahn}}, \bibinfo {author}
  {\bibfnamefont {B.~M.}\ \bibnamefont {Rubenstein}},\ and\ \bibinfo {author}
  {\bibfnamefont {J.~T.}\ \bibnamefont {Krogel}},\ }\href
  {https://arxiv.org/abs/2412.14388} {\enquote {\bibinfo {title} {Identifying
  band inversions in topological materials using diffusion monte carlo},}\ }
  (\bibinfo {year} {2024}),\ \Eprint {https://arxiv.org/abs/2412.14388}
  {arXiv:2412.14388 [cond-mat.str-el]} \BibitemShut {NoStop}%
\bibitem [{\citenamefont {Fumega}\ \emph {et~al.}(2023)\citenamefont {Fumega},
  \citenamefont {Diego}, \citenamefont {Pardo}, \citenamefont {Blanco-Canosa},\
  and\ \citenamefont {Errea}}]{vse2-anharm}%
  \BibitemOpen
  \bibfield  {author} {\bibinfo {author} {\bibfnamefont {A.~O.}\ \bibnamefont
  {Fumega}}, \bibinfo {author} {\bibfnamefont {J.}~\bibnamefont {Diego}},
  \bibinfo {author} {\bibfnamefont {V.}~\bibnamefont {Pardo}}, \bibinfo
  {author} {\bibfnamefont {S.}~\bibnamefont {Blanco-Canosa}},\ and\ \bibinfo
  {author} {\bibfnamefont {I.}~\bibnamefont {Errea}},\ }\bibfield  {title}
  {\enquote {\bibinfo {title} {Anharmonicity reveals the tunability of the
  charge density wave orders in monolayer vse2},}\ }\href
  {https://doi.org/10.1021/acs.nanolett.2c04584} {\bibfield  {journal}
  {\bibinfo  {journal} {Nano Letters}\ }\textbf {\bibinfo {volume} {23}},\
  \bibinfo {pages} {1794--1800} (\bibinfo {year} {2023})}\BibitemShut {NoStop}%
\bibitem [{\citenamefont {Pandey}\ and\ \citenamefont
  {Soni}(2020)}]{PhysRevResearch.2.033118}%
  \BibitemOpen
  \bibfield  {author} {\bibinfo {author} {\bibfnamefont {J.}~\bibnamefont
  {Pandey}}\ and\ \bibinfo {author} {\bibfnamefont {A.}~\bibnamefont {Soni}},\
  }\bibfield  {title} {\enquote {\bibinfo {title} {Electron-phonon interactions
  and two-phonon modes associated with charge density wave in single
  crystalline $1\mathrm{T}\text{\ensuremath{-}}\mathrm{V}{\mathrm{se}}_{2}$},}\
  }\href {https://doi.org/10.1103/PhysRevResearch.2.033118} {\bibfield
  {journal} {\bibinfo  {journal} {Phys. Rev. Res.}\ }\textbf {\bibinfo {volume}
  {2}},\ \bibinfo {pages} {033118} (\bibinfo {year} {2020})}\BibitemShut
  {NoStop}%
\bibitem [{\citenamefont {Lu}\ \emph {et~al.}(2024)\citenamefont {Lu},
  \citenamefont {Li}, \citenamefont {Feng}, \citenamefont {Zheng},
  \citenamefont {Liu}, \citenamefont {Yan}, \citenamefont {Hu},\ and\
  \citenamefont {Xue}}]{Se-ring}%
  \BibitemOpen
  \bibfield  {author} {\bibinfo {author} {\bibfnamefont {W.}~\bibnamefont
  {Lu}}, \bibinfo {author} {\bibfnamefont {Z.}~\bibnamefont {Li}}, \bibinfo
  {author} {\bibfnamefont {M.}~\bibnamefont {Feng}}, \bibinfo {author}
  {\bibfnamefont {L.}~\bibnamefont {Zheng}}, \bibinfo {author} {\bibfnamefont
  {S.}~\bibnamefont {Liu}}, \bibinfo {author} {\bibfnamefont {B.}~\bibnamefont
  {Yan}}, \bibinfo {author} {\bibfnamefont {J.-S.}\ \bibnamefont {Hu}},\ and\
  \bibinfo {author} {\bibfnamefont {D.-J.}\ \bibnamefont {Xue}},\ }\bibfield
  {title} {\enquote {\bibinfo {title} {Structure of amorphous selenium: Small
  ring, big controversy},}\ }\href {https://doi.org/10.1021/jacs.4c00219}
  {\bibfield  {journal} {\bibinfo  {journal} {Journal of the American Chemical
  Society}\ }\textbf {\bibinfo {volume} {146}},\ \bibinfo {pages} {6345--6351}
  (\bibinfo {year} {2024})}\BibitemShut {NoStop}%
\bibitem [{\citenamefont {Goldan}\ \emph {et~al.}(2016)\citenamefont {Goldan},
  \citenamefont {Li}, \citenamefont {Pennycook}, \citenamefont {Schneider},
  \citenamefont {Blom},\ and\ \citenamefont {Zhao}}]{10.1063/1.4962315}%
  \BibitemOpen
  \bibfield  {author} {\bibinfo {author} {\bibfnamefont {A.~H.}\ \bibnamefont
  {Goldan}}, \bibinfo {author} {\bibfnamefont {C.}~\bibnamefont {Li}}, \bibinfo
  {author} {\bibfnamefont {S.~J.}\ \bibnamefont {Pennycook}}, \bibinfo {author}
  {\bibfnamefont {J.}~\bibnamefont {Schneider}}, \bibinfo {author}
  {\bibfnamefont {A.}~\bibnamefont {Blom}},\ and\ \bibinfo {author}
  {\bibfnamefont {W.}~\bibnamefont {Zhao}},\ }\bibfield  {title} {\enquote
  {\bibinfo {title} {{Molecular structure of vapor-deposited amorphous
  selenium}},}\ }\href {https://doi.org/10.1063/1.4962315} {\bibfield
  {journal} {\bibinfo  {journal} {Journal of Applied Physics}\ }\textbf
  {\bibinfo {volume} {120}},\ \bibinfo {pages} {135101} (\bibinfo {year}
  {2016})},\ \Eprint
  {https://arxiv.org/abs/https://pubs.aip.org/aip/jap/article-pdf/doi/10.1063/1.4962315/15185642/135101\_1\_online.pdf}
  {https://pubs.aip.org/aip/jap/article-pdf/doi/10.1063/1.4962315/15185642/135101\_1\_online.pdf}
  \BibitemShut {NoStop}%
\bibitem [{\citenamefont {Guo}\ and\ \citenamefont
  {Lu}(1998)}]{PhysRevB.57.10414}%
  \BibitemOpen
  \bibfield  {author} {\bibinfo {author} {\bibfnamefont {F.~Q.}\ \bibnamefont
  {Guo}}\ and\ \bibinfo {author} {\bibfnamefont {K.}~\bibnamefont {Lu}},\
  }\bibfield  {title} {\enquote {\bibinfo {title} {Microstructural evolution in
  melt-quenched amorphous se during mechanical attrition},}\ }\href
  {https://doi.org/10.1103/PhysRevB.57.10414} {\bibfield  {journal} {\bibinfo
  {journal} {Phys. Rev. B}\ }\textbf {\bibinfo {volume} {57}},\ \bibinfo
  {pages} {10414--10420} (\bibinfo {year} {1998})}\BibitemShut {NoStop}%
\bibitem [{\citenamefont {Kresse}\ and\ \citenamefont
  {Furthm\"uller}(1996)}]{PhysRevB.54.11169}%
  \BibitemOpen
  \bibfield  {author} {\bibinfo {author} {\bibfnamefont {G.}~\bibnamefont
  {Kresse}}\ and\ \bibinfo {author} {\bibfnamefont {J.}~\bibnamefont
  {Furthm\"uller}},\ }\bibfield  {title} {\enquote {\bibinfo {title}
  {{Efficient Iterative Schemes for ab initio Total-energy Calculations Using a
  Plane-wave Basis Set}},}\ }\href {https://doi.org/10.1103/PhysRevB.54.11169}
  {\bibfield  {journal} {\bibinfo  {journal} {Phys. Rev. B}\ }\textbf {\bibinfo
  {volume} {54}},\ \bibinfo {pages} {11169--11186} (\bibinfo {year}
  {1996})}\BibitemShut {NoStop}%
\bibitem [{\citenamefont {Kresse}\ and\ \citenamefont
  {Joubert}(1999)}]{PhysRevB.59.1758}%
  \BibitemOpen
  \bibfield  {author} {\bibinfo {author} {\bibfnamefont {G.}~\bibnamefont
  {Kresse}}\ and\ \bibinfo {author} {\bibfnamefont {D.}~\bibnamefont
  {Joubert}},\ }\bibfield  {title} {\enquote {\bibinfo {title} {{From Ultrasoft
  Pseudopotentials to the Projector Augmented-wave Method}},}\ }\href
  {https://doi.org/10.1103/PhysRevB.59.1758} {\bibfield  {journal} {\bibinfo
  {journal} {Phys. Rev. B}\ }\textbf {\bibinfo {volume} {59}},\ \bibinfo
  {pages} {1758--1775} (\bibinfo {year} {1999})}\BibitemShut {NoStop}%
\bibitem [{\citenamefont {Perdew}\ and\ \citenamefont
  {Zunger}(1981)}]{PhysRevB.23.5048}%
  \BibitemOpen
  \bibfield  {author} {\bibinfo {author} {\bibfnamefont {J.~P.}\ \bibnamefont
  {Perdew}}\ and\ \bibinfo {author} {\bibfnamefont {A.}~\bibnamefont
  {Zunger}},\ }\bibfield  {title} {\enquote {\bibinfo {title} {Self-interaction
  correction to density-functional approximations for many-electron systems},}\
  }\href {https://doi.org/10.1103/PhysRevB.23.5048} {\bibfield  {journal}
  {\bibinfo  {journal} {Phys. Rev. B}\ }\textbf {\bibinfo {volume} {23}},\
  \bibinfo {pages} {5048--5079} (\bibinfo {year} {1981})}\BibitemShut {NoStop}%
\bibitem [{\citenamefont {Ceperley}\ and\ \citenamefont
  {Alder}(1980)}]{PhysRevLett.45.566}%
  \BibitemOpen
  \bibfield  {author} {\bibinfo {author} {\bibfnamefont {D.~M.}\ \bibnamefont
  {Ceperley}}\ and\ \bibinfo {author} {\bibfnamefont {B.~J.}\ \bibnamefont
  {Alder}},\ }\bibfield  {title} {\enquote {\bibinfo {title} {Ground state of
  the electron gas by a stochastic method},}\ }\href
  {https://doi.org/10.1103/PhysRevLett.45.566} {\bibfield  {journal} {\bibinfo
  {journal} {Phys. Rev. Lett.}\ }\textbf {\bibinfo {volume} {45}},\ \bibinfo
  {pages} {566--569} (\bibinfo {year} {1980})}\BibitemShut {NoStop}%
\bibitem [{\citenamefont {Perdew}, \citenamefont {Burke},\ and\ \citenamefont
  {Ernzerhof}(1996)}]{PhysRevLett.77.3865}%
  \BibitemOpen
  \bibfield  {author} {\bibinfo {author} {\bibfnamefont {J.~P.}\ \bibnamefont
  {Perdew}}, \bibinfo {author} {\bibfnamefont {K.}~\bibnamefont {Burke}},\ and\
  \bibinfo {author} {\bibfnamefont {M.}~\bibnamefont {Ernzerhof}},\ }\bibfield
  {title} {\enquote {\bibinfo {title} {Generalized gradient approximation made
  simple},}\ }\href {https://doi.org/10.1103/PhysRevLett.77.3865} {\bibfield
  {journal} {\bibinfo  {journal} {Phys. Rev. Lett.}\ }\textbf {\bibinfo
  {volume} {77}},\ \bibinfo {pages} {3865--3868} (\bibinfo {year}
  {1996})}\BibitemShut {NoStop}%
\bibitem [{\citenamefont {Sun}, \citenamefont {Ruzsinszky},\ and\ \citenamefont
  {Perdew}(2015)}]{PhysRevLett.115.036402}%
  \BibitemOpen
  \bibfield  {author} {\bibinfo {author} {\bibfnamefont {J.}~\bibnamefont
  {Sun}}, \bibinfo {author} {\bibfnamefont {A.}~\bibnamefont {Ruzsinszky}},\
  and\ \bibinfo {author} {\bibfnamefont {J.~P.}\ \bibnamefont {Perdew}},\
  }\bibfield  {title} {\enquote {\bibinfo {title} {Strongly constrained and
  appropriately normed semilocal density functional},}\ }\href
  {https://doi.org/10.1103/PhysRevLett.115.036402} {\bibfield  {journal}
  {\bibinfo  {journal} {Phys. Rev. Lett.}\ }\textbf {\bibinfo {volume} {115}},\
  \bibinfo {pages} {036402} (\bibinfo {year} {2015})}\BibitemShut {NoStop}%
\bibitem [{\citenamefont {Baroni}\ \emph {et~al.}(2001)\citenamefont {Baroni},
  \citenamefont {de~Gironcoli}, \citenamefont {Dal~Corso},\ and\ \citenamefont
  {Giannozzi}}]{RevModPhys.73.515}%
  \BibitemOpen
  \bibfield  {author} {\bibinfo {author} {\bibfnamefont {S.}~\bibnamefont
  {Baroni}}, \bibinfo {author} {\bibfnamefont {S.}~\bibnamefont
  {de~Gironcoli}}, \bibinfo {author} {\bibfnamefont {A.}~\bibnamefont
  {Dal~Corso}},\ and\ \bibinfo {author} {\bibfnamefont {P.}~\bibnamefont
  {Giannozzi}},\ }\bibfield  {title} {\enquote {\bibinfo {title} {Phonons and
  related crystal properties from density-functional perturbation theory},}\
  }\href {https://doi.org/10.1103/RevModPhys.73.515} {\bibfield  {journal}
  {\bibinfo  {journal} {Rev. Mod. Phys.}\ }\textbf {\bibinfo {volume} {73}},\
  \bibinfo {pages} {515--562} (\bibinfo {year} {2001})}\BibitemShut {NoStop}%
\bibitem [{\citenamefont {Togo}\ \emph {et~al.}(2023)\citenamefont {Togo},
  \citenamefont {Chaput}, \citenamefont {Tadano},\ and\ \citenamefont
  {Tanaka}}]{phonopy-phono3py-JPCM}%
  \BibitemOpen
  \bibfield  {author} {\bibinfo {author} {\bibfnamefont {A.}~\bibnamefont
  {Togo}}, \bibinfo {author} {\bibfnamefont {L.}~\bibnamefont {Chaput}},
  \bibinfo {author} {\bibfnamefont {T.}~\bibnamefont {Tadano}},\ and\ \bibinfo
  {author} {\bibfnamefont {I.}~\bibnamefont {Tanaka}},\ }\bibfield  {title}
  {\enquote {\bibinfo {title} {Implementation strategies in phonopy and
  phono3py},}\ }\href {https://doi.org/10.1088/1361-648X/acd831} {\bibfield
  {journal} {\bibinfo  {journal} {J. Phys. Condens. Matter}\ }\textbf {\bibinfo
  {volume} {35}},\ \bibinfo {pages} {353001} (\bibinfo {year}
  {2023})}\BibitemShut {NoStop}%
\bibitem [{\citenamefont {Togo}(2023)}]{phonopy-phono3py-JPSJ}%
  \BibitemOpen
  \bibfield  {author} {\bibinfo {author} {\bibfnamefont {A.}~\bibnamefont
  {Togo}},\ }\bibfield  {title} {\enquote {\bibinfo {title} {First-principles
  phonon calculations with phonopy and phono3py},}\ }\href
  {https://doi.org/10.7566/JPSJ.92.012001} {\bibfield  {journal} {\bibinfo
  {journal} {J. Phys. Soc. Jpn.}\ }\textbf {\bibinfo {volume} {92}},\ \bibinfo
  {pages} {012001} (\bibinfo {year} {2023})}\BibitemShut {NoStop}%
\bibitem [{\citenamefont {Giannozzi}\ \emph {et~al.}(2009)\citenamefont
  {Giannozzi}, \citenamefont {Baroni}, \citenamefont {Bonini}, \citenamefont
  {Calandra}, \citenamefont {Car}, \citenamefont {Cavazzoni}, \citenamefont
  {Ceresoli}, \citenamefont {Chiarotti}, \citenamefont {Cococcioni},
  \citenamefont {Dabo}, \citenamefont {Corso}, \citenamefont {de~Gironcoli},
  \citenamefont {Fabris}, \citenamefont {Fratesi}, \citenamefont {Gebauer},
  \citenamefont {Gerstmann}, \citenamefont {Gougoussis}, \citenamefont
  {Kokalj}, \citenamefont {Lazzeri}, \citenamefont {Martin-Samos},
  \citenamefont {Marzari}, \citenamefont {Mauri}, \citenamefont {Mazzarello},
  \citenamefont {Paolini}, \citenamefont {Pasquarello}, \citenamefont
  {Paulatto}, \citenamefont {Sbraccia}, \citenamefont {Scandolo}, \citenamefont
  {Sclauzero}, \citenamefont {Seitsonen}, \citenamefont {Smogunov},
  \citenamefont {Umari},\ and\ \citenamefont {Wentzcovitch}}]{Giannozzi_2009}%
  \BibitemOpen
  \bibfield  {author} {\bibinfo {author} {\bibfnamefont {P.}~\bibnamefont
  {Giannozzi}}, \bibinfo {author} {\bibfnamefont {S.}~\bibnamefont {Baroni}},
  \bibinfo {author} {\bibfnamefont {N.}~\bibnamefont {Bonini}}, \bibinfo
  {author} {\bibfnamefont {M.}~\bibnamefont {Calandra}}, \bibinfo {author}
  {\bibfnamefont {R.}~\bibnamefont {Car}}, \bibinfo {author} {\bibfnamefont
  {C.}~\bibnamefont {Cavazzoni}}, \bibinfo {author} {\bibfnamefont
  {D.}~\bibnamefont {Ceresoli}}, \bibinfo {author} {\bibfnamefont {G.~L.}\
  \bibnamefont {Chiarotti}}, \bibinfo {author} {\bibfnamefont {M.}~\bibnamefont
  {Cococcioni}}, \bibinfo {author} {\bibfnamefont {I.}~\bibnamefont {Dabo}},
  \bibinfo {author} {\bibfnamefont {A.~D.}\ \bibnamefont {Corso}}, \bibinfo
  {author} {\bibfnamefont {S.}~\bibnamefont {de~Gironcoli}}, \bibinfo {author}
  {\bibfnamefont {S.}~\bibnamefont {Fabris}}, \bibinfo {author} {\bibfnamefont
  {G.}~\bibnamefont {Fratesi}}, \bibinfo {author} {\bibfnamefont
  {R.}~\bibnamefont {Gebauer}}, \bibinfo {author} {\bibfnamefont
  {U.}~\bibnamefont {Gerstmann}}, \bibinfo {author} {\bibfnamefont
  {C.}~\bibnamefont {Gougoussis}}, \bibinfo {author} {\bibfnamefont
  {A.}~\bibnamefont {Kokalj}}, \bibinfo {author} {\bibfnamefont
  {M.}~\bibnamefont {Lazzeri}}, \bibinfo {author} {\bibfnamefont
  {L.}~\bibnamefont {Martin-Samos}}, \bibinfo {author} {\bibfnamefont
  {N.}~\bibnamefont {Marzari}}, \bibinfo {author} {\bibfnamefont
  {F.}~\bibnamefont {Mauri}}, \bibinfo {author} {\bibfnamefont
  {R.}~\bibnamefont {Mazzarello}}, \bibinfo {author} {\bibfnamefont
  {S.}~\bibnamefont {Paolini}}, \bibinfo {author} {\bibfnamefont
  {A.}~\bibnamefont {Pasquarello}}, \bibinfo {author} {\bibfnamefont
  {L.}~\bibnamefont {Paulatto}}, \bibinfo {author} {\bibfnamefont
  {C.}~\bibnamefont {Sbraccia}}, \bibinfo {author} {\bibfnamefont
  {S.}~\bibnamefont {Scandolo}}, \bibinfo {author} {\bibfnamefont
  {G.}~\bibnamefont {Sclauzero}}, \bibinfo {author} {\bibfnamefont {A.~P.}\
  \bibnamefont {Seitsonen}}, \bibinfo {author} {\bibfnamefont {A.}~\bibnamefont
  {Smogunov}}, \bibinfo {author} {\bibfnamefont {P.}~\bibnamefont {Umari}},\
  and\ \bibinfo {author} {\bibfnamefont {R.~M.}\ \bibnamefont {Wentzcovitch}},\
  }\bibfield  {title} {\enquote {\bibinfo {title} {{{QUANTUM} {ESPRESSO}: A
  Modular and Open-source Software Project for Quantum Simulations of
  Materials}},}\ }\href {https://doi.org/10.1088/0953-8984/21/39/395502}
  {\bibfield  {journal} {\bibinfo  {journal} {J. of Phys.: Cond. Matt.}\
  }\textbf {\bibinfo {volume} {21}},\ \bibinfo {pages} {395502} (\bibinfo
  {year} {2009})}\BibitemShut {NoStop}%
\bibitem [{\citenamefont {Krogel}, \citenamefont {Santana},\ and\ \citenamefont
  {Reboredo}(2016)}]{PhysRevB.93.075143}%
  \BibitemOpen
  \bibfield  {author} {\bibinfo {author} {\bibfnamefont {J.~T.}\ \bibnamefont
  {Krogel}}, \bibinfo {author} {\bibfnamefont {J.~A.}\ \bibnamefont
  {Santana}},\ and\ \bibinfo {author} {\bibfnamefont {F.~A.}\ \bibnamefont
  {Reboredo}},\ }\bibfield  {title} {\enquote {\bibinfo {title}
  {Pseudopotentials for quantum {Monte Carlo} studies of transition metal
  oxides},}\ }\href {https://doi.org/10.1103/PhysRevB.93.075143} {\bibfield
  {journal} {\bibinfo  {journal} {Phys. Rev. B}\ }\textbf {\bibinfo {volume}
  {93}},\ \bibinfo {pages} {075143} (\bibinfo {year} {2016})}\BibitemShut
  {NoStop}%
\bibitem [{\citenamefont {Burkatzki}, \citenamefont {Filippi},\ and\
  \citenamefont {Dolg}(2007)}]{doi:10.1063/1.2741534}%
  \BibitemOpen
  \bibfield  {author} {\bibinfo {author} {\bibfnamefont {M.}~\bibnamefont
  {Burkatzki}}, \bibinfo {author} {\bibfnamefont {C.}~\bibnamefont {Filippi}},\
  and\ \bibinfo {author} {\bibfnamefont {M.}~\bibnamefont {Dolg}},\ }\bibfield
  {title} {\enquote {\bibinfo {title} {{Energy-consistent Pseudopotentials for
  Quantum Monte Carlo Calculations}},}\ }\href
  {https://doi.org/10.1063/1.2741534} {\bibfield  {journal} {\bibinfo
  {journal} {J. Chem. Phys.}\ }\textbf {\bibinfo {volume} {126}},\ \bibinfo
  {pages} {234105} (\bibinfo {year} {2007})}\BibitemShut {NoStop}%
\bibitem [{\citenamefont {Kim}\ \emph {et~al.}(2018)\citenamefont {Kim},
  \citenamefont {Baczewski}, \citenamefont {Beaudet}, \citenamefont {Benali},
  \citenamefont {Bennett}, \citenamefont {Berrill}, \citenamefont {Blunt},
  \citenamefont {Borda}, \citenamefont {Casula}, \citenamefont {Ceperley},
  \citenamefont {Chiesa}, \citenamefont {Clark}, \citenamefont {Clay},
  \citenamefont {Delaney}, \citenamefont {Dewing}, \citenamefont {Esler},
  \citenamefont {Hao}, \citenamefont {Heinonen}, \citenamefont {Kent},
  \citenamefont {Krogel}, \citenamefont {Kyl{\"a}np{\"a}{\"a}}, \citenamefont
  {Li}, \citenamefont {Lopez}, \citenamefont {Luo}, \citenamefont {Malone},
  \citenamefont {Martin}, \citenamefont {Mathuriya}, \citenamefont {McMinis},
  \citenamefont {Melton}, \citenamefont {Mitas}, \citenamefont {Morales},
  \citenamefont {Neuscamman}, \citenamefont {Parker}, \citenamefont {Flores},
  \citenamefont {Romero}, \citenamefont {Rubenstein}, \citenamefont {Shea},
  \citenamefont {Shin}, \citenamefont {Shulenburger}, \citenamefont {Tillack},
  \citenamefont {Townsend}, \citenamefont {Tubman}, \citenamefont {Goetz},
  \citenamefont {Vincent}, \citenamefont {Yang}, \citenamefont {Yang},
  \citenamefont {Zhang},\ and\ \citenamefont {Zhao}}]{Kim_2018}%
  \BibitemOpen
  \bibfield  {author} {\bibinfo {author} {\bibfnamefont {J.}~\bibnamefont
  {Kim}}, \bibinfo {author} {\bibfnamefont {A.~D.}\ \bibnamefont {Baczewski}},
  \bibinfo {author} {\bibfnamefont {T.~D.}\ \bibnamefont {Beaudet}}, \bibinfo
  {author} {\bibfnamefont {A.}~\bibnamefont {Benali}}, \bibinfo {author}
  {\bibfnamefont {M.~C.}\ \bibnamefont {Bennett}}, \bibinfo {author}
  {\bibfnamefont {M.~A.}\ \bibnamefont {Berrill}}, \bibinfo {author}
  {\bibfnamefont {N.~S.}\ \bibnamefont {Blunt}}, \bibinfo {author}
  {\bibfnamefont {E.~J.~L.}\ \bibnamefont {Borda}}, \bibinfo {author}
  {\bibfnamefont {M.}~\bibnamefont {Casula}}, \bibinfo {author} {\bibfnamefont
  {D.~M.}\ \bibnamefont {Ceperley}}, \bibinfo {author} {\bibfnamefont
  {S.}~\bibnamefont {Chiesa}}, \bibinfo {author} {\bibfnamefont {B.~K.}\
  \bibnamefont {Clark}}, \bibinfo {author} {\bibfnamefont {R.~C.}\ \bibnamefont
  {Clay}}, \bibinfo {author} {\bibfnamefont {K.~T.}\ \bibnamefont {Delaney}},
  \bibinfo {author} {\bibfnamefont {M.}~\bibnamefont {Dewing}}, \bibinfo
  {author} {\bibfnamefont {K.~P.}\ \bibnamefont {Esler}}, \bibinfo {author}
  {\bibfnamefont {H.}~\bibnamefont {Hao}}, \bibinfo {author} {\bibfnamefont
  {O.}~\bibnamefont {Heinonen}}, \bibinfo {author} {\bibfnamefont {P.~R.~C.}\
  \bibnamefont {Kent}}, \bibinfo {author} {\bibfnamefont {J.~T.}\ \bibnamefont
  {Krogel}}, \bibinfo {author} {\bibfnamefont {I.}~\bibnamefont
  {Kyl{\"a}np{\"a}{\"a}}}, \bibinfo {author} {\bibfnamefont {Y.~W.}\
  \bibnamefont {Li}}, \bibinfo {author} {\bibfnamefont {M.~G.}\ \bibnamefont
  {Lopez}}, \bibinfo {author} {\bibfnamefont {Y.}~\bibnamefont {Luo}}, \bibinfo
  {author} {\bibfnamefont {F.~D.}\ \bibnamefont {Malone}}, \bibinfo {author}
  {\bibfnamefont {R.~M.}\ \bibnamefont {Martin}}, \bibinfo {author}
  {\bibfnamefont {A.}~\bibnamefont {Mathuriya}}, \bibinfo {author}
  {\bibfnamefont {J.}~\bibnamefont {McMinis}}, \bibinfo {author} {\bibfnamefont
  {C.~A.}\ \bibnamefont {Melton}}, \bibinfo {author} {\bibfnamefont
  {L.}~\bibnamefont {Mitas}}, \bibinfo {author} {\bibfnamefont {M.~A.}\
  \bibnamefont {Morales}}, \bibinfo {author} {\bibfnamefont {E.}~\bibnamefont
  {Neuscamman}}, \bibinfo {author} {\bibfnamefont {W.~D.}\ \bibnamefont
  {Parker}}, \bibinfo {author} {\bibfnamefont {S.~D.~P.}\ \bibnamefont
  {Flores}}, \bibinfo {author} {\bibfnamefont {N.~A.}\ \bibnamefont {Romero}},
  \bibinfo {author} {\bibfnamefont {B.~M.}\ \bibnamefont {Rubenstein}},
  \bibinfo {author} {\bibfnamefont {J.~A.~R.}\ \bibnamefont {Shea}}, \bibinfo
  {author} {\bibfnamefont {H.}~\bibnamefont {Shin}}, \bibinfo {author}
  {\bibfnamefont {L.}~\bibnamefont {Shulenburger}}, \bibinfo {author}
  {\bibfnamefont {A.~F.}\ \bibnamefont {Tillack}}, \bibinfo {author}
  {\bibfnamefont {J.~P.}\ \bibnamefont {Townsend}}, \bibinfo {author}
  {\bibfnamefont {N.~M.}\ \bibnamefont {Tubman}}, \bibinfo {author}
  {\bibfnamefont {B.~V.~D.}\ \bibnamefont {Goetz}}, \bibinfo {author}
  {\bibfnamefont {J.~E.}\ \bibnamefont {Vincent}}, \bibinfo {author}
  {\bibfnamefont {D.~C.}\ \bibnamefont {Yang}}, \bibinfo {author}
  {\bibfnamefont {Y.}~\bibnamefont {Yang}}, \bibinfo {author} {\bibfnamefont
  {S.}~\bibnamefont {Zhang}},\ and\ \bibinfo {author} {\bibfnamefont
  {L.}~\bibnamefont {Zhao}},\ }\bibfield  {title} {\enquote {\bibinfo {title}
  {{QMCPACK}: An open source ab initio quantum {Monte Carlo} package for the
  electronic structure of atoms, molecules and solids},}\ }\href
  {https://doi.org/10.1088/1361-648x/aab9c3} {\bibfield  {journal} {\bibinfo
  {journal} {J. of Phys.: Cond. Matter}\ }\textbf {\bibinfo {volume} {30}},\
  \bibinfo {pages} {195901} (\bibinfo {year} {2018})}\BibitemShut {NoStop}%
\bibitem [{\citenamefont {Kent}\ \emph {et~al.}(2020)\citenamefont {Kent},
  \citenamefont {Annaberdiyev}, \citenamefont {Benali}, \citenamefont
  {Bennett}, \citenamefont {Landinez~Borda}, \citenamefont {Doak},
  \citenamefont {Hao}, \citenamefont {Jordan}, \citenamefont {Krogel},
  \citenamefont {Kyl{\"a}np{\"a}{\"a}}, \citenamefont {Lee}, \citenamefont
  {Luo}, \citenamefont {Malone}, \citenamefont {Melton}, \citenamefont {Mitas},
  \citenamefont {Morales}, \citenamefont {Neuscamman}, \citenamefont
  {Reboredo}, \citenamefont {Rubenstein}, \citenamefont {Saritas},
  \citenamefont {Upadhyay}, \citenamefont {Wang}, \citenamefont {Zhang},\ and\
  \citenamefont {Zhao}}]{doi:10.1063/5.0004860}%
  \BibitemOpen
  \bibfield  {author} {\bibinfo {author} {\bibfnamefont {P.~R.~C.}\
  \bibnamefont {Kent}}, \bibinfo {author} {\bibfnamefont {A.}~\bibnamefont
  {Annaberdiyev}}, \bibinfo {author} {\bibfnamefont {A.}~\bibnamefont
  {Benali}}, \bibinfo {author} {\bibfnamefont {M.~C.}\ \bibnamefont {Bennett}},
  \bibinfo {author} {\bibfnamefont {E.~J.}\ \bibnamefont {Landinez~Borda}},
  \bibinfo {author} {\bibfnamefont {P.}~\bibnamefont {Doak}}, \bibinfo {author}
  {\bibfnamefont {H.}~\bibnamefont {Hao}}, \bibinfo {author} {\bibfnamefont
  {K.~D.}\ \bibnamefont {Jordan}}, \bibinfo {author} {\bibfnamefont {J.~T.}\
  \bibnamefont {Krogel}}, \bibinfo {author} {\bibfnamefont {I.}~\bibnamefont
  {Kyl{\"a}np{\"a}{\"a}}}, \bibinfo {author} {\bibfnamefont {J.}~\bibnamefont
  {Lee}}, \bibinfo {author} {\bibfnamefont {Y.}~\bibnamefont {Luo}}, \bibinfo
  {author} {\bibfnamefont {F.~D.}\ \bibnamefont {Malone}}, \bibinfo {author}
  {\bibfnamefont {C.~A.}\ \bibnamefont {Melton}}, \bibinfo {author}
  {\bibfnamefont {L.}~\bibnamefont {Mitas}}, \bibinfo {author} {\bibfnamefont
  {M.~A.}\ \bibnamefont {Morales}}, \bibinfo {author} {\bibfnamefont
  {E.}~\bibnamefont {Neuscamman}}, \bibinfo {author} {\bibfnamefont {F.~A.}\
  \bibnamefont {Reboredo}}, \bibinfo {author} {\bibfnamefont {B.}~\bibnamefont
  {Rubenstein}}, \bibinfo {author} {\bibfnamefont {K.}~\bibnamefont {Saritas}},
  \bibinfo {author} {\bibfnamefont {S.}~\bibnamefont {Upadhyay}}, \bibinfo
  {author} {\bibfnamefont {G.}~\bibnamefont {Wang}}, \bibinfo {author}
  {\bibfnamefont {S.}~\bibnamefont {Zhang}},\ and\ \bibinfo {author}
  {\bibfnamefont {L.}~\bibnamefont {Zhao}},\ }\bibfield  {title} {\enquote
  {\bibinfo {title} {{QMCPACK: Advances in the Development, Efficiency, and
  Application of Auxiliary Field and Real-space Variational and Diffusion
  Quantum Monte Carlo}},}\ }\href {https://doi.org/10.1063/5.0004860}
  {\bibfield  {journal} {\bibinfo  {journal} {J. Chem. Phys.}\ }\textbf
  {\bibinfo {volume} {152}},\ \bibinfo {pages} {174105} (\bibinfo {year}
  {2020})}\BibitemShut {NoStop}%
\bibitem [{\citenamefont {Krogel}(2016)}]{nexus}%
  \BibitemOpen
  \bibfield  {author} {\bibinfo {author} {\bibfnamefont {J.~T.}\ \bibnamefont
  {Krogel}},\ }\bibfield  {title} {\enquote {\bibinfo {title} {Nexus: A modular
  workflow management system for quantum simulation codes},}\ }\href
  {https://doi.org/https://doi.org/10.1016/j.cpc.2015.08.012} {\bibfield
  {journal} {\bibinfo  {journal} {Computer Physics Communications}\ }\textbf
  {\bibinfo {volume} {198}},\ \bibinfo {pages} {154 -- 168} (\bibinfo {year}
  {2016})}\BibitemShut {NoStop}%
\bibitem [{\citenamefont {Slater}(1929)}]{PhysRev.34.1293}%
  \BibitemOpen
  \bibfield  {author} {\bibinfo {author} {\bibfnamefont {J.~C.}\ \bibnamefont
  {Slater}},\ }\bibfield  {title} {\enquote {\bibinfo {title} {The theory of
  complex spectra},}\ }\href {https://doi.org/10.1103/PhysRev.34.1293}
  {\bibfield  {journal} {\bibinfo  {journal} {Phys. Rev.}\ }\textbf {\bibinfo
  {volume} {34}},\ \bibinfo {pages} {1293--1322} (\bibinfo {year}
  {1929})}\BibitemShut {NoStop}%
\bibitem [{\citenamefont {Jastrow}(1955)}]{PhysRev.98.1479}%
  \BibitemOpen
  \bibfield  {author} {\bibinfo {author} {\bibfnamefont {R.}~\bibnamefont
  {Jastrow}},\ }\bibfield  {title} {\enquote {\bibinfo {title} {Many-body
  problem with strong forces},}\ }\href
  {https://doi.org/10.1103/PhysRev.98.1479} {\bibfield  {journal} {\bibinfo
  {journal} {Phys. Rev.}\ }\textbf {\bibinfo {volume} {98}},\ \bibinfo {pages}
  {1479--1484} (\bibinfo {year} {1955})}\BibitemShut {NoStop}%
\bibitem [{\citenamefont {Drummond}, \citenamefont {Towler},\ and\
  \citenamefont {Needs}(2004)}]{PhysRevB.70.235119}%
  \BibitemOpen
  \bibfield  {author} {\bibinfo {author} {\bibfnamefont {N.~D.}\ \bibnamefont
  {Drummond}}, \bibinfo {author} {\bibfnamefont {M.~D.}\ \bibnamefont
  {Towler}},\ and\ \bibinfo {author} {\bibfnamefont {R.~J.}\ \bibnamefont
  {Needs}},\ }\bibfield  {title} {\enquote {\bibinfo {title} {{Jastrow
  Correlation Factor for Atoms, Molecules, and Solids}},}\ }\href
  {https://doi.org/10.1103/PhysRevB.70.235119} {\bibfield  {journal} {\bibinfo
  {journal} {Phys. Rev. B}\ }\textbf {\bibinfo {volume} {70}},\ \bibinfo
  {pages} {235119} (\bibinfo {year} {2004})}\BibitemShut {NoStop}%
\bibitem [{\citenamefont {Umrigar}\ \emph {et~al.}(2007)\citenamefont
  {Umrigar}, \citenamefont {Toulouse}, \citenamefont {Filippi}, \citenamefont
  {Sorella},\ and\ \citenamefont {Hennig}}]{PhysRevLett.98.110201}%
  \BibitemOpen
  \bibfield  {author} {\bibinfo {author} {\bibfnamefont {C.~J.}\ \bibnamefont
  {Umrigar}}, \bibinfo {author} {\bibfnamefont {J.}~\bibnamefont {Toulouse}},
  \bibinfo {author} {\bibfnamefont {C.}~\bibnamefont {Filippi}}, \bibinfo
  {author} {\bibfnamefont {S.}~\bibnamefont {Sorella}},\ and\ \bibinfo {author}
  {\bibfnamefont {R.~G.}\ \bibnamefont {Hennig}},\ }\bibfield  {title}
  {\enquote {\bibinfo {title} {Alleviation of the fermion-sign problem by
  optimization of many-body wave functions},}\ }\href
  {https://doi.org/10.1103/PhysRevLett.98.110201} {\bibfield  {journal}
  {\bibinfo  {journal} {Phys. Rev. Lett.}\ }\textbf {\bibinfo {volume} {98}},\
  \bibinfo {pages} {110201} (\bibinfo {year} {2007})}\BibitemShut {NoStop}%
\bibitem [{\citenamefont {Umrigar}\ and\ \citenamefont
  {Filippi}(2005)}]{PhysRevLett.94.150201}%
  \BibitemOpen
  \bibfield  {author} {\bibinfo {author} {\bibfnamefont {C.~J.}\ \bibnamefont
  {Umrigar}}\ and\ \bibinfo {author} {\bibfnamefont {C.}~\bibnamefont
  {Filippi}},\ }\bibfield  {title} {\enquote {\bibinfo {title} {Energy and
  variance optimization of many-body wave functions},}\ }\href
  {https://doi.org/10.1103/PhysRevLett.94.150201} {\bibfield  {journal}
  {\bibinfo  {journal} {Phys. Rev. Lett.}\ }\textbf {\bibinfo {volume} {94}},\
  \bibinfo {pages} {150201} (\bibinfo {year} {2005})}\BibitemShut {NoStop}%
\bibitem [{\citenamefont {Mitas}, \citenamefont {Shirley},\ and\ \citenamefont
  {Ceperley}(1991)}]{doi:10.1063/1.460849}%
  \BibitemOpen
  \bibfield  {author} {\bibinfo {author} {\bibfnamefont {L.}~\bibnamefont
  {Mitas}}, \bibinfo {author} {\bibfnamefont {E.~L.}\ \bibnamefont {Shirley}},\
  and\ \bibinfo {author} {\bibfnamefont {D.~M.}\ \bibnamefont {Ceperley}},\
  }\bibfield  {title} {\enquote {\bibinfo {title} {{Nonlocal Pseudopotentials
  and Diffusion Monte Carlo}},}\ }\href {https://doi.org/10.1063/1.460849}
  {\bibfield  {journal} {\bibinfo  {journal} {J. Chem. Phys.}\ }\textbf
  {\bibinfo {volume} {95}},\ \bibinfo {pages} {3467--3475} (\bibinfo {year}
  {1991})}\BibitemShut {NoStop}%
\bibitem [{\citenamefont {Annaberdiyev}, \citenamefont {Ganesh},\ and\
  \citenamefont {Krogel}(2024)}]{ta}%
  \BibitemOpen
  \bibfield  {author} {\bibinfo {author} {\bibfnamefont {A.}~\bibnamefont
  {Annaberdiyev}}, \bibinfo {author} {\bibfnamefont {P.}~\bibnamefont
  {Ganesh}},\ and\ \bibinfo {author} {\bibfnamefont {J.~T.}\ \bibnamefont
  {Krogel}},\ }\bibfield  {title} {\enquote {\bibinfo {title} {Enhanced
  twist-averaging technique for magnetic metals: Applications using quantum
  monte carlo},}\ }\href {https://doi.org/10.1021/acs.jctc.4c00058} {\bibfield
  {journal} {\bibinfo  {journal} {Journal of Chemical Theory and Computation}\
  }\textbf {\bibinfo {volume} {20}},\ \bibinfo {pages} {2786--2797} (\bibinfo
  {year} {2024})}\BibitemShut {NoStop}%
\bibitem [{\citenamefont {Lado}\ and\ \citenamefont
  {Fern{\'a}ndez-Rossier}(2017)}]{Lado_2017}%
  \BibitemOpen
  \bibfield  {author} {\bibinfo {author} {\bibfnamefont {J.~L.}\ \bibnamefont
  {Lado}}\ and\ \bibinfo {author} {\bibfnamefont {J.}~\bibnamefont
  {Fern{\'a}ndez-Rossier}},\ }\bibfield  {title} {\enquote {\bibinfo {title}
  {On the origin of magnetic anisotropy in two dimensional cri3},}\ }\href
  {https://doi.org/10.1088/2053-1583/aa75ed} {\bibfield  {journal} {\bibinfo
  {journal} {2D Materials}\ }\textbf {\bibinfo {volume} {4}},\ \bibinfo {pages}
  {035002} (\bibinfo {year} {2017})}\BibitemShut {NoStop}%
\bibitem [{\citenamefont {Torelli}\ and\ \citenamefont
  {Olsen}(2018)}]{Torelli_2019}%
  \BibitemOpen
  \bibfield  {author} {\bibinfo {author} {\bibfnamefont {D.}~\bibnamefont
  {Torelli}}\ and\ \bibinfo {author} {\bibfnamefont {T.}~\bibnamefont
  {Olsen}},\ }\bibfield  {title} {\enquote {\bibinfo {title} {Calculating
  critical temperatures for ferromagnetic order in two-dimensional
  materials},}\ }\href {https://doi.org/10.1088/2053-1583/aaf06d} {\bibfield
  {journal} {\bibinfo  {journal} {2D Materials}\ }\textbf {\bibinfo {volume}
  {6}},\ \bibinfo {pages} {015028} (\bibinfo {year} {2018})}\BibitemShut
  {NoStop}%
\end{thebibliography}%


\providecommand{\noopsort}[1]{}\providecommand{\singleletter}[1]{#1}%
\begin{thebibliography}{8}%
\makeatletter
\providecommand \@ifxundefined [1]{%
 \@ifx{#1\undefined}
}%
\providecommand \@ifnum [1]{%
 \ifnum #1\expandafter \@firstoftwo
 \else \expandafter \@secondoftwo
 \fi
}%
\providecommand \@ifx [1]{%
 \ifx #1\expandafter \@firstoftwo
 \else \expandafter \@secondoftwo
 \fi
}%
\providecommand \natexlab [1]{#1}%
\providecommand \enquote  [1]{``#1''}%
\providecommand \bibnamefont  [1]{#1}%
\providecommand \bibfnamefont [1]{#1}%
\providecommand \citenamefont [1]{#1}%
\providecommand \href@noop [0]{\@secondoftwo}%
\providecommand \href [0]{\begingroup \@sanitize@url \@href}%
\providecommand \@href[1]{\@@startlink{#1}\@@href}%
\providecommand \@@href[1]{\endgroup#1\@@endlink}%
\providecommand \@sanitize@url [0]{\catcode `\\12\catcode `\$12\catcode
  `\&12\catcode `\#12\catcode `\^12\catcode `\_12\catcode `\%12\relax}%
\providecommand \@@startlink[1]{}%
\providecommand \@@endlink[0]{}%
\providecommand \url  [0]{\begingroup\@sanitize@url \@url }%
\providecommand \@url [1]{\endgroup\@href {#1}{\urlprefix }}%
\providecommand \urlprefix  [0]{URL }%
\providecommand \Eprint [0]{\href }%
\providecommand \doibase [0]{http://dx.doi.org/}%
\providecommand \selectlanguage [0]{\@gobble}%
\providecommand \bibinfo  [0]{\@secondoftwo}%
\providecommand \bibfield  [0]{\@secondoftwo}%
\providecommand \translation [1]{[#1]}%
\providecommand \BibitemOpen [0]{}%
\providecommand \bibitemStop [0]{}%
\providecommand \bibitemNoStop [0]{.\EOS\space}%
\providecommand \EOS [0]{\spacefactor3000\relax}%
\providecommand \BibitemShut  [1]{\csname bibitem#1\endcsname}%
\let\auto@bib@innerbib\@empty
\bibitem [{\citenamefont {Wines}\ \emph {et~al.}(2023)\citenamefont {Wines},
  \citenamefont {Tiihonen}, \citenamefont {Saritas}, \citenamefont {Krogel},\
  and\ \citenamefont {Ataca}}]{vse2-wines}%
  \BibitemOpen
  \bibfield  {author} {\bibinfo {author} {\bibfnamefont {D.}~\bibnamefont
  {Wines}}, \bibinfo {author} {\bibfnamefont {J.}~\bibnamefont {Tiihonen}},
  \bibinfo {author} {\bibfnamefont {K.}~\bibnamefont {Saritas}}, \bibinfo
  {author} {\bibfnamefont {J.~T.}\ \bibnamefont {Krogel}}, \ and\ \bibinfo
  {author} {\bibfnamefont {C.}~\bibnamefont {Ataca}},\ }\bibfield  {title}
  {\enquote {\bibinfo {title} {A quantum monte carlo study of the structural,
  energetic, and magnetic properties of two-dimensional h and t phase vse2},}\
  }\href {\doibase 10.1021/acs.jpclett.3c00497} {\bibfield  {journal} {\bibinfo
   {journal} {The Journal of Physical Chemistry Letters}\ }\textbf {\bibinfo
  {volume} {14}},\ \bibinfo {pages} {3553--3560} (\bibinfo {year}
  {2023})}\BibitemShut {NoStop}%
\bibitem [{\citenamefont {Yin}\ \emph {et~al.}(2022)\citenamefont {Yin},
  \citenamefont {Berlijn}, \citenamefont {Juneja}, \citenamefont {Lindsay},\
  and\ \citenamefont {Parker}}]{PhysRevB.106.085117}%
  \BibitemOpen
  \bibfield  {author} {\bibinfo {author} {\bibfnamefont {L.}~\bibnamefont
  {Yin}}, \bibinfo {author} {\bibfnamefont {T.}~\bibnamefont {Berlijn}},
  \bibinfo {author} {\bibfnamefont {R.}~\bibnamefont {Juneja}}, \bibinfo
  {author} {\bibfnamefont {L.}~\bibnamefont {Lindsay}}, \ and\ \bibinfo
  {author} {\bibfnamefont {D.~S.}\ \bibnamefont {Parker}},\ }\bibfield  {title}
  {\enquote {\bibinfo {title} {Competing magnetic and nonmagnetic states in
  monolayer ${\mathrm{vse}}_{2}$ with charge density wave},}\ }\href {\doibase
  10.1103/PhysRevB.106.085117} {\bibfield  {journal} {\bibinfo  {journal}
  {Phys. Rev. B}\ }\textbf {\bibinfo {volume} {106}},\ \bibinfo {pages}
  {085117} (\bibinfo {year} {2022})}\BibitemShut {NoStop}%
\bibitem [{\citenamefont {{\v{S}}abani}, \citenamefont {Bacaksiz},\ and\
  \citenamefont {Milo{\v{s}}evi{\'c}}(2020)}]{vsabani2020ab}%
  \BibitemOpen
  \bibfield  {author} {\bibinfo {author} {\bibfnamefont {D.}~\bibnamefont
  {{\v{S}}abani}}, \bibinfo {author} {\bibfnamefont {C.}~\bibnamefont
  {Bacaksiz}}, \ and\ \bibinfo {author} {\bibfnamefont {M.}~\bibnamefont
  {Milo{\v{s}}evi{\'c}}},\ }\bibfield  {title} {\enquote {\bibinfo {title} {Ab
  initio methodology for magnetic exchange parameters: Generic four-state
  energy mapping onto a heisenberg spin hamiltonian},}\ }\href@noop {}
  {\bibfield  {journal} {\bibinfo  {journal} {Physical Review B}\ }\textbf
  {\bibinfo {volume} {102}},\ \bibinfo {pages} {014457} (\bibinfo {year}
  {2020})}\BibitemShut {NoStop}%
\bibitem [{\citenamefont {Momma}\ and\ \citenamefont
  {Izumi}(2011)}]{Momma:db5098}%
  \BibitemOpen
  \bibfield  {author} {\bibinfo {author} {\bibfnamefont {K.}~\bibnamefont
  {Momma}}\ and\ \bibinfo {author} {\bibfnamefont {F.}~\bibnamefont {Izumi}},\
  }\bibfield  {title} {\enquote {\bibinfo {title} {{{\it VESTA3} for
  three-dimensional visualization of crystal, volumetric and morphology
  data}},}\ }\href {\doibase 10.1107/S0021889811038970} {\bibfield  {journal}
  {\bibinfo  {journal} {Journal of Applied Crystallography}\ }\textbf {\bibinfo
  {volume} {44}},\ \bibinfo {pages} {1272--1276} (\bibinfo {year}
  {2011})}\BibitemShut {NoStop}%
\bibitem [{\citenamefont {Rigoult}\ \emph {et~al.}(1982)\citenamefont
  {Rigoult}, \citenamefont {Guidi-Morosini}, \citenamefont {Tomas},\ and\
  \citenamefont {Molinie}}]{Rigoult:a21216}%
  \BibitemOpen
  \bibfield  {author} {\bibinfo {author} {\bibfnamefont {J.}~\bibnamefont
  {Rigoult}}, \bibinfo {author} {\bibfnamefont {C.}~\bibnamefont
  {Guidi-Morosini}}, \bibinfo {author} {\bibfnamefont {A.}~\bibnamefont
  {Tomas}}, \ and\ \bibinfo {author} {\bibfnamefont {P.}~\bibnamefont
  {Molinie}},\ }\bibfield  {title} {\enquote {\bibinfo {title} {{An accurate
  refinement of 1{\it T}-VSe${\sb 2}$ at room temperature}},}\ }\href {\doibase
  10.1107/S0567740882006360} {\bibfield  {journal} {\bibinfo  {journal} {Acta
  Crystallographica Section B}\ }\textbf {\bibinfo {volume} {38}},\ \bibinfo
  {pages} {1557--1559} (\bibinfo {year} {1982})}\BibitemShut {NoStop}%
\bibitem [{\citenamefont {Lu}\ \emph {et~al.}(2024)\citenamefont {Lu},
  \citenamefont {Li}, \citenamefont {Feng}, \citenamefont {Zheng},
  \citenamefont {Liu}, \citenamefont {Yan}, \citenamefont {Hu},\ and\
  \citenamefont {Xue}}]{Se-ring}%
  \BibitemOpen
  \bibfield  {author} {\bibinfo {author} {\bibfnamefont {W.}~\bibnamefont
  {Lu}}, \bibinfo {author} {\bibfnamefont {Z.}~\bibnamefont {Li}}, \bibinfo
  {author} {\bibfnamefont {M.}~\bibnamefont {Feng}}, \bibinfo {author}
  {\bibfnamefont {L.}~\bibnamefont {Zheng}}, \bibinfo {author} {\bibfnamefont
  {S.}~\bibnamefont {Liu}}, \bibinfo {author} {\bibfnamefont {B.}~\bibnamefont
  {Yan}}, \bibinfo {author} {\bibfnamefont {J.-S.}\ \bibnamefont {Hu}}, \ and\
  \bibinfo {author} {\bibfnamefont {D.-J.}\ \bibnamefont {Xue}},\ }\bibfield
  {title} {\enquote {\bibinfo {title} {Structure of amorphous selenium: Small
  ring, big controversy},}\ }\href {\doibase 10.1021/jacs.4c00219} {\bibfield
  {journal} {\bibinfo  {journal} {Journal of the American Chemical Society}\
  }\textbf {\bibinfo {volume} {146}},\ \bibinfo {pages} {6345--6351} (\bibinfo
  {year} {2024})}\BibitemShut {NoStop}%
\bibitem [{\citenamefont {Goldan}\ \emph {et~al.}(2016)\citenamefont {Goldan},
  \citenamefont {Li}, \citenamefont {Pennycook}, \citenamefont {Schneider},
  \citenamefont {Blom},\ and\ \citenamefont {Zhao}}]{10.1063/1.4962315}%
  \BibitemOpen
  \bibfield  {author} {\bibinfo {author} {\bibfnamefont {A.~H.}\ \bibnamefont
  {Goldan}}, \bibinfo {author} {\bibfnamefont {C.}~\bibnamefont {Li}}, \bibinfo
  {author} {\bibfnamefont {S.~J.}\ \bibnamefont {Pennycook}}, \bibinfo {author}
  {\bibfnamefont {J.}~\bibnamefont {Schneider}}, \bibinfo {author}
  {\bibfnamefont {A.}~\bibnamefont {Blom}}, \ and\ \bibinfo {author}
  {\bibfnamefont {W.}~\bibnamefont {Zhao}},\ }\bibfield  {title} {\enquote
  {\bibinfo {title} {{Molecular structure of vapor-deposited amorphous
  selenium}},}\ }\href {\doibase 10.1063/1.4962315} {\bibfield  {journal}
  {\bibinfo  {journal} {Journal of Applied Physics}\ }\textbf {\bibinfo
  {volume} {120}},\ \bibinfo {pages} {135101} (\bibinfo {year} {2016})},\
  \Eprint
  {http://arxiv.org/abs/https://pubs.aip.org/aip/jap/article-pdf/doi/10.1063/1.4962315/15185642/135101\_1\_online.pdf}
  {https://pubs.aip.org/aip/jap/article-pdf/doi/10.1063/1.4962315/15185642/135101\_1\_online.pdf}
  \BibitemShut {NoStop}%
\bibitem [{\citenamefont {Guo}\ and\ \citenamefont
  {Lu}(1998)}]{PhysRevB.57.10414}%
  \BibitemOpen
  \bibfield  {author} {\bibinfo {author} {\bibfnamefont {F.~Q.}\ \bibnamefont
  {Guo}}\ and\ \bibinfo {author} {\bibfnamefont {K.}~\bibnamefont {Lu}},\
  }\bibfield  {title} {\enquote {\bibinfo {title} {Microstructural evolution in
  melt-quenched amorphous se during mechanical attrition},}\ }\href {\doibase
  10.1103/PhysRevB.57.10414} {\bibfield  {journal} {\bibinfo  {journal} {Phys.
  Rev. B}\ }\textbf {\bibinfo {volume} {57}},\ \bibinfo {pages} {10414--10420}
  (\bibinfo {year} {1998})}\BibitemShut {NoStop}%
\end{thebibliography}%

\end{document}


\preprint{AIP/123-QED}

\title{Supporting Information: Quantum Monte Carlo and Density Functional Theory Study of Strain and Magnetism in 2D 1T-VSe$_2$ with Charge Density Wave States}

\author{Daniel Wines*}
\email{daniel.wines@nist.gov}
\affiliation{Material Measurement Laboratory, National Institute of Standards and Technology (NIST),
Gaithersburg, Maryland 20899, USA}

\author{Akram Ibrahim}
\affiliation{%
Department of Physics, University of Maryland Baltimore County, Baltimore, Maryland 21250, USA
}%

\author{Nishwanth Gudibandla}
\affiliation{Material Measurement Laboratory, National Institute of Standards and Technology (NIST),
Gaithersburg, MD 20899, USA}
\affiliation{%
Department of Physics, University of Maryland Baltimore County, Baltimore, Maryland 21250, USA
}%

\author{Tehseen Adel}
\affiliation{Physical Measurement Laboratory, National Institute of Standards and Technology (NIST),
Gaithersburg, Maryland 20899, USA}
\affiliation{ Department of Physical Sciences, University of Findlay, Findlay, Ohio 45840, USA}

\author{Frank M. Abel}
\affiliation{Material Measurement Laboratory, National Institute of Standards and Technology (NIST),
Gaithersburg, Maryland 20899, USA}
\affiliation{United States Naval Academy, Annapolis, Maryland 21402, USA}

\author{Sharadh Jois}
\affiliation{Laboratory for Physical Sciences,
College Park, Maryland 20740, USA}

\author{Kayahan Saritas}%

\affiliation{ 
Material Science and Technology Division, Oak Ridge National Laboratory, Oak Ridge, Tennessee 37831, USA
}%

\author{Jaron T. Krogel}%

\affiliation{ 
Material Science and Technology Division, Oak Ridge National Laboratory, Oak Ridge, Tennessee 37831, USA
}%

\author{Li Yin}%
\affiliation{ 
Department of Physics and Engineering Physics, Tulane University, New Orleans, Louisiana 70118, USA
}%

\author{Tom Berlijn}%

\affiliation{ 
Center for Nanophase Materials Sciences, Oak Ridge National Laboratory, Oak Ridge, Tennessee 37831, USA
}%

\author{Aubrey T. Hanbicki}
\affiliation{Laboratory for Physical Sciences,
College Park, Maryland 20740, USA}

\author{Gregory M. Stephen}
\affiliation{Laboratory for Physical Sciences,
College Park, Maryland 20740, USA}

\author{Adam L. Friedman}
\affiliation{Laboratory for Physical Sciences,
College Park, Maryland 20740, USA}

\author{Sergiy Krylyuk}
\affiliation{Material Measurement Laboratory, National Institute of Standards and Technology (NIST),
Gaithersburg, Maryland 20899, USA}

\author{Albert V. Davydov}
\affiliation{Material Measurement Laboratory, National Institute of Standards and Technology (NIST),
Gaithersburg, Maryland 20899, USA}

\author{Brian Donovan}
\affiliation{United States Naval Academy, Annapolis, Maryland 21402, USA}

\author{Michelle E. Jamer}
\affiliation{United States Naval Academy, Annapolis, Maryland 21402, USA}

\author{Angela R. Hight Walker}
\affiliation{Physical Measurement Laboratory, National Institute of Standards and Technology (NIST),
Gaithersburg, Maryland 20899, USA}

\author{Kamal Choudhary}
\affiliation{Material Measurement Laboratory, National Institute of Standards and Technology (NIST),
Gaithersburg, Maryland 20899, USA}

\author{Francesca Tavazza}
\affiliation{Material Measurement Laboratory, National Institute of Standards and Technology (NIST),
Gaithersburg, Maryland 20899, USA}

\author{Can Ataca*}
 \email{ataca@umbc.edu}
\affiliation{%
Department of Physics, University of Maryland Baltimore County, Baltimore, Maryland 21250, USA
}%

\date{\today}

\maketitle

\section{Additional Discussion}

\begin{figure*}
    \centering
    \includegraphics[trim={0. 0cm 0 0cm},clip,width=0.8\textwidth]{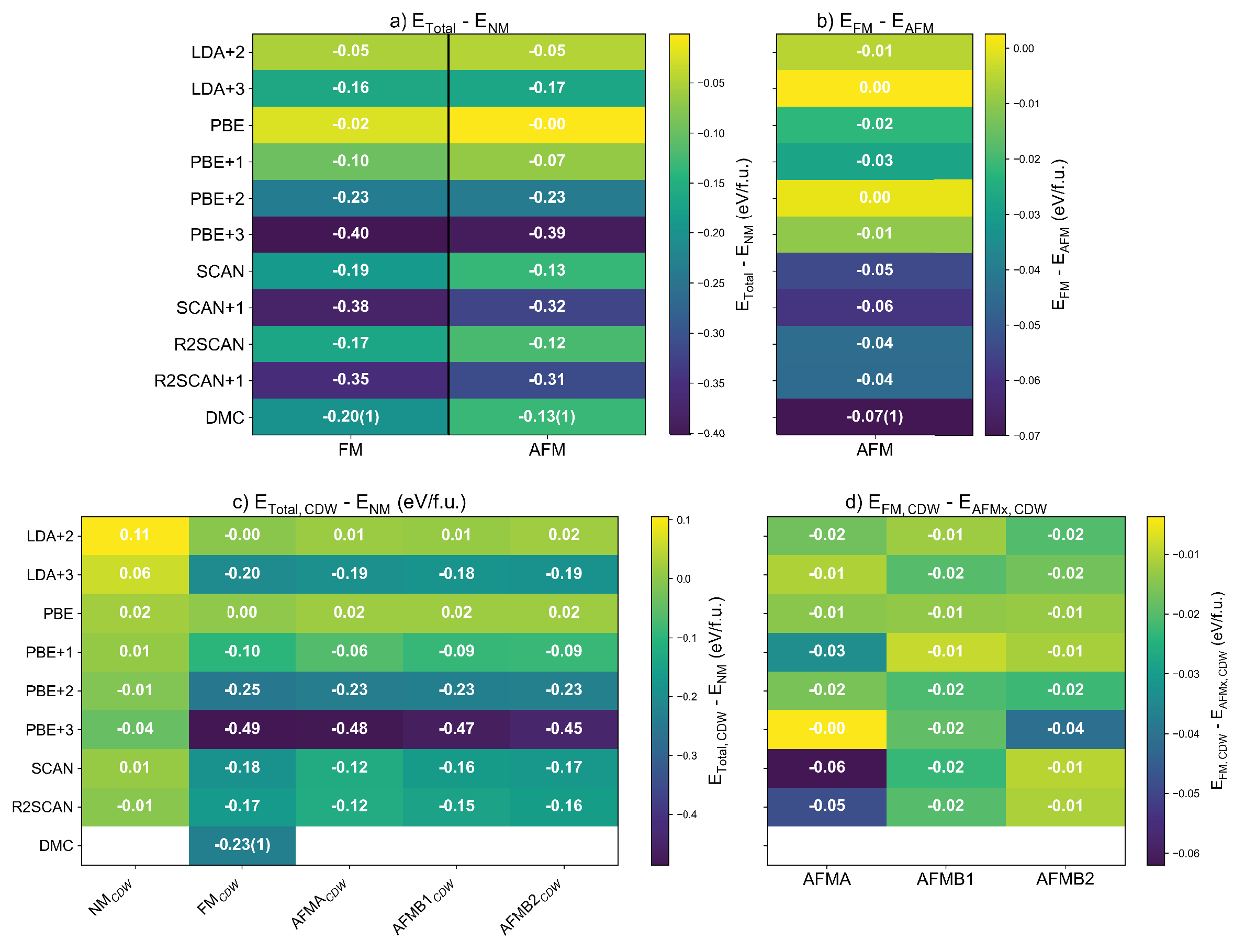}
    \caption{A depiction of DFT and DMC benchmarking calculations using various DFT functionals and Hubbard U corrections. a) depicts the total energy of the undistorted FM and AFM structures with respect to the undistorted NM structure while b) depicts the difference in energy between undistorted FM and undistorted AFM. c) depicts the total energy of the distorted (CDW) NM, FM, AFMA, AFMB1, and AFMB2 structures with respect to the undistorted NM structure while d) depicts the energy difference between distorted (CDW) FM and distorted (CDW) AFMA, AFMB1, AFMB2. DMC uncertainties (standard error of the mean) are given in shorthand form. Reported energies are per f.u. A portion of this data is presented in Table 1.}
    \label{dft-benchmark}
\end{figure*}

In addition to benchmarking the energy differences with Density Functional Theory (DFT) and Diffusion Monte Carlo (DMC) for the distorted charge density wave (CDW) structures with various magnetic orientations, we decided to study the impact that each DFT functional has on the structural distortions (initially displaced atoms). Interestingly, we found that when the atomic coordinates of some of the distorted CDW structures (with specific DFT functionals and Hubbard corrections) are relaxed, the distortions are significantly reduced and almost disappear. This is most apparent for the distorted AFM-A orientation (See Fig. 1 for atomic ordering of all discussed structues). For example, when SCAN, r$^2$SCAN, PBE+$U$ ($U$ = 1, 2, 3 eV) and LDA+$U$ ($U$ = 2, 3 eV) are used to relax the AFM-A structure, the distortions disappear completely and the structure relaxes to the normal (undistorted) AFM (stripy) structure. This implies that the stripy AFM ordering cannot exist in the distorted structure. Interestingly, for all DFT functionals considered in this work, the distortions did not disappear from the NM-CDW structure. 
In contrast, we see the distortion disappear in CDW-AFM-B1 for LDA+$U$ and PBE+$U$ ($U$ = 3 eV) and in CDW-AFM-B2 for LDA+$U$ ($U$ = 3 eV). Regarding the CDW-FM structure, we see a significant reduction of the distortion for LDA+$U$ ($U$ = 0, 3 eV) and PBE+$U$ ($U$ = 0, 2, 3 eV). The disappearance of the CDW distortion may be due in part to performing these DFT calculations at a fixed lattice constant predicted by DMC (from Ref. \onlinecite{vse2-wines}) that is different than the lattice constant computed by each respective functional, which can induce strain. The fact that the geometry predicted by SCAN/r$^2$SCAN and PBE+$U$ are closest to the DMC value could be an explanation for why the distortion does not disappear for those functionals (since minimal strain is induced). From these results, we can infer that FM in the distorted-CDW structure is highly sensitive to strain (which is discussed in the main text). This highlights the competitive nature and interplay of structural parameters, strain mechanisms and magnetic ordering that can occur within the 1T-VSe$_2$ monolayer.

For comparison sake, we analyzed the distortions in the FM-CDW structure obtained from SCAN/r$^2$SCAN (fixed to the scaled DMC lattice constant) to the distortions obtained from PBE calculations in Ref. \cite{PhysRevB.106.085117} (where the lattice constant was fixed to 3.36 \AA). These comparisons are given in Table S1, where we display the deviation in nearest neighbor bond distances (V-V) with respect to the undistorted bond distances (calculated with SCAN, r$^2$SCAN and PBE respectively) for all of the V-V bonds in the honeycomb lattice (6 bond distances for 5 different V atoms in the CDW supercell). We find that the distortions are reduced in the SCAN/r$^2$SCAN calculations when compared to the previous PBE results \cite{PhysRevB.106.085117}, which can again be due in part to fixing the lattice constant and inducing strain. Nonetheless, our DMC energies indicate that undistorted FM and CDW-FM are quite close in energy (-0.20(1) eV vs. -0.23(1) eV, with respect to the undistorted NM phase). We propose that the -0.20 eV is coming from the magnetic instability while the -0.03 eV is coming from the CDW, which is consistent with the fact that the magnetic moments (SCAN) between undistorted FM and CDW-FM are nearly identical.

\section{Theoretical Details}

The Heisenberg spin Hamiltonian of the 1T-VSe$_2$ monolayer can be written as, 
\begin{equation}
\label{eq:hamiltonian}
H = -\frac{J}{2} \sum_{i,j} \bar{S}_i \cdot \bar{S}_j - \frac{\lambda}{2} \sum_{i,j} S_i^z S_j^z - A \sum_i (S_i^{z})^2
\end{equation}
where the indices $i$ and $j$ iterate over all magnetic sites and their corresponding first nearest neighbor magnetic sites, respectively. It is important to note that the magnetic moments are assumed to be localized at the V sites.  
\\

The spin orientations are described classically in a three-dimensional spherical coordinate system as follows, 
\begin{align}
S_i^x &= S \sin \theta_i \cos \phi_i \\
S_i^y &= S \sin \theta_i \sin \phi_i \notag \\
S_i^z &= S \cos \theta_i \notag
\end{align}
where $\theta$ represents the polar angle measured from the positive 
$z$-axis $(0 \leq \theta \leq \pi)$, and $\phi$ denotes the azimuthal angle in the $xy$-plane, measured from the positive $x$-axis $(0 \leq \phi < 2\pi)$. 

\begin{align}
\begin{aligned}
\label{eq:s_z2}
\bar{S}_i \cdot \bar{S}_j &= S_i^x S_j^x + S_i^y S_j^y + S_i^z S_j^z \\
&= S^2 \left(\sin \theta_i \sin \theta_j \cos \phi_i \cos \phi_j + \sin \theta_i \sin \theta_j \sin \phi_i \sin \phi_j\right. \\
&\phantom{=} + \left.\cos \theta_i \cos \theta_j\right) \\
&= S^2 \left(\sin \theta_i \sin \theta_j \cos (\phi_i - \phi_j) + \cos \theta_i \cos \theta_j\right) \\
S_i^{z} S_j^{z} &= S^2 \cos \theta_i \cos \theta_j \\
(S_i^{z})^2 &= S^2 \cos^2 \theta_i 
\end{aligned}
\end{align}

The spin Hamiltonian can then be written as,
\begin{equation}
\begin{aligned}
\label{eq:final_hamiltonian}
H = & -\frac{J}{2} S^2 \sum_{i,j} \left(\sin \theta_i \sin \theta_j \cos (\phi_i-\phi_j) + \cos \theta_i \cos \theta_j\right) \\
& -\frac{\lambda}{2} S^2 \sum_{i,j} (\cos \theta_i \cos \theta_j) - A S^2 \sum_i (\cos \theta_i)^2
\end{aligned}
\end{equation}
The interaction parameters $J$, $\lambda$, and $A$ are obtained using the methodology of four-state energy mapping onto the Heisenberg spin Hamiltonian \cite{vsabani2020ab}. Since the on-site magnetic moment of V is closest to 1 $\mu_B$ in the distorted and undistorted cases, we take $S=1/2$.\\

For the undistorted 1T-VSe$_2$ phase, we utilize $2 \times 2 \times 1$ supercells with \(\text{FM}_x\), \(\text{FM}_z\), \(\text{AFM}_x\), and \(\text{AFM}_z\) spin configurations. Each spin in the FM configuration has $6$ parallel first nearest neighbor spins, whereas for the AFM (single stripe), each spin has $2$ parallel and $4$ anti-parallel neighbors. Hence, the energies per formula unit (f.u.) are given by

\begin{equation}
\begin{aligned}
&E_{\text{FM}_x} = -\frac{J}{2}(6S^2) \\
&E_{\text{AFM}_x} = -\frac{J}{2}(-2S^2) \\
&\Delta E_{\text{FM}} = - (-\frac{\lambda}{2} * 6 - A)S^2 \\
&\Delta E_{\text{AFM}} = - (-\frac{\lambda}{2} * -2 - A)S^2 
\end{aligned}
\end{equation}
Noting that $S=1/2$ for 1T-VSe$_2$, we get

\begin{equation}
\begin{aligned}
J &= E_{\text{AFM}_x} - E_{\text{FM}_x} \\
\lambda &= \Delta E_{\text{FM}} - \Delta E_{\text{AFM}} \\
A &= \Delta E_{\text{FM}} + 3 \Delta E_{\text{AFM}}
\end{aligned}
\end{equation}
where $\Delta E_{\text{FM}} = E_{\text{FM}_x} - E_{\text{FM}_z}$ and $\Delta E_{\text{AFM}} = E_{\text{AFM}_x} - E_{\text{AFM}_z}$.

For the CDW 1T-VSe$_2$ phase, we use the \(\text{FM}_x\) and \(\text{FM}_z\) supercells in addition to the incommensurate $\sqrt{3} \times \sqrt{7} \times 1$ supercell for the \(\text{AFM-B2}_x\) and \(\text{AFM-B2}_z\) spin configurations (AFM-B2 is the ground-state AFM spin configuration). In the AFM-B2 supercell, $2$ spins have $(6,0)$, $4$ spins have $(5,1)$, and $4$ spins have $(3,3)$ parallel and anti-parallel neighbors, respectively. Hence, the energies per formula unit (f.u.) are given by

\begin{equation}
\begin{aligned}
&E_{\text{FM}_x} = -\frac{J_{\text{cdw}}}{2}(6S^2) \\
&E_{\text{AFM-B2}_x} = \frac{1}{10}*-\frac{J_{cdw}}{2}[2(6)+4(4)]S^2 \\
&\Delta E_{\text{FM}} = - (-\frac{\lambda_{\text{cdw}}}{2} * 6 - A_{\text{cdw}})S^2 \\
&\Delta E_{\text{AFM-B2}} = - \frac{1}{10}*(-\frac{\lambda_{\text{cdw}}}{2}[2(6)+4(4)]-10A)S^2
\end{aligned}
\end{equation}

Using $S=1/2$, we get

\begin{equation}
\begin{aligned}
J_{\text{cdw}} &= \frac{5}{2}(E_{\text{AFM-B2}_x} - E_{\text{FM}_x}) \\
\lambda_{\text{cdw}} &= \frac{5}{2}(\Delta E_{\text{FM}} - \Delta E_{\text{AFM-B2}}) \\
A_{\text{cdw}} &= -\frac{7}{2} \Delta E_{\text{FM}} +\frac{15}{2} \Delta E_{\text{AFM-B2}}
\end{aligned}
\end{equation}

where $\Delta E_{\text{AFM-B2}} = E_{\text{AFM-B2}_x} - E_{\text{AFM-B2}_z}$.

\begin{table}[!ht]
    \centering
    \caption{The deviation in nearest neighbor bond distances (V-V) with respect to the undistorted bond distances (calculated with SCAN, r$^2$SCAN and PBE respectively) for all of the V-V bonds in the honeycomb lattice (6 bond distances for 5 different V atoms in the FM-CDW supercell). Units are in \AA.}
    \label{distortions}
    \begin{tabular}{|l|l|l|l|l|l|l|}
    \hline
        SCAN & a$_1$ & a$_2$ & a$_3$ & a$_4$ & a$_5$ & a$_6$ \\ \hline
        V$_1$ & 0.02 & 0.14 & -0.02 & -0.03 & -0.07 & -0.06 \\ \hline
        V$_2$ & 0.10 & -0.14 & 0.06 & 0.02 & -0.10 & 0.17 \\ \hline
        V$_3$ & -0.01 & -0.10 & -0.06 & -0.04 & 0.14 & 0.06 \\ \hline
        V$_4$ & -0.04 & -0.07 & -0.11 & 0.10 & 0.19 & -0.02 \\ \hline
        V$_5$ & -0.03 & 0.19 & 0.17 & -0.01 & -0.14 & -0.11 \\ \hline
        r$^2$SCAN & a$_1$ & a$_2$ & a$_3$ & a$_4$ & a$_5$ & a$_6$ \\ \hline
        V$_1$ & 0.02 & 0.14 & -0.02 & -0.03 & -0.07 & -0.06 \\ \hline
        V$_2$ & 0.10 & -0.13 & 0.06 & 0.02 & -0.10 & 0.17 \\ \hline
        V$_3$ & -0.01 & -0.10 & -0.06 & -0.03 & 0.14 & 0.06 \\ \hline
        V$_4$ & -0.03 & -0.07 & -0.11 & 0.10 & 0.19 & -0.02 \\ \hline
        V$_5$ & -0.03 & 0.19 & 0.17 & -0.01 & -0.13 & -0.11 \\ \hline
        PBE \cite{PhysRevB.106.085117} & a$_1$ & a$_2$ & a$_3$ & a$_4$ & a$_5$ & a$_6$ \\ \hline
        V$_1$ & 0.16 & 0.06 & 0.04 & 0.16 & -0.13 & -0.13 \\ \hline
        V$_2$ & -0.03 & -0.12 & -0.19 & 0.16 & 0.41 & -0.12 \\ \hline
        V$_3$ & -0.04 & 0.41 & -0.13 & -0.16 & 0.06 & -0.19 \\ \hline
        V$_4$ & -0.16 & -0.13 & 0.43 & -0.03 & -0.19 & 0.04 \\ \hline
        V$_5$ & 0.16 & -0.19 & -0.12 & -0.04 & -0.12 & 0.43 \\ \hline
    \end{tabular}
\end{table}

\begin{table*}[]
\caption{Total energies for undistorted structures calculated using various functionals and U values (in eV/f.u.)}
\centering
\begin{tabular}{c|c|c|c}
\hline
 & NM (eV/f.u.) & FM (eV/f.u.) & AFM (eV/f.u.) \\ \hline
\multicolumn{4}{c}{LDA} \\ \hline
U = 2 & -17.986 & -18.038 & -18.033 \\
U = 3 & -16.873 & -17.037 & -17.039 \\ \hline
\multicolumn{4}{c}{PBE} \\ \hline
U = 0 & -17.961 & -17.984 & -17.962 \\
U = 1 & -16.846 & -16.942 & -16.914 \\
U = 2 & -15.745 & -15.979 & -15.979 \\
U = 3 & -14.662 & -15.064 & -15.054 \\ \hline
\multicolumn{4}{c}{SCAN} \\ \hline
U = 0 & -59.666 & -59.853 & -59.799 \\
U = 1 & -58.564 & -58.943 & -58.884 \\ \hline
\multicolumn{4}{c}{r$^2$SCAN} \\ \hline
U = 0 & -44.931 & -45.097 & -45.052 \\
U = 1 & -43.831 & -44.186 & -44.141 \\ \hline
\end{tabular}
\label{tab:undistorted}
\end{table*}

\begin{table*}[]
\caption{Total energies for distorted-CDW structures calculated using various functionals and U values (in eV/f.u.)}
\centering
\begin{tabular}{c|c|c|c|c|c}
\hline
 & CDW-NM & CDW-FM & CDW-AFMA & CDW-AFMB1 & CDW-AFMB2 \\ \hline
\multicolumn{6}{c}{LDA} \\ \hline
U = 2 & -17.881 & -17.988 & -17.972 & -17.976 & -17.968 \\
U = 3 & -16.812 & -17.076 & -17.065 & -17.056 & -17.059 \\ \hline
\multicolumn{6}{c}{PBE} \\ \hline
U = 0 & -17.941 & -17.958 & -17.944 & -17.944 & -17.945 \\
U = 1 & -16.839 & -16.944 & -16.911 & -16.935 & -16.933 \\
U = 2 & -15.757 & -15.996 & -15.979 & -15.975 & -15.973 \\
U = 3 & -14.702 & -15.148 & -15.144 & -15.129 & -15.107 \\ \hline
\multicolumn{6}{c}{SCAN} \\ \hline
U = 0 & -59.656 & -59.850 & -59.788 & -59.827 & -59.840 \\ \hline
\multicolumn{6}{c}{r$^2$SCAN} \\ \hline
U = 0 & -44.936 & -45.099 & -45.051 & -45.079 & -45.087 \\ \hline
\end{tabular}
\label{tab:distorted_cdw}
\end{table*}

\begin{table*}[]
\caption{DMC total energies and uncertainties for FM, AFM, NM, and CDW-FM configurations (in eV/f.u.) at each supercell size (where $N$ is the number of atoms) in addition to the finite-size extrapolated values.}
\centering
\begin{tabular}{c|c|c}
\hline
 & Total Energy (eV/f.u.) & Error (eV/f.u.) \\ \hline
\multicolumn{3}{c}{FM} \\ \hline
N = 27 & -2459.906 & 0.005 \\
N = 48 & -2459.713 & 0.003 \\
Extrap. & -2459.465 & 0.007 \\ \hline
\multicolumn{3}{c}{AFM} \\ \hline
N = 36 & -2459.726 & 0.005 \\
N = 72 & -2459.560 & 0.004 \\
Extrap. & -2459.395 & 0.007 \\ \hline
\multicolumn{3}{c}{NM} \\ \hline
N = 36 & -2459.523 & 0.006 \\
N = 72 & -2459.396 & 0.005 \\
Extrap. & -2459.269 & 0.008 \\ \hline
\multicolumn{3}{c}{CDW-FM} \\ \hline
N = 30 & -2459.806 & 0.005 \\
N = 60 & -2459.653 & 0.003 \\
Extrap. & -2459.501 & 0.006 \\ \hline
\end{tabular}
\label{tab:dmc_energies}
\end{table*}

\begin{table}[h!]
\caption{Comparison of magnetic exchange and anisotropy parameters for undistorted and distorted structures calculated with DMC and SCAN. DMC uncertainty is given in parenthesis. }
\centering
\begin{tabular}{c|c|c}
\hline
Parameter & Undistorted (meV) & Distorted (meV) \\ \hline
$J$ (SCAN)      & 54                & 21              \\
$J$ (DMC)      &   70(10)              &               \\
$A$ (SCAN)      & -0.77             & 0.27            \\
$\lambda$ (SCAN) & -0.57             & -0.77           \\ \hline
\end{tabular}
\label{tab:undistorted_distorted}
\end{table}

\section{Experimental Details}

VSe$_2$ crystals were grown by the Chemical Vapor Transport (CVT) method using stoichiometric amounts of V and Se powders and a small quantity (about 4 mg/cm$^3$) of iodine transport agent. The precursors vacuum-sealed in a quartz ampoule were slowly heated to 850 $^{\circ}$C (60 $^{\circ}$C/h), kept in a 850 $^{\circ}$C - 820 $^{\circ}$C temperature gradient for 5 days and then cooled to 500 $^{\circ}$C at a 10 $^{\circ}$C/h rate followed by natural cooling to room temperature. Powder and bulk single crystal X-ray diffraction (XRD) was performed using a Bruker D8 utilizing Cu K$_\alpha$ X-rays on large bulk crystals grown in the cold zone of the ampoule confirmed the 1T phase of VSe$_2$ along with traces of selenium, primarily observed in ground powder sample. The residual selenium can be effectively removed by annealing the VSe$_2$ flakes in a dynamic vacuum at 210 $^{\circ}$C for several hours. XRD of a bulk single crystal is shown in Figure S2. Compositional and morphological analysis was done using scanning electron microscopy (SEM) and energy dispersive X-ray spectroscopy (EDXS) using a JEOL JSM-7100F field emission electron microscope. The average composition was determined to be about 34-35 at. $\%$ V and 65-66 at. $\%$ Se, corresponding to slightly vanadium rich VSe$_2$ phase for the majority of regions observed. Some other compositional regions were seen but likely represent minority phases based on XRD. Raman measurements were collected at ambient conditions using a HORIBA LabRam HR Evolution confocal microscope system using 532 nm wavelength excitation (1 mW power). The inelastic scatter was collected in a 180-degree back-scatter geometry (1800 lines/mm grating, 100 $\mu$m confocal hole) through a 100x objective (0.9 numerical aperture). 

The bulk VSe$_2$ crystals were cleaved in an Ar filled glovebox with $<$ 0.1 ppm O2 and $<$ 0.1 ppm H$_2$O using standard tape-based techniques. VSe$_2$ flakes were exfoliated on 275 nm SiO$_2$ thermal oxide wafers. The identification and encapsulation of thin flakes in h-BN was performed in the automated 2D Factory system inside the Ar glovebox. The stamps used for encapsulation consisted of a film of polypropylene carbonate (PPC) draped over poly-dimethyl siloxane (PDMS) mounted on a glass slide. An optical image of several VSe$_2$ flakes of different thickness sandwiched in h-BN on 90 nm SiO$_2$ substrate is shown in Figure S3(a). The thickness of the stack was measured by atomic force microscope (AFM) using a Nanosurf FlexAFM which is also located inside the glovebox. Figure S3(b) shows the height image of the region around the thinnest flake. The white annotated line shows the direction along which the height profile in Figure S3(c) was taken.

\begin{figure*}
    \centering
    \includegraphics[trim={0. 0cm 0 0cm},clip,width=0.6\textwidth]{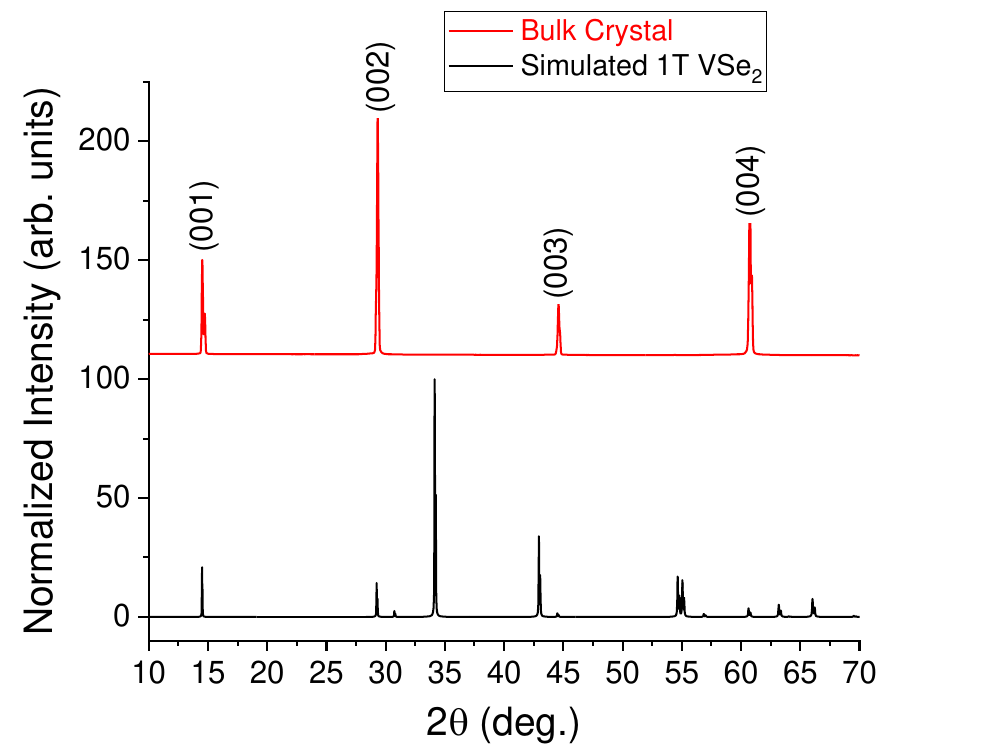}
    \caption{X-ray diffraction (XRD) of bulk crystal sample showing (00$l$) orientation with corresponding simulation of the 1T phase performed with Vesta \cite{Momma:db5098} using crystallographic information file (CIF) reported by Ref. \onlinecite{Rigoult:a21216}.}
    \label{exp-vse2-s1}
\end{figure*}

\begin{figure*}
    \centering
    \includegraphics[trim={0. 0cm 0 0cm},clip,width=0.6\textwidth]{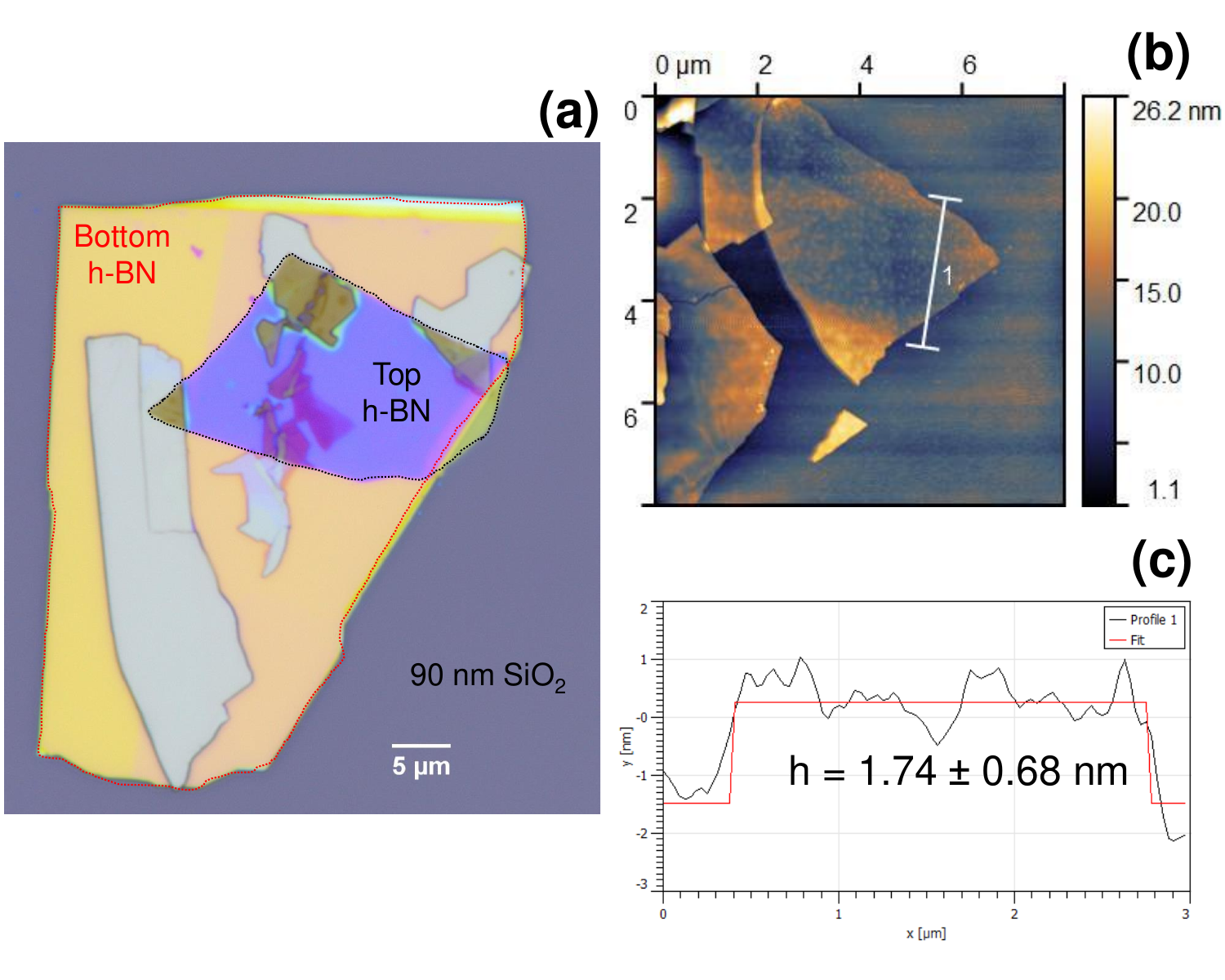}
    \caption{Optical image of h-BN encapsulated VSe$_2$ sample used for Raman measurements (a), atomic force microscopy of the thinnest region (b), and thickness measurement corresponding to annotated white line labeled with 1 in (b) (c). }
    \label{exp-vse2-s2}
\end{figure*}

\begin{figure*}
    \centering
    \includegraphics[trim={0. 0cm 0 0cm},clip,width=0.6\textwidth]{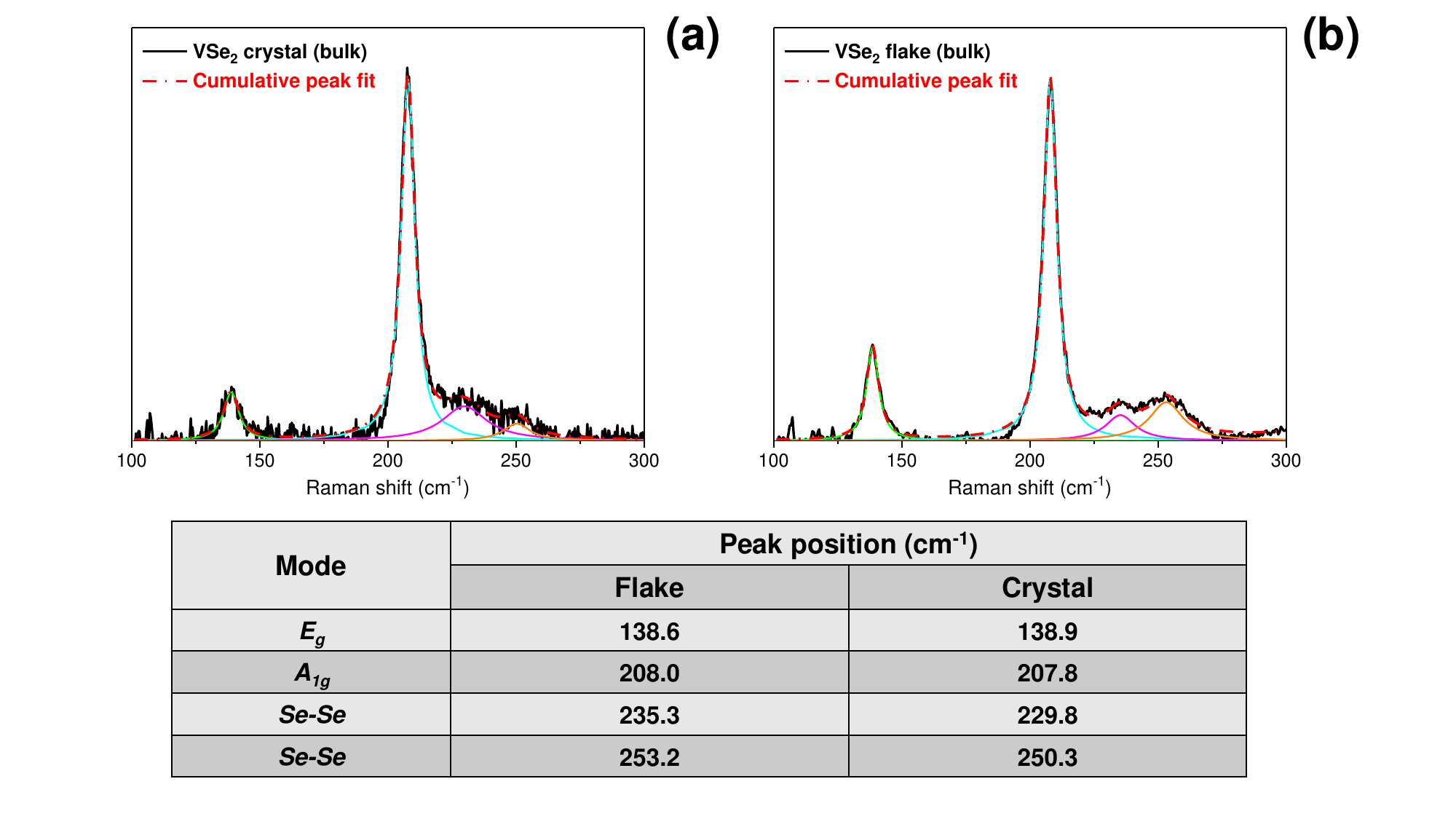}
    \caption{Fitted Raman spectra from the bulk crystal, the same crystal as used for XRD measurement (a), and from the bulk-like region encapsulated in h-BN corresponding to position 1 labeled in the main text figure (b). The peak positions of the two phonons from 1T-VSe$_2$ phase and additional two modes which are identified as selenium \cite{Se-ring,10.1063/1.4962315,PhysRevB.57.10414}.}
    \label{exp-vse2-s3}
\end{figure*}

\section*{References}
\bibliography{si}